\documentclass{article}
\usepackage{amsmath}
\usepackage{amssymb}
\usepackage{caption, subcaption}
\usepackage{color}
\usepackage{float}
\usepackage{graphicx}
\usepackage[top=1in, bottom=1in, left=1in, right=1in]{geometry}
\usepackage[section]{placeins}

\newcommand{\Dt}{\Delta t}
\newcommand{\br}[3]{\left\{#1\right\}_{#2}^{#3}}

\newcommand{\pr}[1]{\mathbb{P}\left(#1\right)}

\newcommand{\ind}[1]{\mathbb{I}_{\{#1\}}}
\newcommand{\cl}[1]{\lceil #1 \rceil}
\newcommand{\fl}[1]{\lfloor #1 \rfloor}
\newcommand{\eq}[1]{Equation~\eqref{eq:#1}}
\newcommand{\ex}{\mathbb{E}}

\newcommand{\spc}{\mbox{ }}
\newcommand{\Binom}{\mbox{BinCDF}}

\title{A Hierarchy of Linear Threshold Models for the Spread of Political Revolutions on Social Networks}
\date{\today}
\author{J.C. Lang and H. De Sterck}

\begin{document}
\maketitle

\begin{abstract}
We study a linear threshold agent-based model (ABM) for the spread of political revolutions on social networks using empirical network data. We propose new techniques for building a hierarchy of simplified ordinary differential equation (ODE) based models that aim to capture essential features of the ABM, including effects of the actual networks, and give insight in the parameter regime transitions of the ABM. We relate the ABM and the hierarchy of models to a population-level compartmental ODE model 
that we proposed previously for the spread of political revolutions \cite{LangDeSterck14}, 
which is shown to be mathematically consistent with the proposed ABM and provides a way to analyze the global behaviour of the ABM.
This consistency with the linear threshold ABM also provides further justification a posteriori for the compartmental model of \cite{LangDeSterck14}. Extending concepts from epidemiological modelling, we define a basic reproduction number $R_0$ for the linear threshold ABM and apply it to predict ABM behaviour on empirical networks. In small-scale numerical tests we investigate experimentally the differences in spreading behaviour that occur under the linear threshold ABM model when applied to some empirical online and offline social networks, searching for quantitative evidence that political revolutions may be facilitated by the modern online social networks of social media.
\end{abstract}

\section{Introduction}
\label{Sec:Intro}
In this paper, we propose a linear threshold agent-based model (ABM) for the spread of political revolutions on social networks in dictatorial regimes. In our ABM, nodes of the network can be in two states: active, i.e., participating in the revolution, or inactive. Transitions from the inactive to the active state are governed by a growth process that uses a traditional linear threshold mechanism: an individual $v$ may change from the inactive to the active state if the fraction of neighbours in the active state exceeds the linear threshold $\theta$. 

Linear threshold models have been studied before in the context of influence maximization \cite{KempeKleinbergTardos03}, enforcement of unpopular norms \cite{CentolaEtAl05}, and general ``complex contagion'' processes \cite{CentolaEguiluzMacy07, CentolaMacy07, Watts02}. The development and analysis of linear threshold processes has been highly influenced by more established epidemiology \cite{AmesEtAl11, Bansal07, Hethcote00} and rumour spreading \cite{LindEtAl07, ZhaoEtAl11} models, with one key difference: whereas a contact with a single ``infected'' individual is enough to spread a contagion in most epidemiology or rumour spreading applications, i.e. in ``simple contagion'' processes, multiple contacts with infected individuals are generally necessary to spread a contagion in a linear threshold process. Indeed, epidemiological and rumour spreading models can be considered to be a special case of linear threshold models where the threshold parameter is chosen sufficiently small such that one infected neighbour enables propagation. Most of the linear threshold models that have been studied in the context of social spreading or contagion processes \cite{CentolaEguiluzMacy07, CentolaMacy07, CentolaEtAl05, KempeKleinbergTardos03, Watts02, Granovetter78} consider evolution in discrete time. In contrast, our ABM is chosen to evolve in continuous time since that facilitates the comparison of the ABM with simplified population-level models, which is a major goal of this paper.

The main justification for our linear threshold modelling approach in the context of political revolutions is that the decision to join a political revolution can be assumed to be a collective action problem \cite{Kuran91}: if individuals act unilaterally they are subject to retaliation by the regime, whereas if they act collectively then the regime loses the ability to punish due to a lack of resources. The linear threshold modelling approach is consistent with the collective action principle, since an individual transitions from an inactive to an active state only after the number of neighbours in the active state has been observed to reach the critical fraction $\theta$, and the individual deems the revolution of sufficient size to consider participation. In order to incorporate communication network structure explicitly, we represent interactions between individuals $v$ by edges $e$ in an undirected graph $G=G(V,E)$. This contrasts with previous attempts to model political revolutions using linear threshold models that assume homogeneous mixing \cite{Granovetter78, Kuran91}, i.e. that assume that $G=G(V,E)$ is the complete graph. 

The goals and contributions of this paper are as follows. 
We start by presenting the linear threshold ABM for the spread of political revolutions on social networks, and then derive a hierarchy of simplified ordinary differential-equation (ODE) models of varying degree of sophistication that characterize the solutions of the linear threshold ABM, see Table~\ref{tab:hierarchy}. 
Although the linear threshold ABM allows us to explicitly model the effects of the communication network on the dynamics of the political revolution it models, the inherent complexity of modelling the state of all individuals simultaneously makes this approach difficult to analyze and expensive to simulate.
For these reasons there is significant interest in simplified aggregate or population-level dynamical models, which can provide more cost-effective dynamical simulations, and can give insight in the qualitative dynamics of the ABM and its parameter regimes, through dynamical analysis of the simplified model. It is a significant challenge to incorporate actual network structure in the ODE models, and this is often important since details of the network structure may determine the qualitative behaviour of ABM solutions as the ABM parameters are changed. In this paper, we present two new effective ways to incorporate network structure into the one-compartmental ODE that approximates the dynamical evolution of the expected fraction of the population that participates in the revolution in the linear threshold ABM model. These approaches make use of the degree distribution of the graph or samplings of the network, and we call the resulting population-level ODE models the \emph{binomial visibility function} (BVF) and the \emph{empirical visibility function} (EVF) models.
The EVF model is significantly less expensive than the BVF, but we demonstrate its equivalence with the BVF in the limit of large network and sample size. Extensive numerical tests show that these ODE models perform remarkably well to predict qualitative behaviour of the ABM; in fact, they often perform as well as or better than a previously specified higher-order model that has many compartments and is much more expensive computationally \cite{NekoveeEtAl07}, and that we refer to as the \emph{degree approximation} (DA) model. 

\begin{table}[hp]
	\centering
	\caption{Summary of hierarchy of models from highest (network-level) to lowest (population-level) complexity.}
	\begin{tabular}{llllc}
		\hline
		Model  	& Abbreviation	& Section	& Equation(s)		& Reference(s) \\ \hline
		Linear Threshold Agent-Based Model	& ABM 	& \ref{Sec:Spec_ABM}		& \eqref{eq:0to1} and \eqref{eq:1to0}		& 
		\\
		Degree Approximation Model		& DA		& \ref{Sec:HOT}		& \eqref{eq:ApproxDA}				& \cite{NekoveeEtAl07}\\
		Binomial Visibility Function Model	& BVF	& \ref{Sec:BVF}		& \eqref{eq:Approx} and \eqref{eq:BVF}	&  \\
		Empirical Visibility Function Model	& EVF	& \ref{Sec:EVF}		& \eqref{eq:Approx} and \eqref{eq:EVF}	& \\
		Step Visibility Function Model		& SVF	& \ref{Sec:SVF}		& \eqref{eq:SVF}					& \cite{LangDeSterck14}\\ \hline
	\end{tabular}
	\label{tab:hierarchy}
\end{table}

Similar to the population-level one-compartmental ODE model that we proposed in \cite{LangDeSterck14} for the spread of a political revolution, the linear threshold ABM proposed in this paper applies to dictatorial regimes that employ censorship and police repression.
We make the simplifying assumption that the population is uniform in its dislike of the regime (such that all nodes can potentially become active in the revolution), and we assume that the regime censors the communication between individuals except for the links in the social network. A decay process models police repression that is limited by the regime's finite police capacity. The model features parameters that describe timescales of growth and decay, a linear threshold parameter for the growth process, and a parameter for the finite police capacity. The dynamics of our ABM is governed by stochastic transitions in continuous time, because this allows us to relate the ABM mathematically to simplifying time-dependent ODE models in limits of expectation over the aggregate population. 
These simplified ODE-based models aim to capture essential features of the linear threshold ABM at the population level, including effects of the actual networks, and give insight in the parameter regime transitions of the ABM.

We also relate the ABM to the population-level one-compartmental ODE model for the spread of political revolutions that was proposed in \cite{LangDeSterck14} and that indirectly takes network structure into account by a single parameter that describes from which fraction of participants the growth term is switched on discontinuously. As such, it is a precursor of the BVF/EVF models introduced in this paper. In \cite{LangDeSterck14} this single parameter was called the visibility parameter of the revolution, with the interpretation that people will only join the revolution when the revolution is of sufficiently large size to be visible to the population, while repressive regimes attempt to make unrest invisible through censorship. We call the model from \cite{LangDeSterck14} the \emph{step-visibility function} (SVF) model in this paper. This ODE model was justified in \cite{LangDeSterck14} by a simplified network model with significant further assumptions; in this paper, we show that our ABM is mathematically consistent with the model from \cite{LangDeSterck14} and we test the assumptions from \cite{LangDeSterck14} with the BVF/EVF models and the ABM model on real networks, corroborating a posteriori the assumptions from \cite{LangDeSterck14}, and showing that the model from \cite{LangDeSterck14} reproduces the qualitative behaviour of the ABM and the spread of a revolution under a linear threshold model. This provides further justification for the ODE model of \cite{LangDeSterck14}, and at the same time provides inexpensive ways to predict the parameter regime behaviour of the ABM presented in this paper using simple compartmental models like the SVF and the new BVF/EVF models. We emphasize that the broad applicability of the linear threshold process implies that the framework, methods, and results presented in this paper are relevant to the study of many related social spreading processes.

The final contribution of this paper relates to an interesting application of our linear threshold ABM. We investigate experimentally the differences in spreading behaviour that occur under the linear threshold model when applied to some empirical online and offline social networks, searching for quantitative evidence that political revolutions may be facilitated by the online social networks of social media.
Indeed, it is often assumed that the connectivity of modern online social networks has greatly facilitated the spread of political revolutions in the past decade, e.g., in the Arab Spring revolutions of 2011 \cite{eltantawy2011arab,khondker2011role,tufekci2012social,
KhamisVaughn11, 
LangDeSterck14}, while the traditional offline social networks of the pre-Internet era (using in-person physical contact, or mail or phone interaction for safe communication) had a different connectivity structure that was often severely restricted by regime censorship.
As a starting point to investigate differences in propagation properties that may arise between online and offline social networks within the linear threshold propagation model, we investigate propagation on two empirical networks of modest size using our linear threshold ABM: we consider a small Facebook network \cite{McAuleyLeskovec12} as a representative of an online social network, and a small physical contact network between individuals \cite{SalatheEtAl10} as a representative of an offline social network. These two networks are also used for all validation and comparison simulations for the ABM, DA, BVF/EVF and SVF models throughout the paper. We also extend the concept of basic reproduction number $R_0$ from epidemiological modelling \cite{Hethcote00} to characterize the potential of networks to spread the revolution. The basic reproduction number for our ABM with a linear threshold process is easy to compute, and we show that it gives useful predictions. Our results indicate that the offline social network indeed is less conducive to spreading the revolution than the online social network: it has a smaller basic reproduction number, and in simulations a larger initial population of revolutionaries is required to spread the revolution. This provides some initial quantitative evidence that the spread of revolutions under a linear threshold process may occur more easily on modern online social networks than on traditional offline networks, but we also comment on limitations in our approach and further investigations that are required to address this intriguing but complex question more comprehensively.

\begin{figure}[h]
	\centering
	\spc\hfill
	\begin{subfigure}[t]{0.4\linewidth}
		\includegraphics[width=\linewidth]{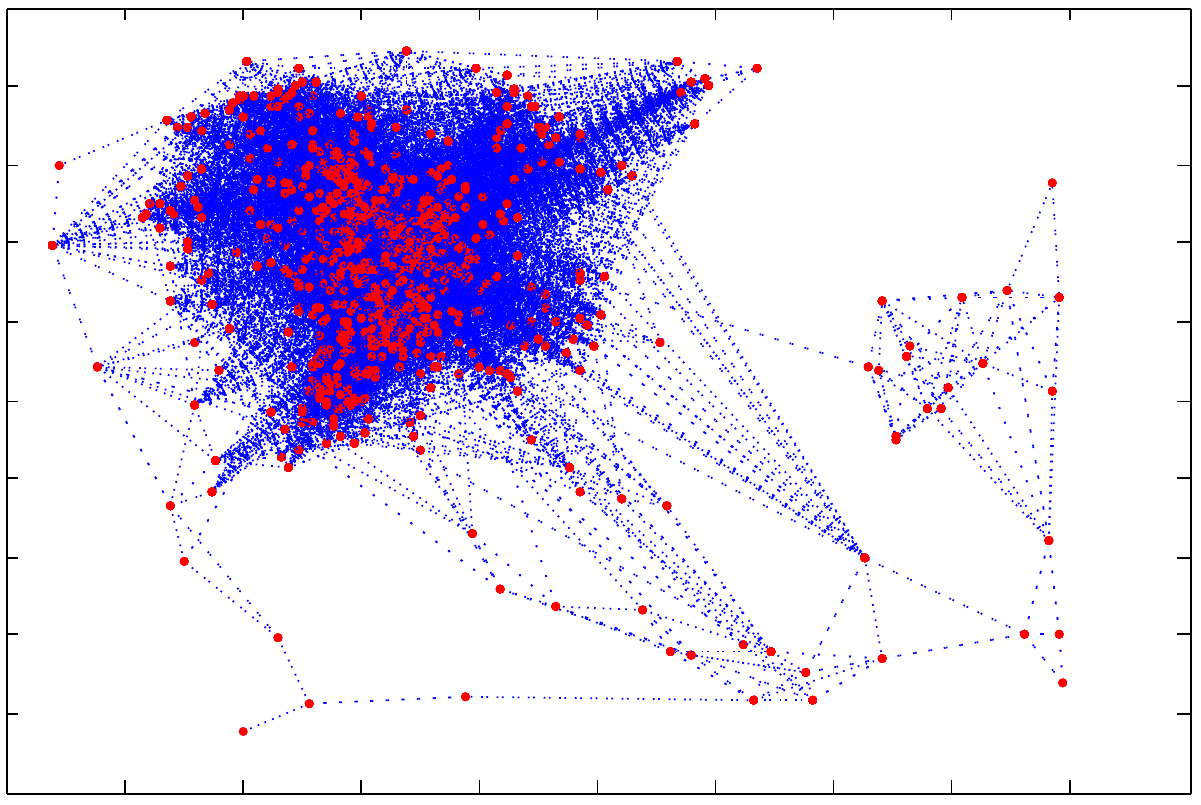}
		\caption{Physical contact network ($G_P$)}
		\label{fig:1118_vis}
	\end{subfigure}
	\hfill
	\begin{subfigure}[t]{0.4\linewidth}
		\includegraphics[width=\linewidth]{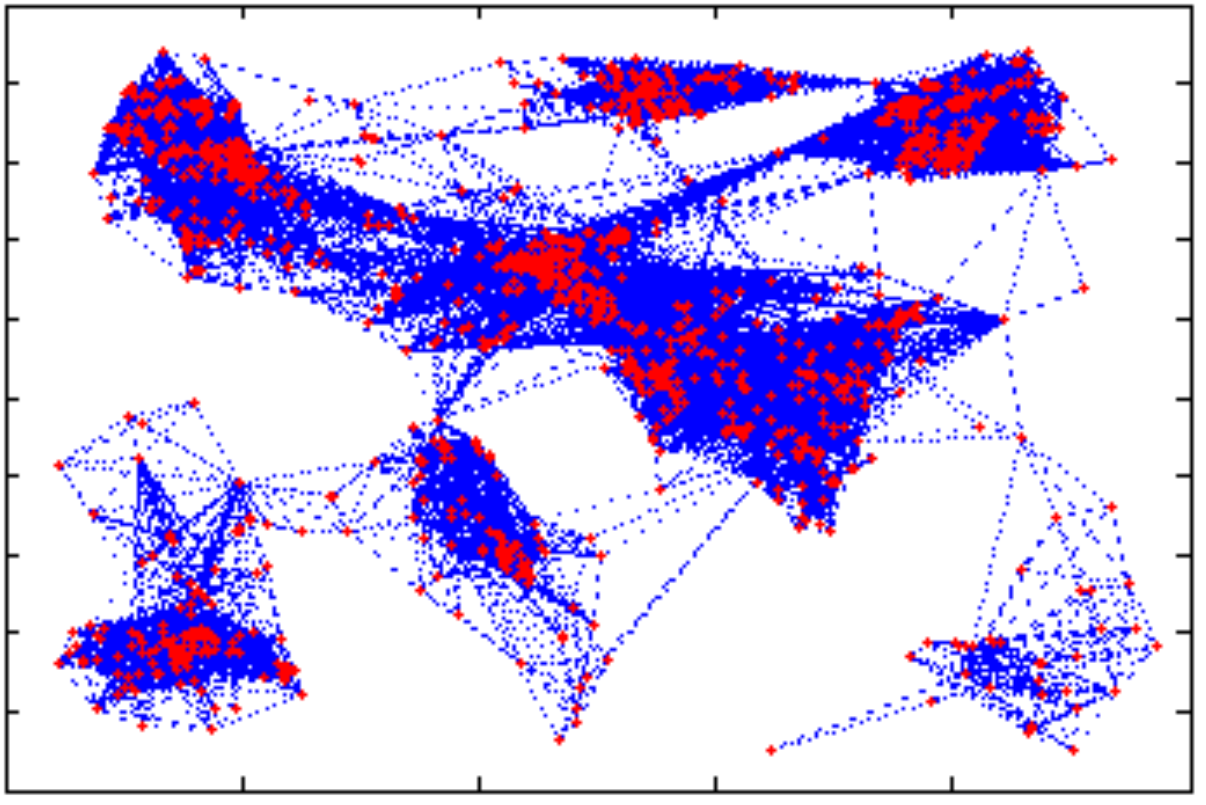}
		\caption{Facebook social network ($G_F$)}
		\label{fig:FB_vis}		
	\end{subfigure}
	\hfill\spc
	\caption{Visualization of physical contact ($G_P$) and Facebook ($G_F$) social networks using the Gursoy-Atun algorithm \cite{gursoy} as implemented in the Matlab BGL package.}
	\label{fig:vis}
\end{figure}

The rest of this paper is structured as follows.
Section \ref{Sec:NewMed_Proxy} describes the Facebook and physical contact empirical networks on which we test our ABM and approximations.
Section \ref{Sec:ABM} specifies the linear threshold ABM in detail and briefly reviews the SVF model presented in \cite{LangDeSterck14}. Section \ref{Sec:ABM} then establishes the consistency between these two approaches through mathematical derivation and numerical simulation on the networks described in Section \ref{Sec:NewMed_Proxy}, i.e., we show that the SVF model can be considered to approximate the aggregate behaviour of the ABM. 
Section \ref{Sec:AVF} describes the BVF and EVF methods to improve the SVF model by changing the functional form of the visibility function, explicitly incorporating the structure of the network without increasing the dimensionality of the SVF model. 
Section \ref{Sec:HOT} confirms the usefulness of the BVF and EVF models by showing that they perform no worse than the higher order degree approximation model of \cite{NekoveeEtAl07}, while being much less expensive to evaluate.
Section \ref{Sec:R0} extends the concept of basic reproduction number from epidemiology \cite{Hethcote00} to the linear threshold process specified in Section \ref{Sec:ABM} and illustrates how it can be interpreted in terms of the BVF/EVF.
Finally, Section \ref{Sec:Interp} explores how our methods can be used to study
the differences in spreading behaviour of political revolutions on online and offline social networks
under the linear threshold mechanism.

\section{Network Data}
\label{Sec:NewMed_Proxy}


In this section we describe the two empirical social networks that will be used in this paper to validate and compare the ABM, DA, BVF/EVF and SVF models for the spread of political revolutions on social networks under the linear threshold model.

The first network we consider is the physical contact network presented in \cite{SalatheEtAl10}. It was constructed by distributing wireless sensors to students, teachers, and staff at a U.S. high school during a one day period. When two wireless sensors are in proximity of one another, i.e. when they are less than approximately 3m apart, they register an interaction with a temporal resolution of 20s. Therefore, the communication network we extract from this data is referred to as the \emph{physical contact network $G_P$}. 
The second network we consider is the Facebook subnetwork presented in \cite{McAuleyLeskovec12}. It was constructed by combining the \emph{ego-networks}\footnote{For a network $G=G(V,E)$ the \emph{ego-network} of an individual $v\in V$ is a subnetwork of the overall network composed of individual $v$ (called the \emph{ego-node}), the neighbours of individual $v$, and all of the connections between these individuals.} of ten individuals and then taking the largest connected component of the resulting network. We refer to this network as the \emph{Facebook social network} $G_F$. For details on the network extraction protocol for $G_P$ we refer to Appendix \ref{Sec:ext_net}. To facilitate the comparison of experimental results, the minimal contact duration to register an edge in $G_P$ is chosen such that $G_P$ and $G_F$ have approximately the same average degree, see Appendix \ref{Sec:ext_net}.

\begin{figure}[h]
	\centering
	\includegraphics[width=0.48\linewidth]{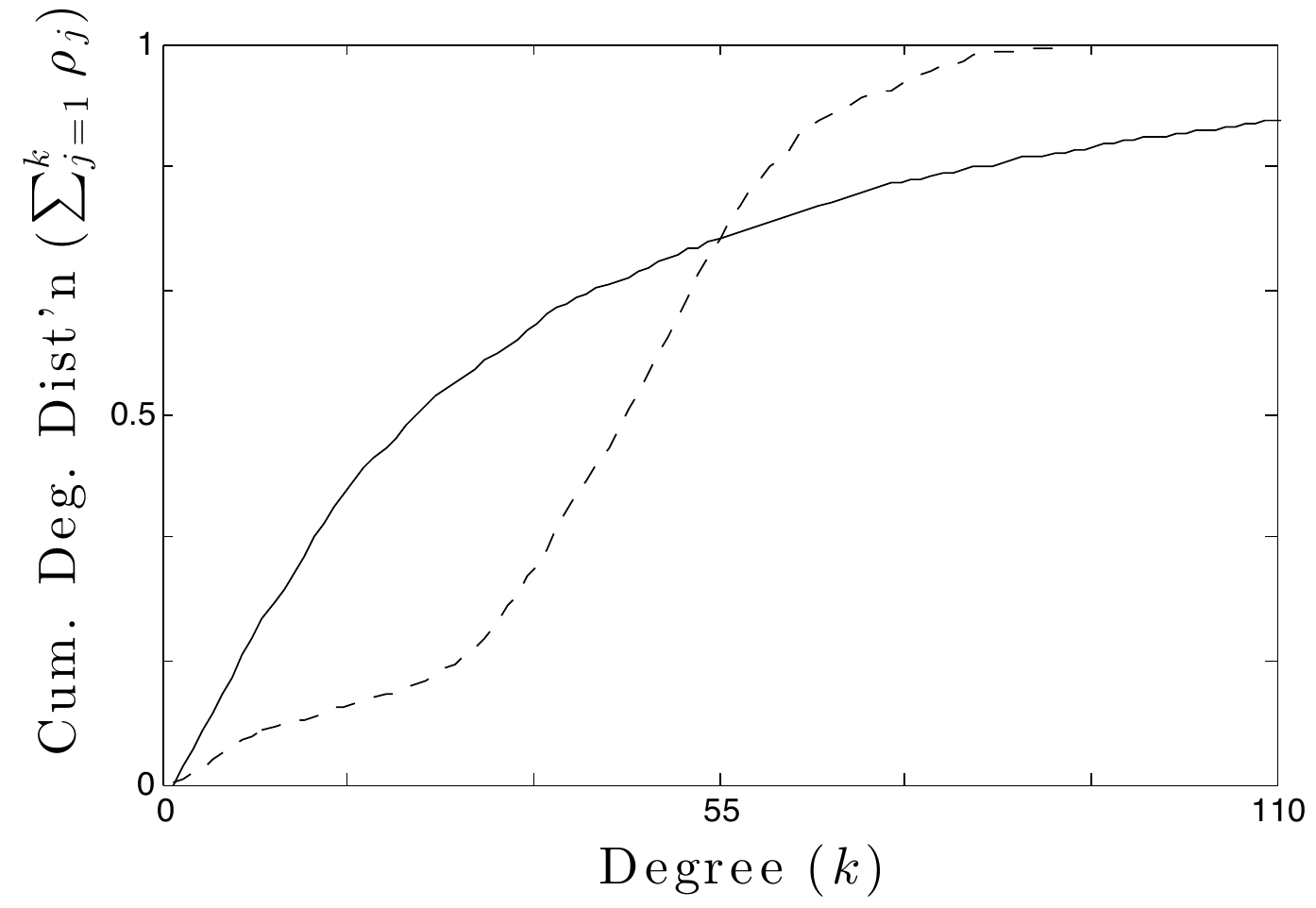}
	\caption{Cumulative degree distribution of physical contact (dashed) and Facebook (solid) social networks.}
	\label{fig:deg_distn}
\end{figure}

\begin{figure}[h]
	\centering
	\spc\hfill
	\begin{subfigure}[t]{0.4\linewidth}
		\includegraphics[width=\linewidth]{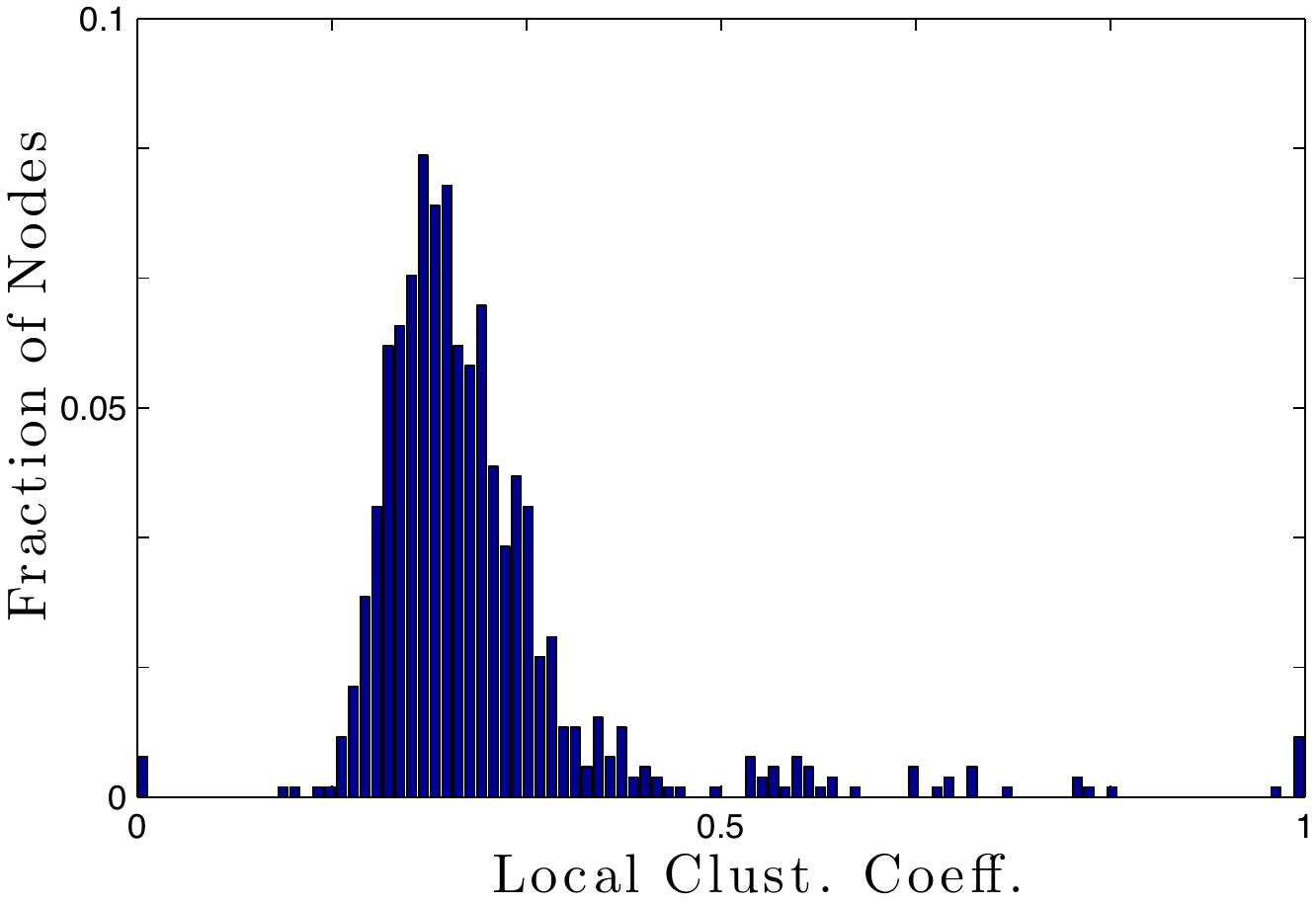}
		\caption{Physical contact  network ($G_P)$}
		\label{fig:1118_clust_coeff}
	\end{subfigure}
	\hfill
	\begin{subfigure}[t]{0.4\linewidth}
		\includegraphics[width=\linewidth]{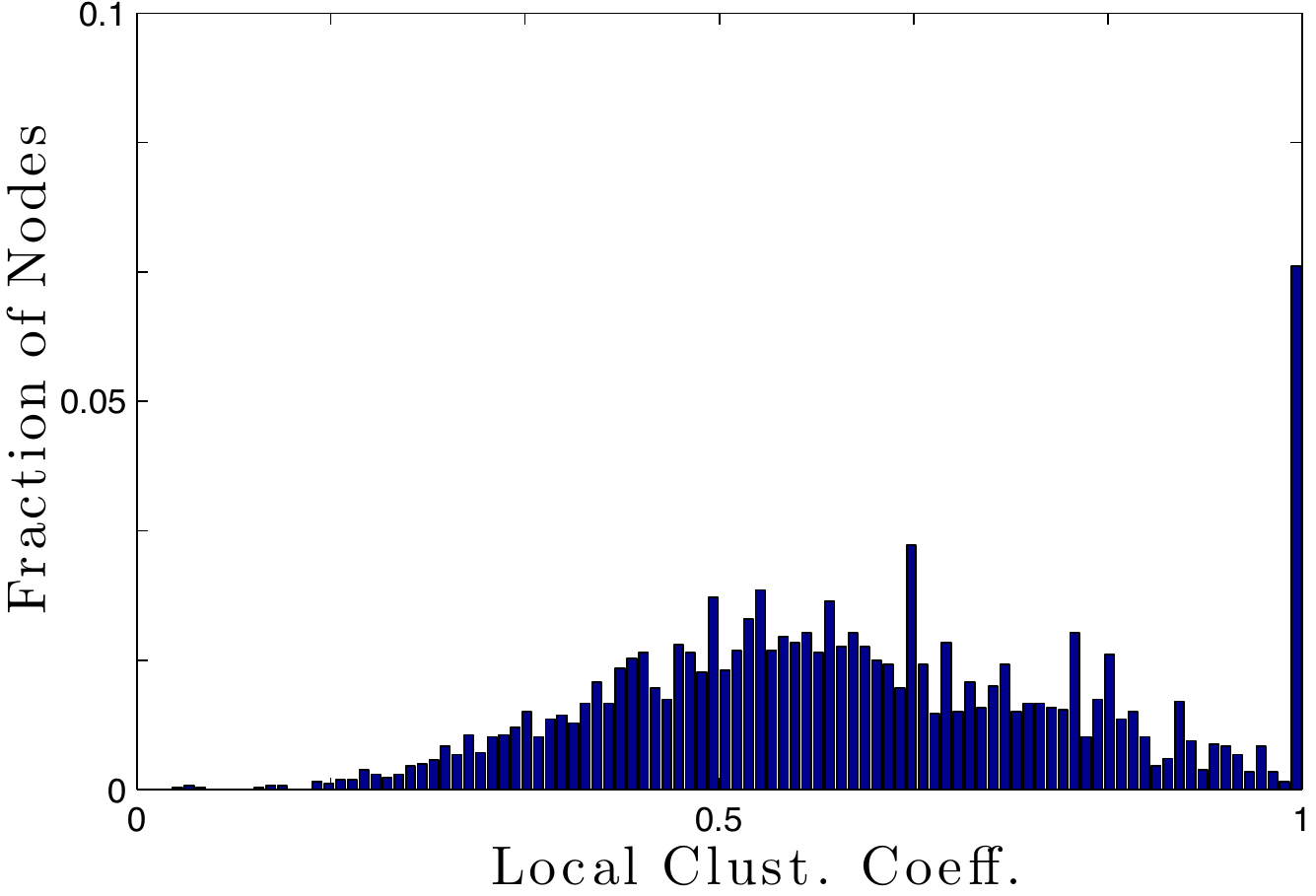}
		\caption{Facebook social network ($G_F$)}
		\label{fig:FB_clust_coeff}		
	\end{subfigure}
	\hfill\spc
	\caption{Distribution of local clustering coefficient for physical contact ($G_P$) and Facebook social ($G_F$)  networks.}
	\label{fig:clust_coeff}
\end{figure}

The physical contact network $G_P$ has $N=776$ nodes with sparsity\footnote{The Sparsity of a network is defined as the fraction of possible edges that are present in the network.} $S=0.06$, and is visualized in Fig.~\ref{fig:1118_vis}. 
The Facebook network $G_F$ has $N=3,963$ nodes with sparsity $S=0.01$, and is visualized in Fig.~\ref{fig:FB_vis}.
The cumulative degree distributions\footnote{If $\rho_k$ is the fraction of nodes with degree $k$ (i.e., the degree distribution) of the graph, then the cumulative degree distribution is the function $F(k) = \sum_{j=1}^k \rho_j$. The cumulative degree distribution is displayed in place of the degree distribution because the degree distribution is subject to significantly more noise than the cumulative degree distribution. This is a common approach for empirical networks.} 
are displayed in Fig.~\ref{fig:deg_distn}. The distributions of local clustering coefficients are displayed in  Fig.~\ref{fig:clust_coeff}.

\begin{table}[h]
	\centering
	\caption{Summary of common network measures for physical contact and Facebook social networks.}
	\begin{tabular}{lcc}
		\hline
		Quantity  & Physical Contact Network & Facebook Social Network \\ \hline
		Number of Nodes ($N$)							& 776			& 3,963 \\
		Average Degree ($\mu_k\pm\sigma_k$)			& $44.8\pm18.2$	& $44.5\pm52.4$\\
		Minimum/Maximum Degree	 ($k_{\min}/k_{\max}$)	& 1/109			& 2/1034\\
		Sparsity ($S$)								& 0.06			& 0.01\\
		Diameter ($D$)								& 8				& 8\\
		Average Path Length($\mu_l\pm\sigma_l$)			& $2.37\pm0.82$	& $3.77\pm1.29$\\
		Acerage Local Clust. Coeff. ($\mu_c\pm\sigma_c$)	& $0.29\pm0.12$	& $0.62\pm0.20$\\ \hline
	\end{tabular}
	\label{tab:sum_stats}
\end{table}

It is interesting to consider how the physical contact and Facebook networks differ on some common network measures, see Table \ref{tab:sum_stats}.
Comparing the physical contact and Facebook social networks along these admittedly limited measures nevertheless highlights substantial structural differences between these networks. The physical contact network appears more homogeneous than the Facebook social network, in the sense that the Facebook social network can be visually grouped into distinct communities, whereas the physical contact network cannot (Fig.~\ref{fig:vis}). The network measures presented in Figs.~\ref{fig:deg_distn}-\ref{fig:clust_coeff} and Table~\ref{tab:sum_stats} are generally supportive of this observation. For example, the cumulative degree distributions displayed in Fig.~\ref{fig:deg_distn} show that the physical contact network has an approximately normal distribution with a relatively small standard deviation, and hence, a thin tail. In contrast, the Facebook social network has a much broader distribution (possibly scale-free or exponential, although this cannot be determined conclusively with such a small network size) with a relatively large standard deviation, and hence, a fat tail. Similarly, it is known that online social networks often have larger clustering coefficients than offline social networks. In Section \ref{Sec:Interp} we use the Facebook and physical contact networks as representatives of online and offline social networks, respectively, to investigate differences in propagation properties that may arise between online and offline social networks within the linear threshold propagation model.

\section{Agent-Based Model}
\label{Sec:ABM}
\subsection{Specification of the Agent-Based Model}
\label{Sec:Spec_ABM}
Consider a population of individuals who are represented by nodes $V = \{v_i\}_{i=1}^N$, and whose interactions are represented by edges $E = \{e_i\}_{i=1}^M$, of the graph $G=G(V,E)$ with degree distribution $\rho_k$. 
An individual $v\in V$ at time $t$ can be in one of two states, i.e. $s_v(t) = 0$ (\emph{inactive} in the revolution) or $s_v(t) = 1$ (\emph{active} in the revolution).
We assume that the network $G$ is static so that the dynamics of the ABM can be fully specified by providing rules for the transition of individuals from an inactive state to an active state and vice versa, together with parameter values and an initial condition. For the moment, we set aside the issue of the choice of parameters and initial conditions and restrict the remainder of this section to the specification and justification of the transition rules.

\subsubsection{Growth Process: Inactive ($s_v=0$) to Active ($s_v=1$)}
\label{Sec:0to1}
To specify the growth process of the ABM, we use the standard linear threshold model that has been used before in the context of social spreading or contagion processes such as opinion formation, technology adoption, marketing, rioting, and political movements \cite{CentolaEguiluzMacy07, CentolaMacy07, CentolaEtAl05, KempeKleinbergTardos03, Watts02, Granovetter78}.
As mentioned in Section \ref{Sec:Intro}, the choice of individuals to join a revolution is a collective action problem \cite{Kuran91}: individuals are averse to unilateral action against the regime for fear of severe retaliation, but are willing to take action collectively in the belief that the regime will lack sufficient resources to punish the entire collective. Thus, an individual $v\in V$ will decide to join the revolution if he or she believes that it has grown sufficiently large to reduce the risk of retaliation from the regime to an acceptable level. 
It is reasonable, therefore, to assume that individuals will decide to join the revolution only after a large enough fraction of their neighbours in the social network have done so. This behaviour is captured by the following transition rule: if individual $v\in V$ has $k$ neighbours $\{w_j\}_{j=1}^k$, if $s_v(t) = 0$, and if
\begin{equation}
	\sum_{j=1}^k s_{w_j}(t) \geq \theta_v k, \label{eq:0to1}
\end{equation}
i.e., if $v$ is inactive at time $t$ and has at least a fraction $\theta_v$ of its neighbours that are active, then node $v$ transitions from state 0 to state 1 at time $t' = t + \xi_1$, where $\xi_1>0$ is the first arrival time of a Poisson process with rate $c_1$. 
We say that nodes that satisfy \eq{0to1} can ``see'' the revolution, i.e. the revolution is visible to them. Alternatively, we say that these nodes are ``considering joining the revolution''. 
Parameter $c_1$ determines the timescale of the growth process.

While most of the linear threshold models that have been studied in the context of social spreading or contagion processes \cite{CentolaEguiluzMacy07, CentolaMacy07, CentolaEtAl05, KempeKleinbergTardos03, Watts02, Granovetter78} consider evolution in discrete time by choosing $\xi_1$ constant, we have specified $\xi_1$ so that \eq{0to1} evolves in continuous time. Specifying $\xi_1$ in this way is often considered in biology and epidemiology \cite{AmesEtAl11, Bansal07}, since it (a) facilitates the comparison of the ABM with population-level models (as the SVF model described in Section \ref{Sec:consist_ABM_SVF}), and (b) eliminates the problem of choosing a suitable time-step for iterating the discrete-time process. Furthermore, we note that by choosing $\xi_1$ to be the first arrival time of a Poisson process, we are assuming that the decision making process is a Markov, or memoryless, process. Specifically, conditioned on the state of the system at time $t$, we assume that the likelihood of an individual joining the revolution in the time interval $[t, t+\Dt]$ is independent of the state of the system at any time $\tilde{t}<t$. In words, the future is independent of the past, given the present. 

\subsubsection{Decay Process: Active ($s_v=1$) to Inactive ($s_v=0$)}
We assume that once an individual $v\in V$ has become active the regime will attempt to arrest or disperse him or her, thus returning $v$ to an inactive state. As in the previous section, we assume that this is a memoryless process. We further assume that the regime can only arrest or disperse protesters so long as the total fraction of active protesters remains less than the regime's finite police capacity $\beta\in(0,1)$. The transition rule can then be characterized by the following: if $s_v(t) = 1 $ and
\begin{equation}
	\label{eq:1to0}
	\frac{1}{N}\sum_{w\in V} s_w(t) < \beta,
\end{equation}
i.e., if $v$ is active at time $t$ and the fraction of active individuals is less than the regime's police capacity $\beta$, then the node $v$ transitions from state 1 to state 0 at time $t'=t+\xi_2$, where $\xi_2>0$ is the first arrival time of a Poisson process with rate $c_2$,
which determines the timescale of the decay process.
This fully specifies the evolution of the ABM, which we simulate using Gillespie's algorithm, see Appendix \ref{Sec:Gillespie}.

\subsection{Brief Review of the SVF Population-Level Compartmental Model of \cite{LangDeSterck14}}
\label{Sec:SVF}
%
\begin{figure}[h]
	\centering
	\begin{subfigure}[t]{0.48\linewidth}
		\includegraphics[width=\linewidth]{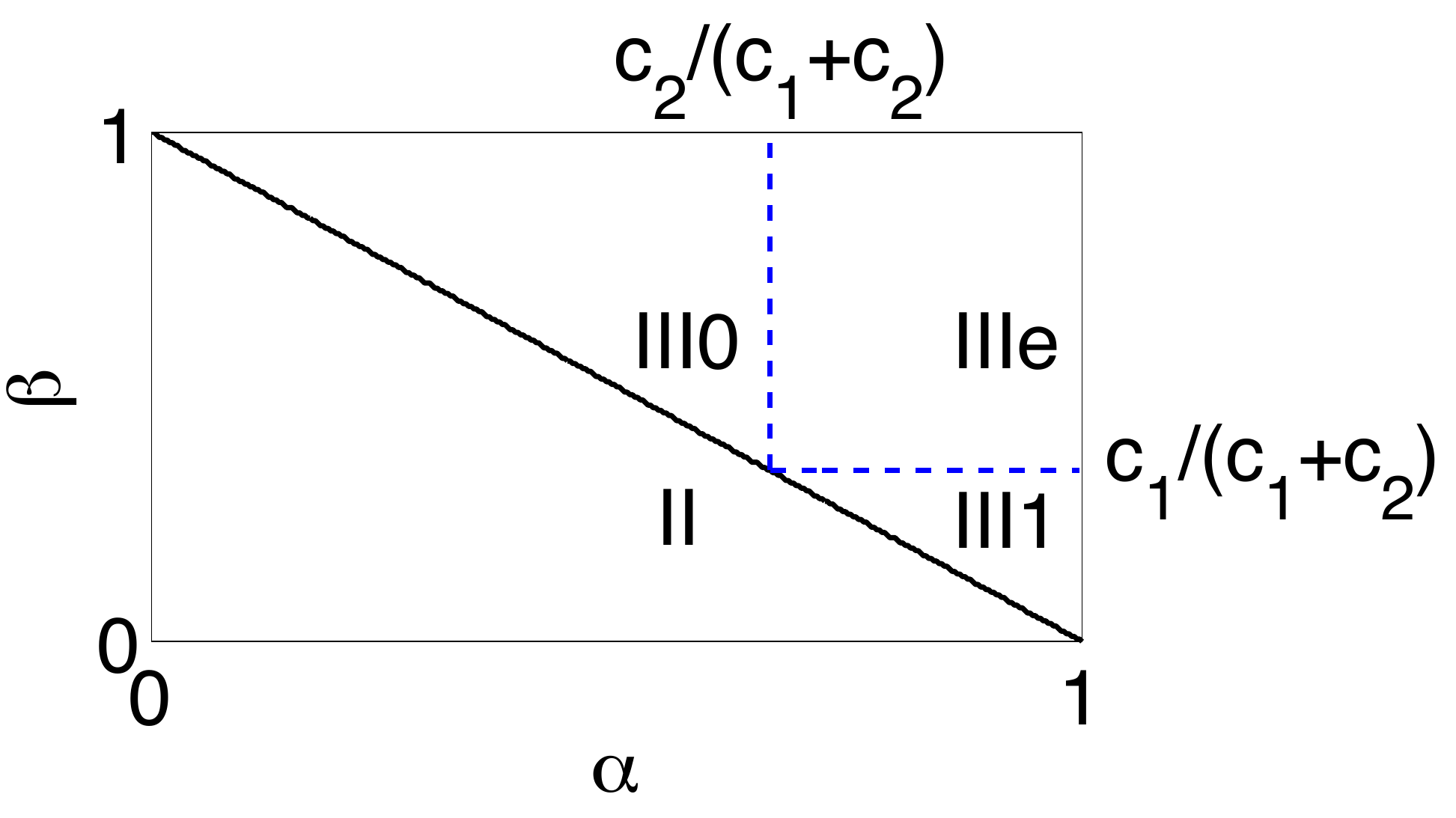}
		\caption{Division of $\alpha-\beta$ parameter space of the SVF model into Regions I, II, IIIe, III0, and III1.}
		\label{fig:SVF_regimes}
	\end{subfigure}
	\hfill
	\begin{subfigure}[t]{0.48\linewidth}
		\includegraphics[width=\linewidth]{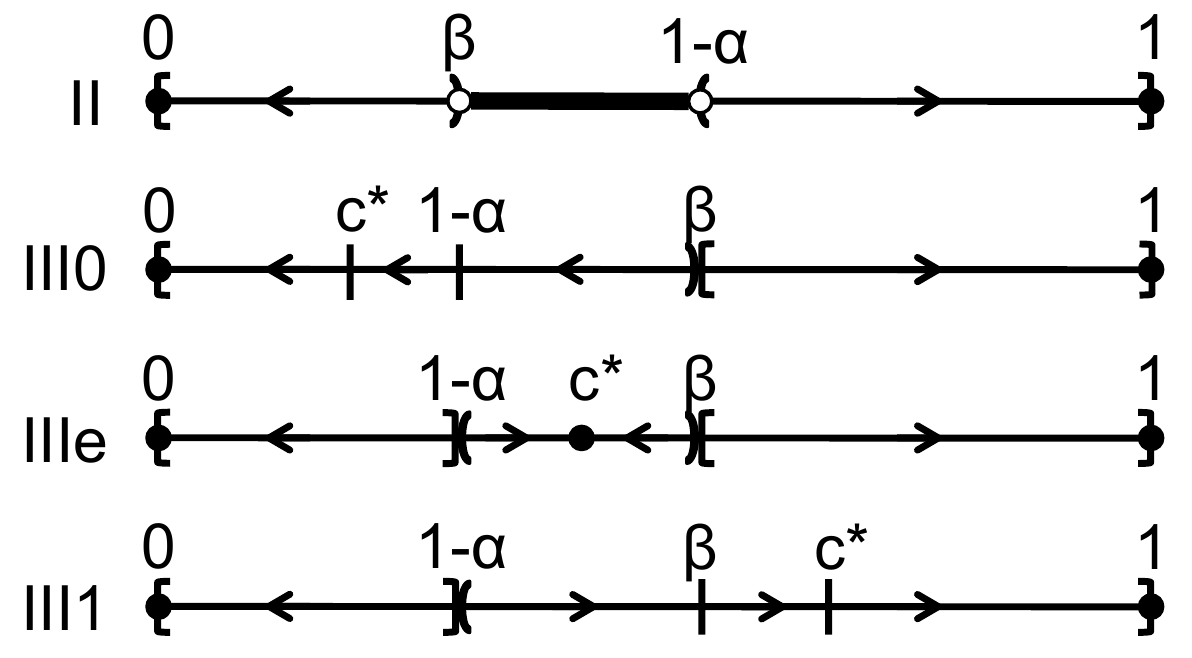}
		\caption{Equilibria, stability and basins of attraction on the $r$-axis ($r \in [0, 1]$) for parameters $\alpha$, $\beta$, $c_1$, and $c_2$ of the SVF model lying in Regions I, III0, IIIe and III1. Closed (open) circles represent locally asymptotically stable (unstable) equilibria. Left (right) arrows indicate regions where $\dot{r} < 0$ ($\dot{r} > 0$).}
		\label{fig:SVF_stability}
	\end{subfigure}
	\caption{Summary of the potential dynamical regimes of the SVF model of \cite{LangDeSterck14}, given in \eq{SVF}, with parameters $\alpha$, $\beta$, $c_1$, and $c_2$.}
	\label{fig:SVF_summary}
\end{figure}
Before deriving a population-level ODE for the linear threshold ABM and establishing the consistency of the SVF model from  \cite{LangDeSterck14} with the ABM in the following section, it is useful to first review some basic properties of the SVF model.
For example, we will show later that the parameter regimes of the SVF are predictive of the parameter regime transitions in the ABM.
The SVF model is given by the equation
\begin{equation}
	\label{eq:SVF}
	\dot{r} = \underbrace{c_1 \spc (1-r) \spc v_s(r;\alpha)}_{g(r)} - \underbrace{c_2 \spc r \spc p(r;\beta)}_{d(r)},
\end{equation}
with parameters $\alpha,\beta\in(0,1)$ and $c_1,c_2>0$. The growth and decay of the fraction of protesters $r(t)$ are modelled by the \emph{growth} and \emph{decay} functions $g,d:[0,1]\rightarrow\mathbb{R}^+$, respectively. Given their roles in determining the rate of growth and decay, the parameters $c_1$ and $c_2$ are called the \emph{protesters' enthusiasm} and \emph{police efficiency} parameters in \cite{LangDeSterck14}. The growth and decay in the number of protesters are modulated by the \emph{visibility} and \emph{policing} terms that are taken to be step functions $v_s(r;\alpha) = \mathbb{I}_{\{r > 1-\alpha\}}$ and $p(r;\beta) = \mathbb{I}_{\{r<\beta\}}$, where the indicator function is defined as
$$
	\mathbb{I}_{\{X\}} = \left\{\begin{array}{ll}  1 & \mbox{if $X$ is true} \\ 0 & \mbox{otherwise} \end{array}\right..
$$ 
These functions are parametrized by \emph{visibility} and \emph{police capacity} parameters $\alpha$ and $\beta$, respectively. In words, the fraction of protesters can grow only when the protest movement is sufficiently large to be ``visible'' to the general population, i.e. when protests are large enough that their existence cannot be masked by the regime through censorship. Conversely, the fraction of protesters can decay only so long as the fraction of protesters is less than the regime's finite capacity to disperse or arrest protesters. This is consistent with the idea that individuals who consider joining the revolution are faced with a collective action problem \cite{Kuran91}: if they choose to act unilaterally then they face severe retaliation from the regime, so they wait to act until the revolution becomes visible and if they act collectively then the regime loses the ability to punish due to a lack of resources.

Depending on the values of the parameters $\alpha$, $\beta$, $c_1$, and $c_2$, the dynamics of the SVF fall into one of four main regimes, see Fig.~\ref{fig:SVF_summary}. The equilibrium of \emph{total state control} $r=0$ and the equilibrium of the \emph{realized revolution} $r=1$ are present in all parameter regimes and are always locally asymptotically stable. Different parameter regimes are distinguished by the behaviour of \eq{SVF} when $r$ lies between $1-\alpha$ and $\beta$. Region II ($\beta<1-\alpha$) is characterized by an interval of stable equilibria $r\in(\beta,1-\alpha)$ bounded by two unstable equilibria $r\in\{\beta,1-\alpha\}$. Since parameters in Region II are characterized by a relatively weak regime (low $\beta$) and relatively small visibility (low $\alpha$), Region II is interpreted in \cite{LangDeSterck14} as the \emph{failed state} parameter region. In contrast, Regions III0, IIIe, and III1 ($1-\alpha<\beta$) are characterized by a relatively strong regime (large $\beta$) and relatively large visibility (large $\alpha$) and differ in the value of the quantity $c^* = c_1/(c_1+c_2)$, i.e. in the ratio of protester's enthusiam to the sum of protester's enthusiasm and police efficiency. In Region III0, i.e. when $c^*\leq 1-\alpha <\beta$, the interval $(1-\alpha, \beta)$ lies in the basin of attraction of the equilibrium of total state control $r=0$. Analogously, in Region III1, i.e. when $1-\alpha <\beta\leq c^*$, the interval $(1-\alpha, \beta)$ lies in the basin of attraction of the equilibrium of the realized revolution $r=1$. Finally, in Region IIIe, i.e. when $1-\alpha < c^* < \beta$, the interval $(1-\alpha,\beta)$ becomes the basin of attraction of a new equilibrium $r=c^*$, which is called the \emph{equilibrium of civil unrest}
in \cite{LangDeSterck14}. Because of the contribution of the interval $(1-\alpha, \beta)$ to the stability of the equilibrium of total state control, of the realized revolution, or of civil unrest, Regions III0, III1, and IIIe are called the \emph{stable police state}, \emph{unstable police state}, and \emph{meta-stable police state}, respectively. For a full description, interpretation, and analysis of the SVF model we refer the reader to \cite{LangDeSterck14}.

\subsection{Population-Level ODE for ABM and Relation to SVF Model}
\label{Sec:consist_ABM_SVF}
\subsubsection{Mathematical Analysis}
To show consistency of the ABM proposed in Section \ref{Sec:Spec_ABM} with the SVF model of \cite{LangDeSterck14} we begin by defining $r_a(t) = r_a(t|t_0)$ to be the fraction of nodes (in the ABM model) that are expected to be in state 1, i.e. the expected fraction of active nodes at time $t$ conditioned on information at time $t_0$:
$$
	 r_a(t) = \frac{1}{N}\sum_{i=1}^N \ex[ s_{v_i}(t) | t_0].
$$
The fraction of nodes that are expected to be in state 0 (inactive) at time $t$, conditioned on information at time $t_0$, is then $1-r_a(t|t_0)$. We now write the change in $r_a$ from time $t_0$ to time $t$ as
\begin{equation}
	\label{eq:preApprox}
	\Delta r_a(t) = r_a(t) - r_a(t_0) = g(t|t_0) - d(t|t_0),
\end{equation}
where $\Delta t = t-t_0$, and the expected growth and decay of the fraction of active nodes are modelled by the non-negative growth and decay functions $g(t) = g(t|t_0)$, and $d(t) = d(t|t_0)$, respectively. In order to obtain the one-compartment model we will need to approximate the quantities $g(t)$ and $d(t)$ in terms of $r_a(t_0)$, $\{\theta_{v_i}\}_{i=1}^N$, and $\beta$.

For notational convenience we begin by considering the case where $\forall v\in V:\theta_v= \theta$. The pool of individuals that can go from active to inactive at time $t_0$ is $r_a(t_0)$. Active nodes become inactive at the first arrival time of a Poisson process with rate $c_2$, provided that the fraction of active nodes does not exceed the regime's police capacity, i.e. $r_a(t_0)<\beta$. Thus,
\begin{equation}
	\label{eq:A1}
	d(t|t_0) = [c_2\spc r_a(t_0) \spc\Delta t + o(\Delta t) ] \spc p(r_a;\beta) = c_2 \spc r_a(t_0) \spc p(r_a(t_0); \beta) \spc \Delta t+ o(\Delta t),
\end{equation}
where we take $p(r;\beta) = \mathbb{I}_{\{r<\beta\}}$, as in \eq{SVF}. Next, let
\begin{equation}
	\label{eq:nu}
	\nu(r_a(t_0);\theta) = \frac{1}{N} \left[\sum_{v\in V} \mathbb{I}_{\left\{\sum_{j=1}^{k_v} s_{w_{v,j}}(t_0) \geq \theta k_v \right\}}\right],
\end{equation}
where $\br{w_{v,j}}{j=1}{k_v}$ denotes the neighbours of individual $v\in V$. In words, we let $\nu(r_a(t_0);\theta)$ be the fraction of the total population at time $t_0$ 
that can see the revolution. 
We call $\nu(r_a(t_0);\theta)$ the visibility function of the linear threshold ABM.
In the chosen notation for $\nu$ it is emphasized that $\nu$ depends on $r_a(t_0)$, i.e. on the fraction of nodes that are active at time $t_0$. At time $t=t_0$, the pool of individuals that can go from inactive to active, i.e. that are considering joining the revolution, is therefore approximately $(1-r_a(t_0))\spc \nu(r_a(t_0);\theta)$. Since inactive nodes that can see the revolution become active at the first arrival time of a Poisson process with rate $c_1$, we have
\begin{equation}
	\label{eq:A0}
	g(t|t_0) = c_1 \spc (1 - r_a(t_0)) \spc \nu(r_a(t_0);\theta) \spc \Delta t+ o(\Delta t).
\end{equation}

Combining Equations \eqref{eq:preApprox}-\eqref{eq:A0} and dividing by $\Delta t$ gives
$$
	\frac{\Delta r_a}{\Delta t} = c_1\mbox{ }(1 - r_a(t_0)) \spc \nu(r_a(t_0);\theta) - c_2 \mbox{ } r_a(t_0) \mbox{ }p(r_a(t_0);\beta) + o(1).
$$
Taking the limit as $\Delta t\rightarrow 0$ yields
\begin{equation}
	\label{eq:Approx}
	\frac{dr_a}{dt} = c_1 \mbox{ }(1 - r_a)\mbox{ }\nu(r_a;\theta)- c_2 \mbox{ }r_a \mbox{ }p(r_a;\beta).
\end{equation}
This establishes a general population-level ODE model that approximates the linear threshold ABM, with visibility function $\nu(r_a;\theta)$. To close the model, the visibility function has to be specified, guided by Equation (\ref{eq:nu}).
Observe that if we substitute $\nu(r_a;\theta)$ by the step function $v_s(r_a;\alpha(\theta))$, then this exactly recovers the SVF model of \cite{LangDeSterck14}, see \eq{SVF}. So, provided that the step function $v_s(r;\alpha(\theta))$ is a suitable choice for $\nu(r;\theta)$, which we argued in the appendix of \cite{LangDeSterck14} based on considering a simplified averaged threshold process that ignores correlations, this confirms the consistency of the ABM and the SVF. The crux of the argument used in \cite{LangDeSterck14}, which we discuss further in Sec.\ \ref{Sec:BVF},
is that, for real networks $\nu(r;\theta)$ can be expected to be a steep sigmoidal function of $r$, which can be approximated by a step function. In this paper, we test these assumptions explicitly on the empirical physical contact and Facebook social networks discussed above using ABM simulations, and we derive new improved ODE models that form a more accurate representation of $\nu(r;\theta)$ than a step function, taking into account the actual network structure of the empirical networks. 
The step function assumption was chosen in \cite{LangDeSterck14} because it is useful when little information is known about the underlying network structure. For example, without having to specify an underlying communication network, the SVF model provides a potential mechanism by which networks with increased visibility $\alpha$ are more susceptible to political revolution.

The following section explores the relationship between the ABM and the SVF model numerically for empirical networks by using a fitting procedure to attempt to estimate the network-dependent $\alpha = \alpha(\theta)$ for which $v_s(r;\alpha(\theta))$ best approximates $\nu(r;\theta)$. Given this estimate for $\alpha=\alpha(\theta)$, we compare ABM simulations to solutions of \eq{SVF} in order to verify the consistency of the ABM and the SVF numerically and to investigate the ABM behaviour in the parameter regimes of the SVF model as described in Fig.\ \ref{fig:SVF_summary}. Section \ref{Sec:AVF} then expands on this by proposing and analyzing a more sophisticated approximation for the
visibility function $\nu(r;\theta)$. We conclude this section by noting that the above derivation can easily be modified for the case where the $\theta_v$ are not uniform. In this case the growth and decay functions $g$ and $d$, respectively, can be computed by averaging over the distribution of the $\theta_v$'s.

\begin{figure}[h]
	\centering
	\begin{subfigure}[t]{\linewidth}
		\includegraphics[width=0.48\linewidth]{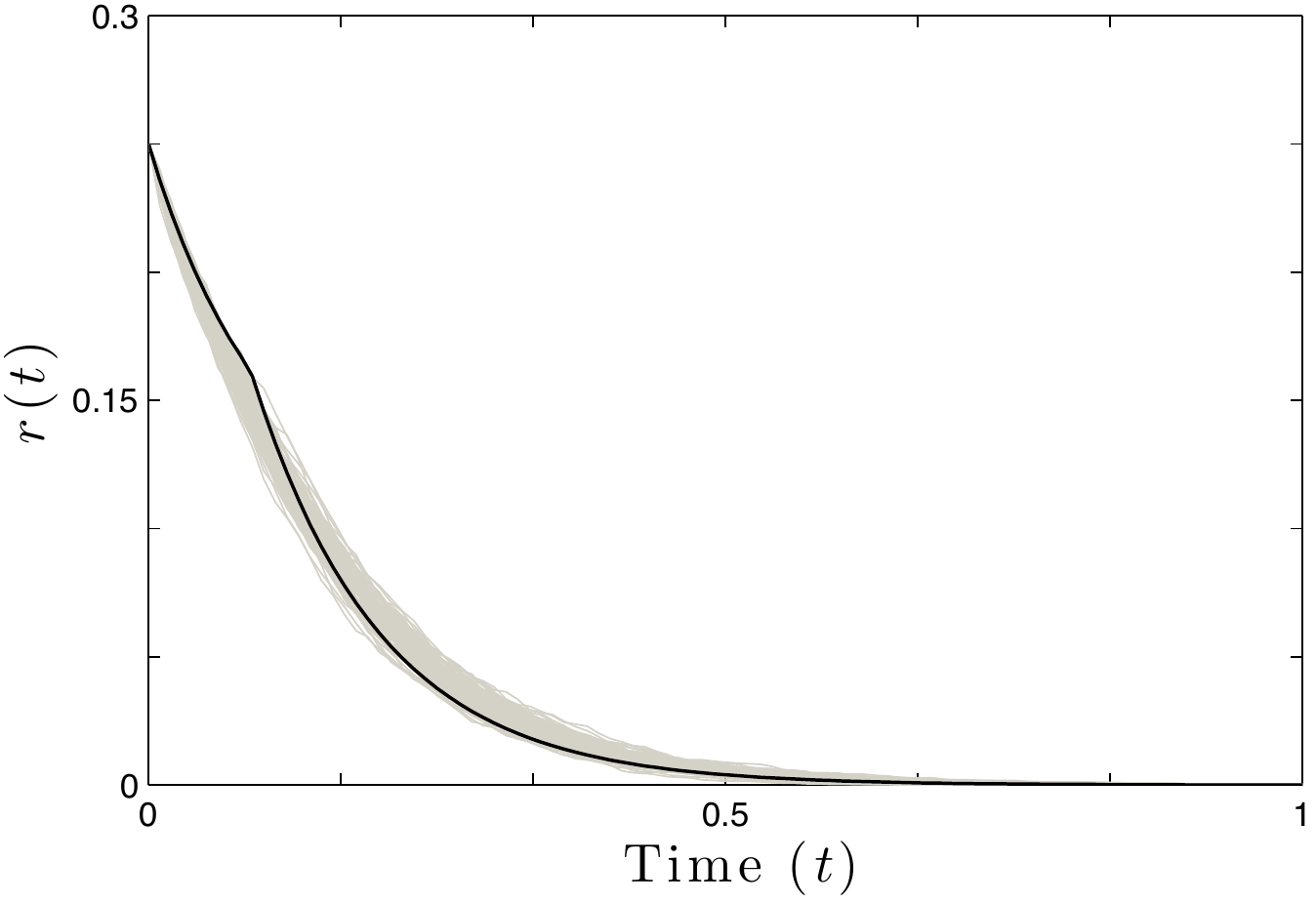}\hfill
		\includegraphics[width=0.48\linewidth]{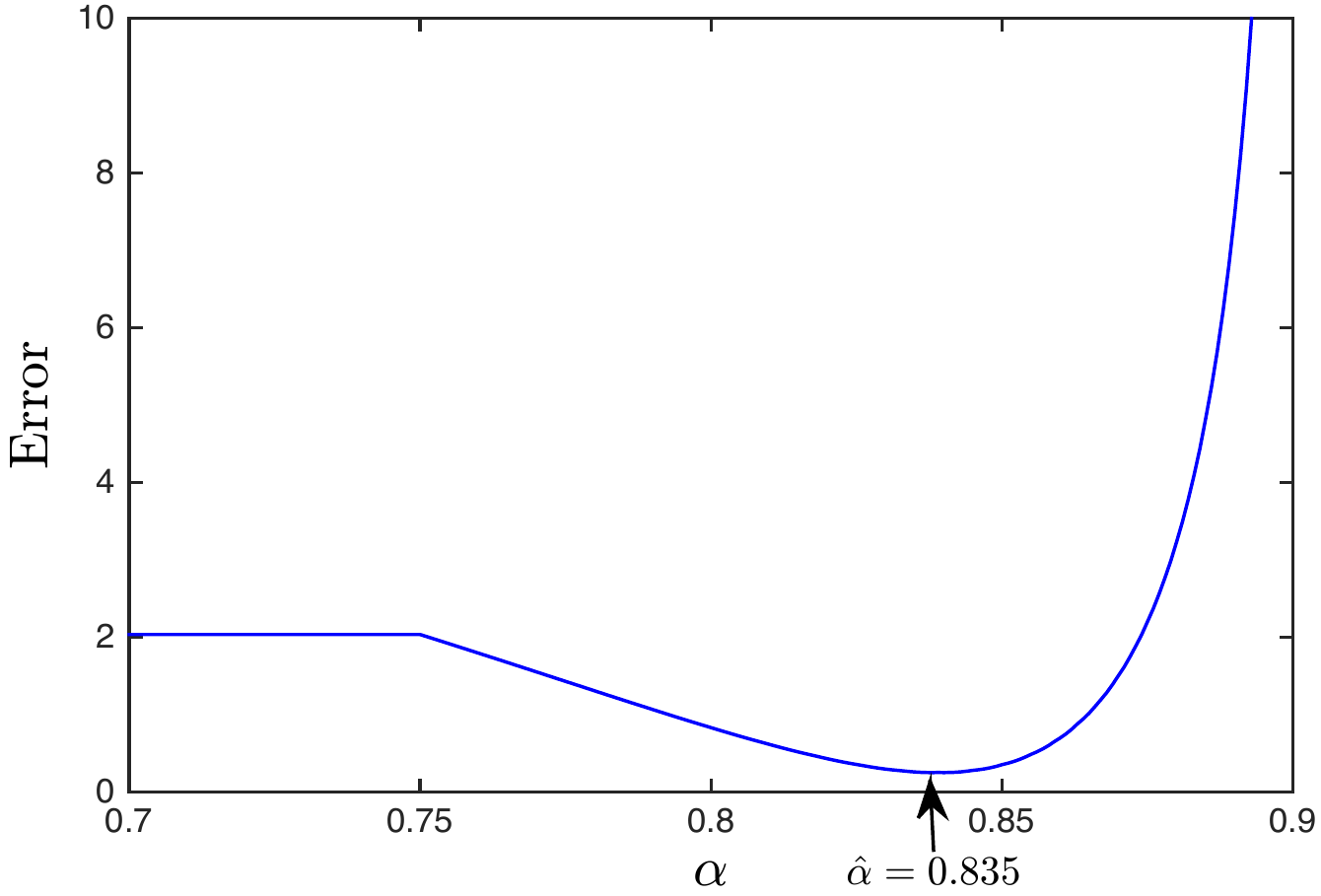}
		\caption{Facebook social network $G_F$ ($\hat{\alpha} = 0.835$)}
		\label{fig:fit_alpha_ODE_online}
	\end{subfigure}
	\\
	\begin{subfigure}[t]{\linewidth}
		\includegraphics[width=0.48\linewidth]{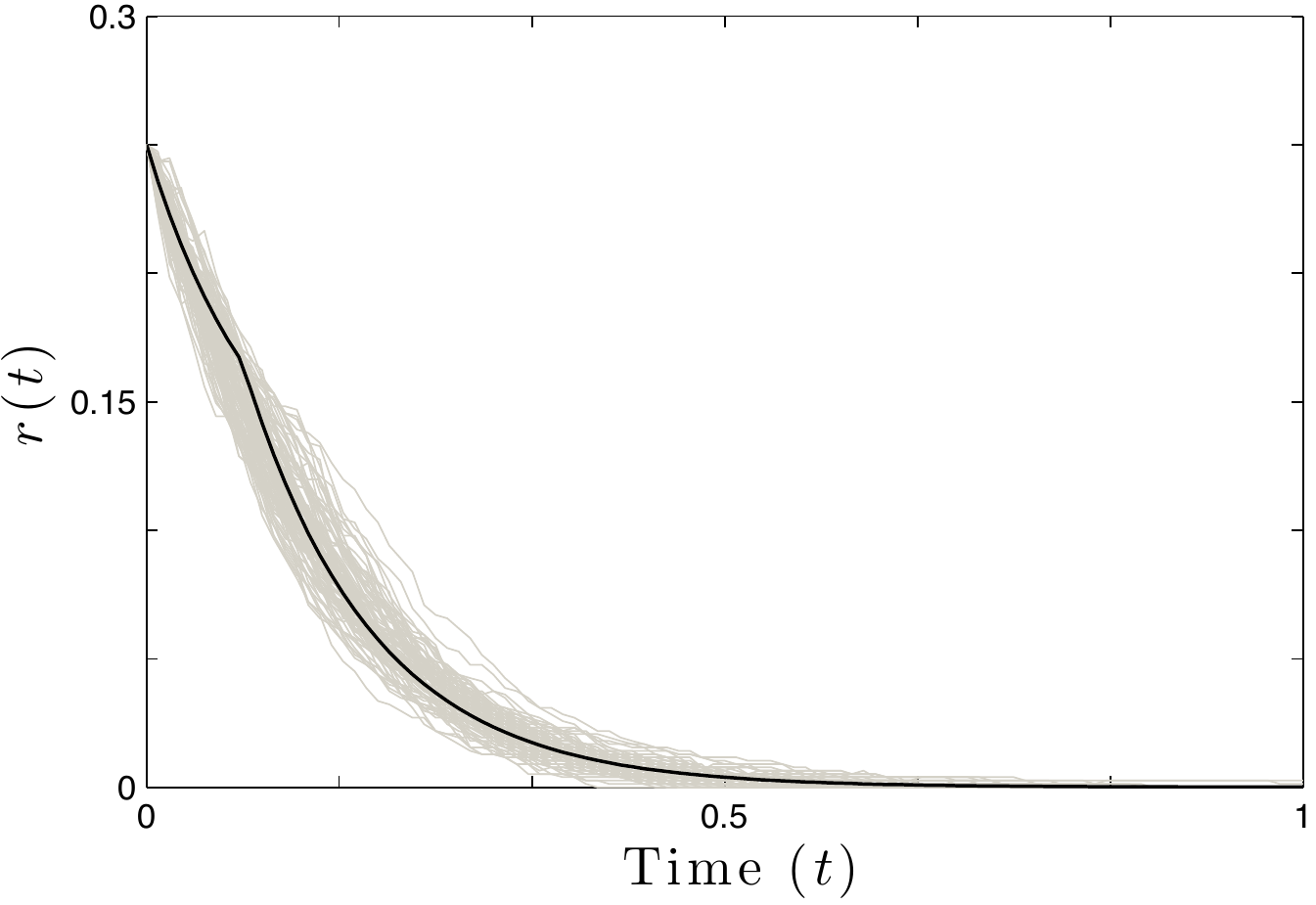}\hfill
		\includegraphics[width=0.48\linewidth]{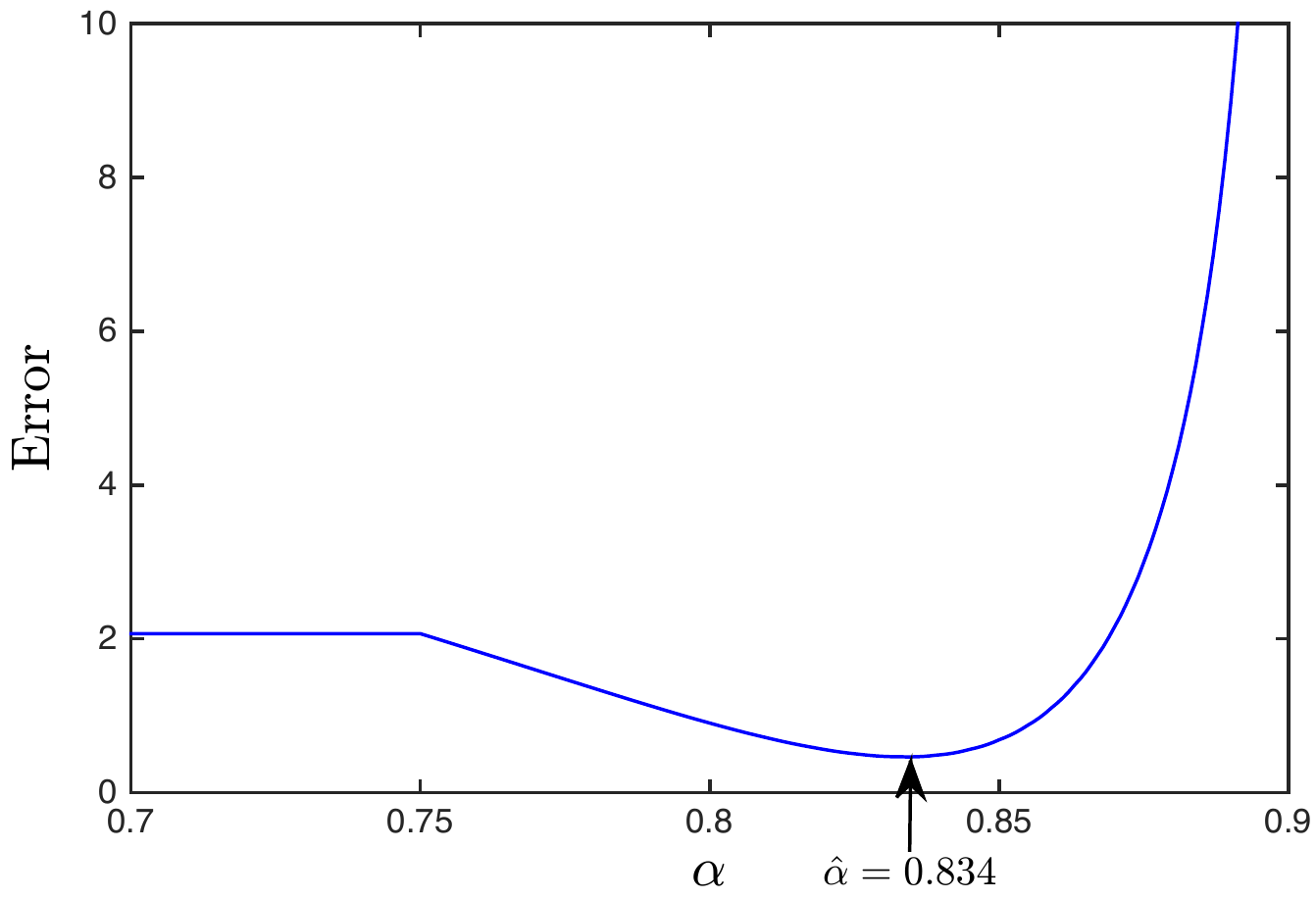}
		\caption{Physical contact network $G_P$ ($\hat{\alpha} = 0.834$)}
		\label{fig:fit_alpha_ODE_offline}
	\end{subfigure}
	\caption{
	Parameter Region III0.
	(Left) Time traces of ABM simulations with optimally fitted $\hat{\alpha}$ (grey) and solution to \eq{SVF} (black). (Right) $L_2^2$ error between ABM realizations and the solution to \eq{SVF} versus $\alpha$.
	The optimal $\hat{\alpha}$ is found by fitting \eq{SVF} (the SVF model) to $rep=100$ ABM simulations with parameters $\theta=0.15$, $\beta=0.3$, $c_1=1$, $c_2=9$, and $r_0=0.25$ (Region III0, $c^* \leq 1-\hat{\alpha} <\beta$). 
	}
	\label{fig:fit_alpha_ODE}
\end{figure}
%
\begin{figure}[h]
	\centering
		\includegraphics[width=0.5\linewidth]{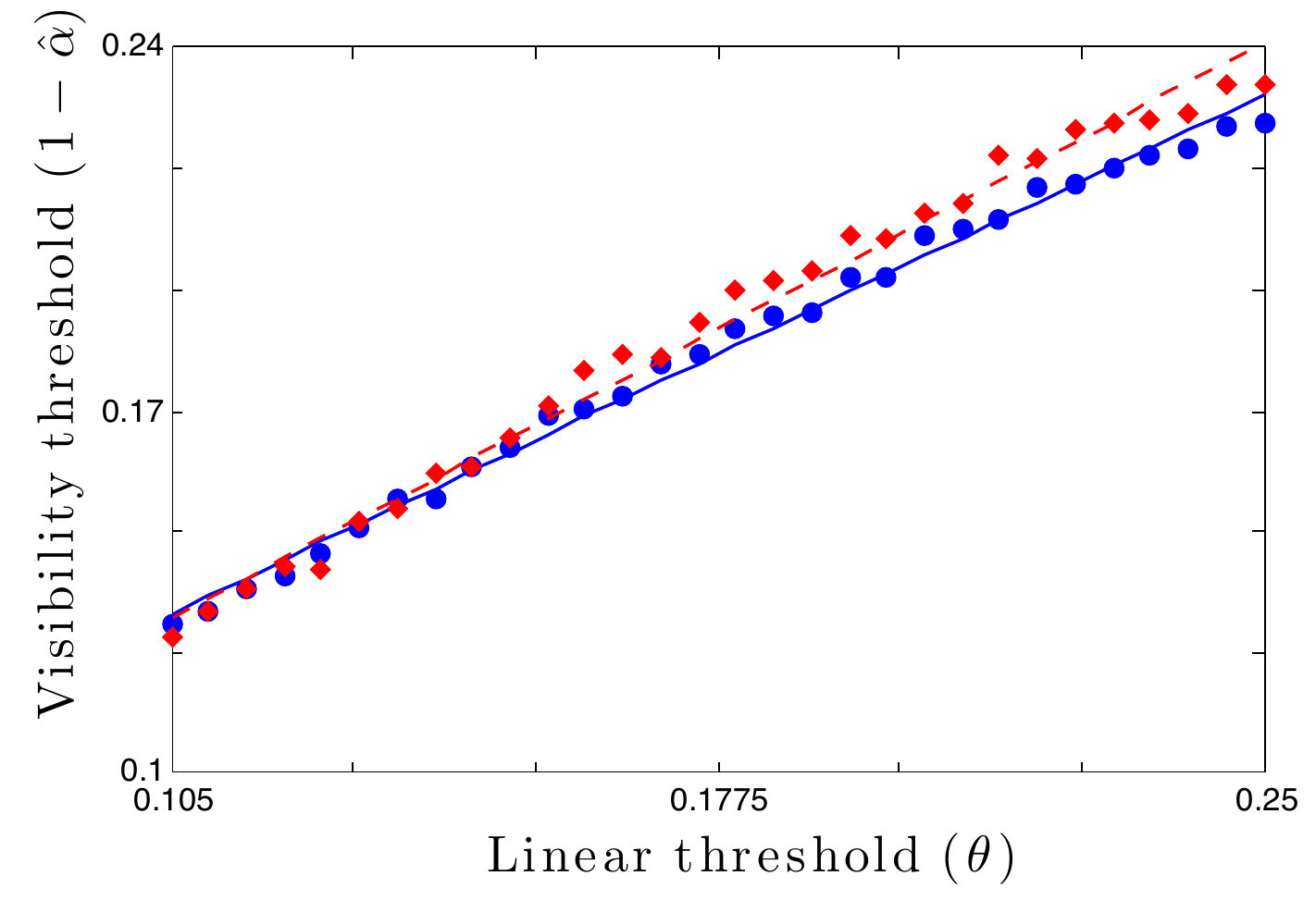}
	\caption{Parameter Region III0. Result of finding $\hat{\alpha}$ as a function of $\theta$ by fitting \eq{SVF} (the SVF) to $rep=100$ ABM simulations with parameters $\beta=0.3$, $c_1=1$, $c_2=9$, and $r_0=0.25$ (Region III0, $c^*\leq 1-\hat{\alpha} < \beta$). (Blue circles) Facebook social network. (Red diamonds) Physical contact network. Line of best fit for networks given by solid ($m=0.690\pm0.022$, $b=0.058\pm0.008$) and dashed ($m=0.764\pm0.032$, $b=0.050\pm0.006$) lines, respectively.}
	\label{fig:fit_multi_theta_alpha}
\end{figure}
\begin{figure}[h]
	\centering
	\begin{subfigure}[t]{\linewidth}
		\includegraphics[width=0.48\linewidth]{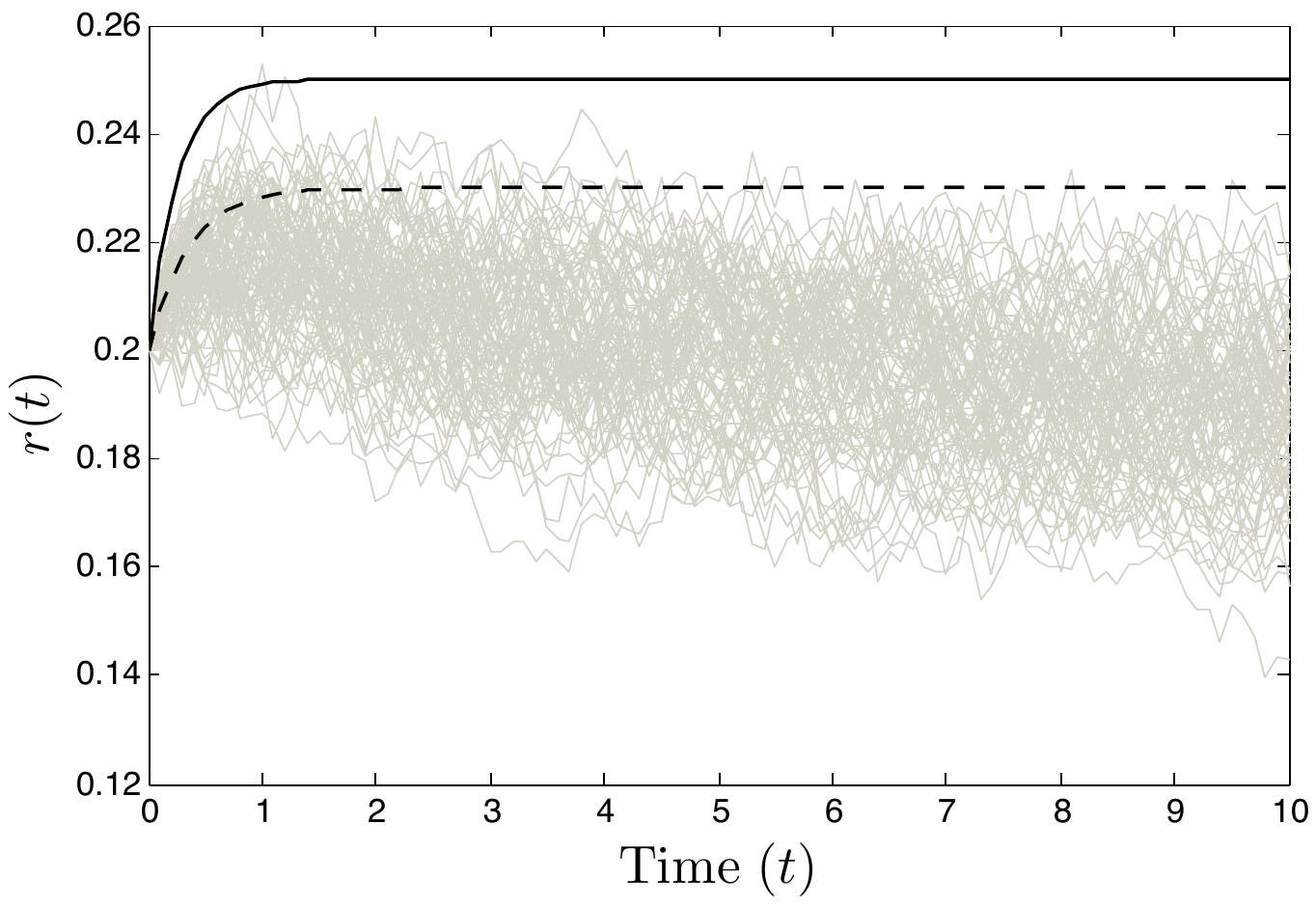}\hfill
		\includegraphics[width=0.48\linewidth]{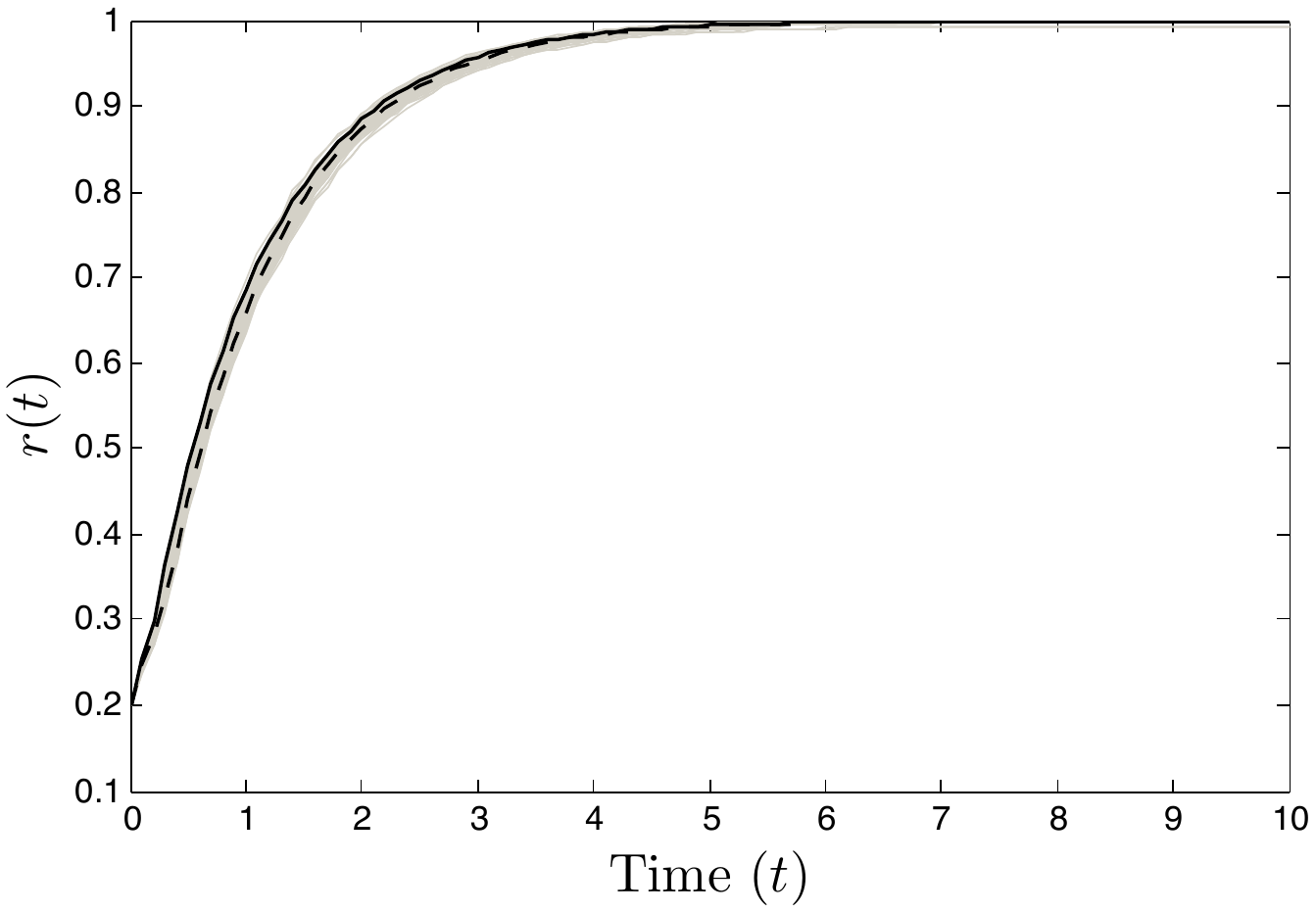}
		\caption{Facebook social network $G_F$ ($1-\hat{\alpha} = 0.131$)}
		\label{fig:fit_alpha_ODE_online}
	\end{subfigure}
	\\
	\begin{subfigure}[t]{\linewidth}
		\includegraphics[width=0.48\linewidth]{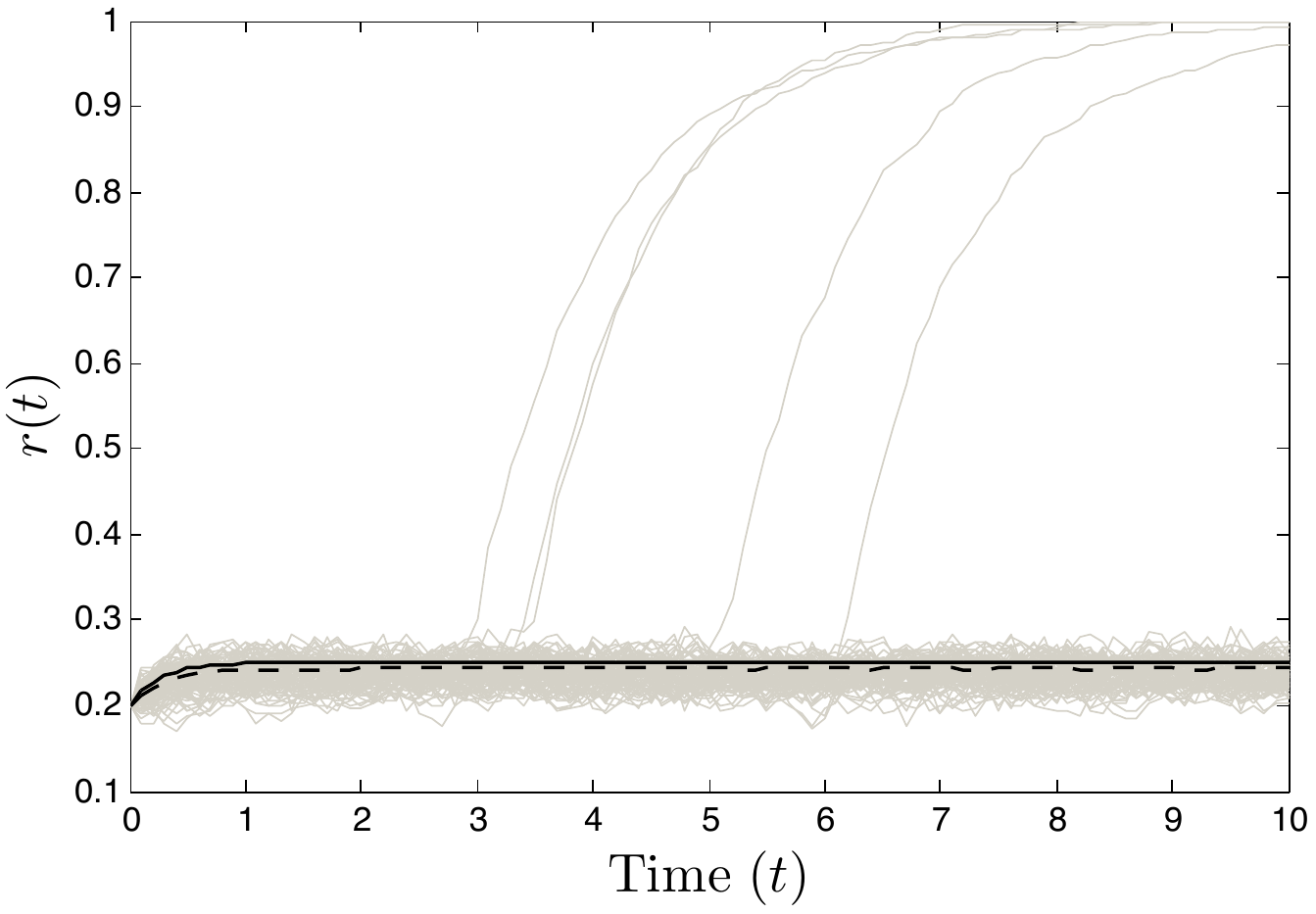}\hfill
		\includegraphics[width=0.48\linewidth]{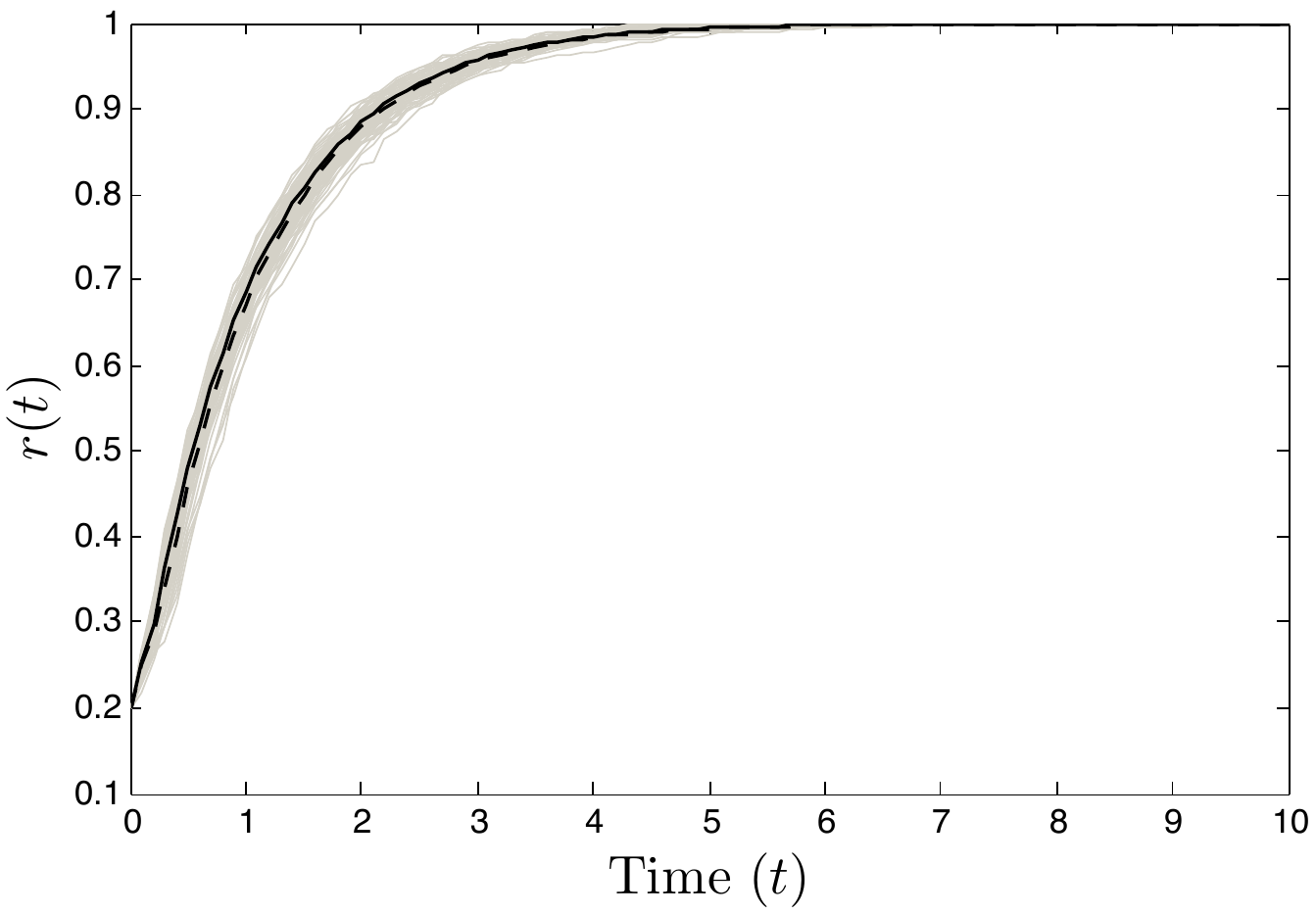}
		\caption{Physical contact network $G_P$ ($1-\hat{\alpha} = 0.134$)}
		\label{fig:fit_alpha_ODE_offline}
	\end{subfigure}
	\caption{(Left) Parameter Region IIIe ($c_2=3$, $1-\hat{\alpha}< c^*< \beta$). (Right) Parameter Region III1 ($c_2=1$, $1-\hat\alpha< \beta \leq c^*$). Time traces of $rep=100$ ABM simulations (grey), solution to the SVF model (solid black), and BVF model (dashed black) with parameters $\theta=0.11$, $\beta=0.3$, $c_1=1$, and $r_0=0.2$.}
	\label{fig:fit_alpha_IIIe1}
\end{figure}
%

\subsubsection{Numerical Simulation}
\label{Sec:NumSim_ABM_SVF}

In this section we simulate the ABM on the physical contact and Facebook networks for a fixed set of model parameters and find the optimal $\alpha$ for \eq{SVF} that minimizes the difference between the ABM realizations and the output of \eq{SVF}. The closeness of the output of \eq{SVF} and the ABM realizations for broad ranges of parameter values
demonstrates that the SVF model does indeed approximate the ABM, at least for non-pathological parameter choices. We begin by fixing the initial condition $r_a(0) = r_0$\footnote{For simulations of the ABM an initial condition $r_a(0) = r_0$ is implemented by choosing $\lceil r_0N\rceil$ nodes uniformly at random to be active.} and the parameters $\theta_v\equiv\theta$, $\beta$, $c_1$, and $c_2$. To compare realizations of the ABM with the solution to \eq{SVF} we use Gillespie's Algorithm (summarized in Appendix \ref{Sec:Gillespie}) to simulate $rep=100$ realizations of the ABM, recording $r_a$ at each $t\in\{0,0.01,0.02,\ldots,1\}$, and then find the $\alpha$ that minimizes the difference (in $L_2$ norm) between the ABM realizations and the solution to \eq{SVF} (see Appendix \ref{Sec:SolnSVF}). The results of this procedure for ABM simulations performed on the physical contact and Facebook social networks with parameters $r_0=0.25$, $\theta=0.15$, $\beta=0.3$, $c_1=1$, and $c_2=9$ are given in Fig.~\ref{fig:fit_alpha_ODE}. For these parameters we find the optimal $\alpha$ is given by $\hat{\alpha} = 0.835$ and $\hat{\alpha} = 0.834$ for the Facebook and physical contact networks, respectively. Equivalently, the fitted value for the visibility threshold $1-\alpha$ is given by $1-\hat{\alpha} = 0.165$ and $1-\hat{\alpha} = 0.166$, respectively. 
Holding the initial condition $r_0=0.25$ and parameters $\beta=0.3$, $c_1=1$ and $c_2=9$ constant, we are able to repeat the fitting procedure to find fitted values of $\alpha$ for $\theta\in\{0.105,0.11,0.115,\ldots,0.25\}$. These results are displayed in Fig.~\ref{fig:fit_multi_theta_alpha} and illustrate that for a large range of individual thresholds $\theta$ the Facebook social network has a slightly lower fitted visibility threshold $1 - \hat{\alpha}$ than does the physical contact network.

\begin{figure}[h]
	\centering
	\begin{subfigure}[t]{\linewidth}
		\includegraphics[width=0.48\linewidth]{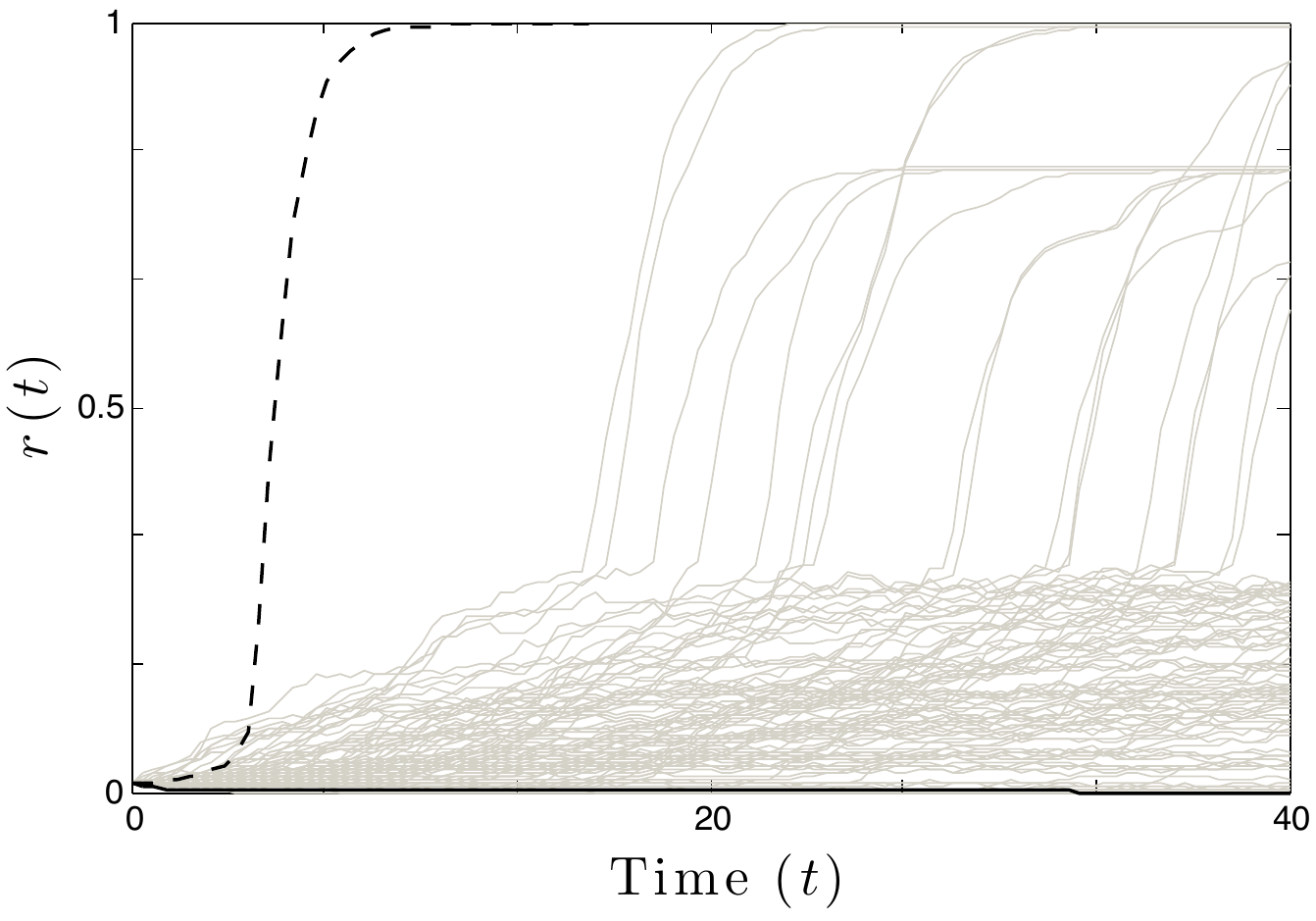}\hfill
		\includegraphics[width=0.48\linewidth]{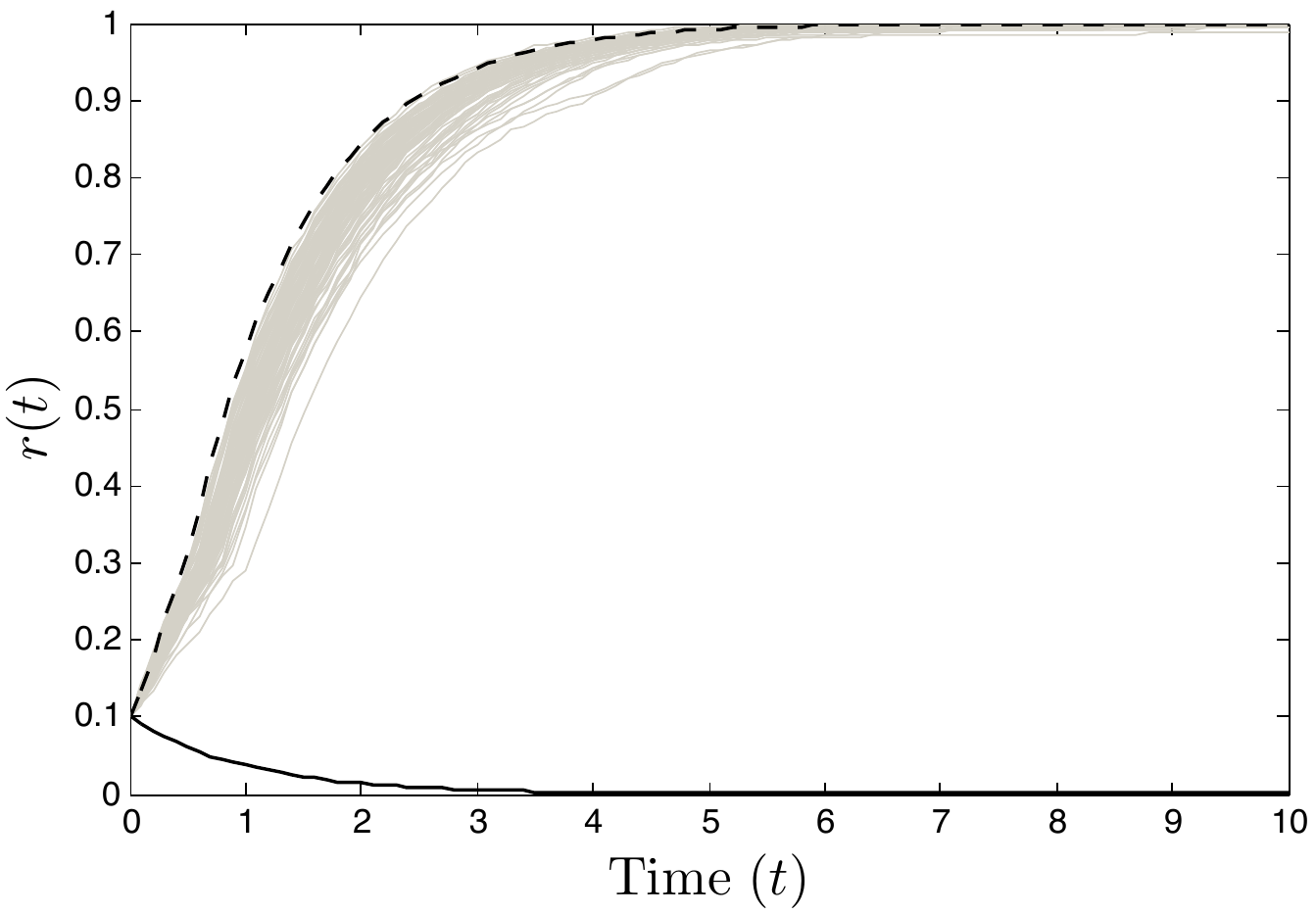}
		\caption{Facebook social network $G_F$ ($1-\hat{\alpha} = 0.131$)}
	\end{subfigure}
	\\
	\begin{subfigure}[t]{\linewidth}
		\includegraphics[width=0.48\linewidth]{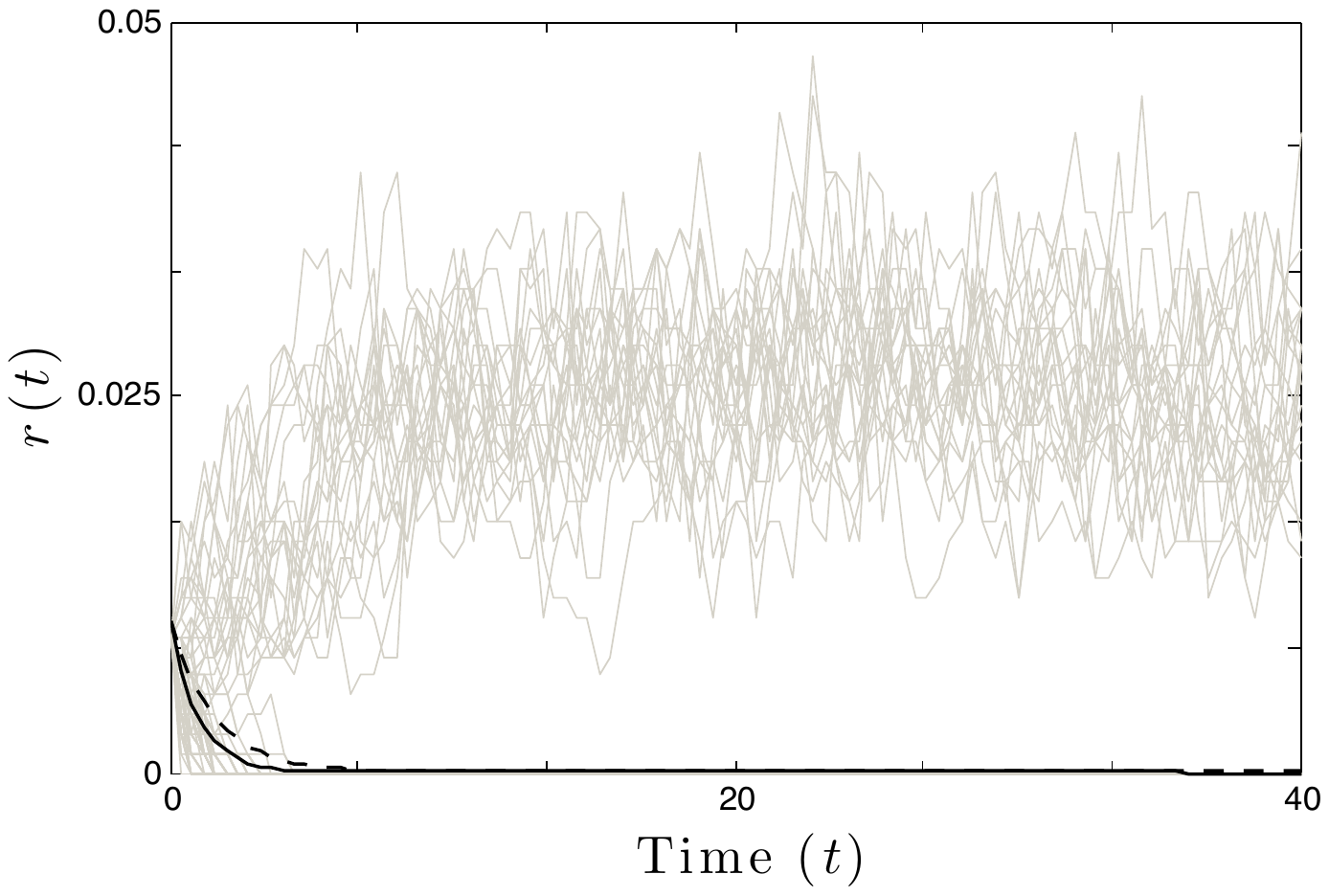}\hfill
		\includegraphics[width=0.48\linewidth]{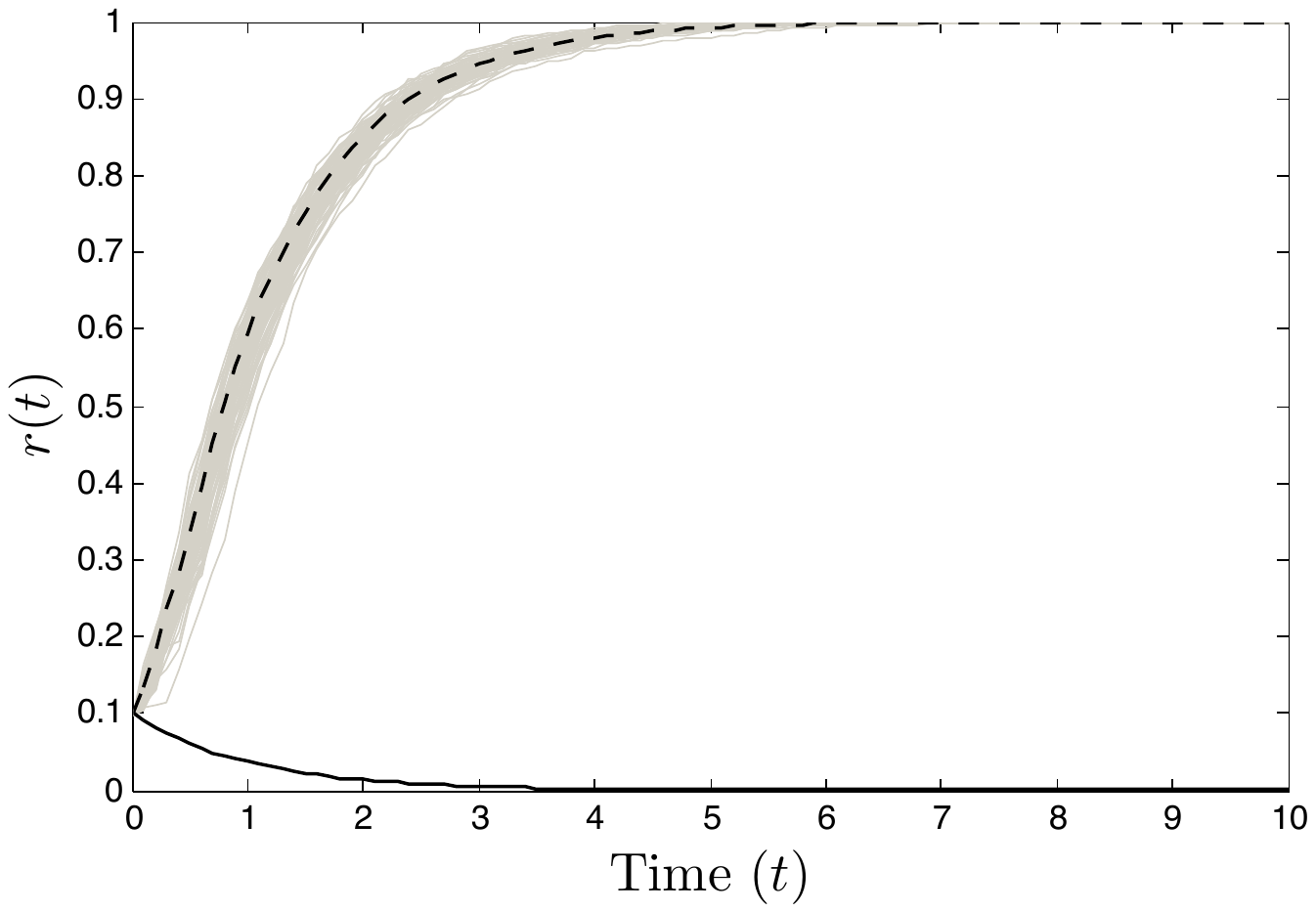}
		\caption{Physical contact network $G_P$ ($1-\hat{\alpha}  = 0.134$)}
	\end{subfigure}
	\caption{The SVF model can fail to approximate the ABM when (left) the initial condition is small ($r_0=0.01$), or (right) the initial condition is near the boundary between the basins of attraction of $r=0$ and $r=1$ ($r_0=0.1\approx 1-\hat{\alpha}$). Time traces of $rep=100$ ABM simulations (grey), solution to the SVF model (solid black), and solution to the BVF model (black dashed) with parameters $\theta=0.11$, $\beta=0.3$ and $c_1=c_2=1$ (Region III1, $1-\hat\alpha <\beta\leq c^*$). In the right panels, the BVF model performs much better than the SVF model.}
	\label{fig:fit_fail_r0_alpha_IIIe1}
\end{figure}

\begin{figure}[h]
	\centering
	\includegraphics[width=0.48\linewidth]{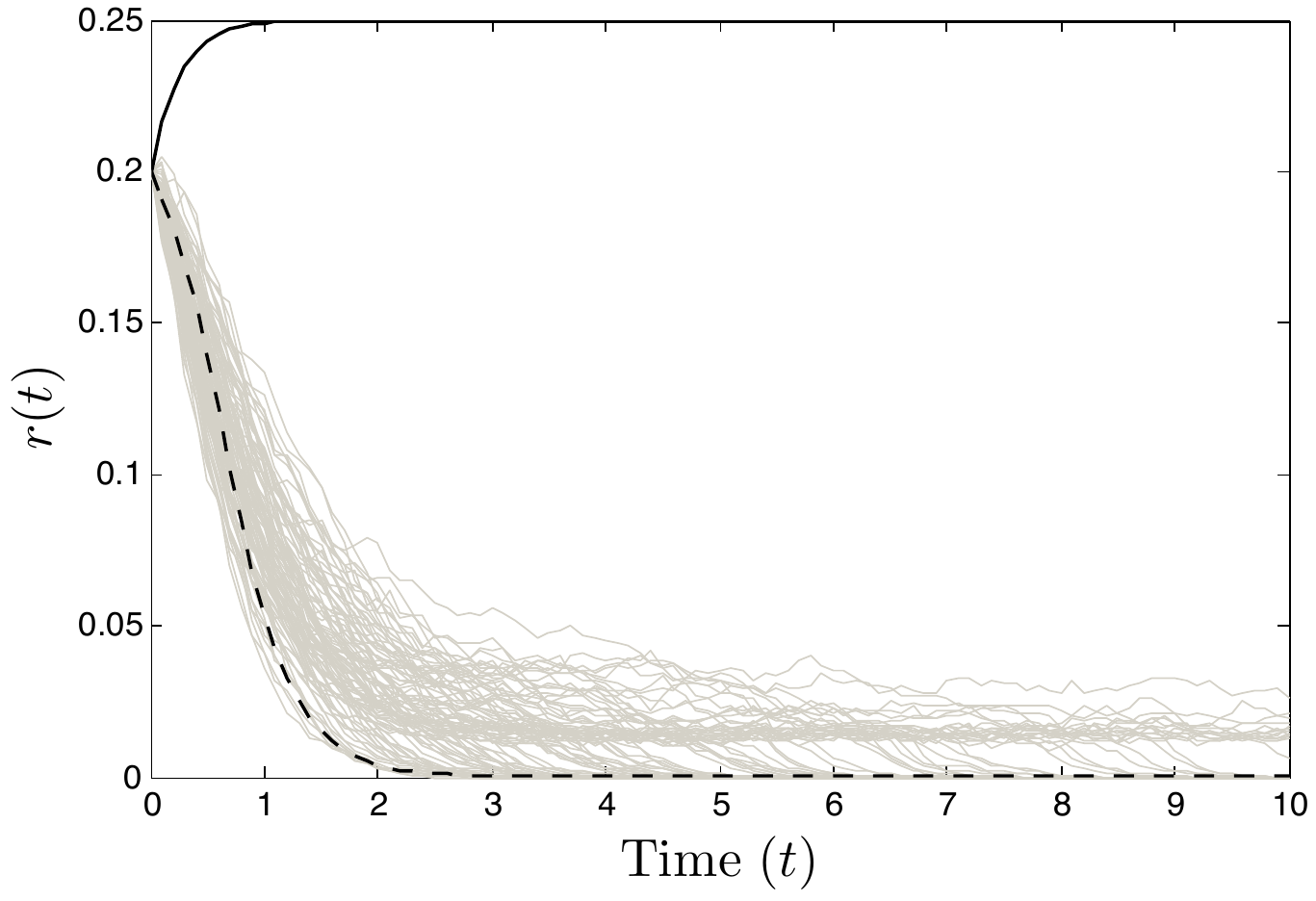}\hfill
	\includegraphics[width = 0.48\linewidth]{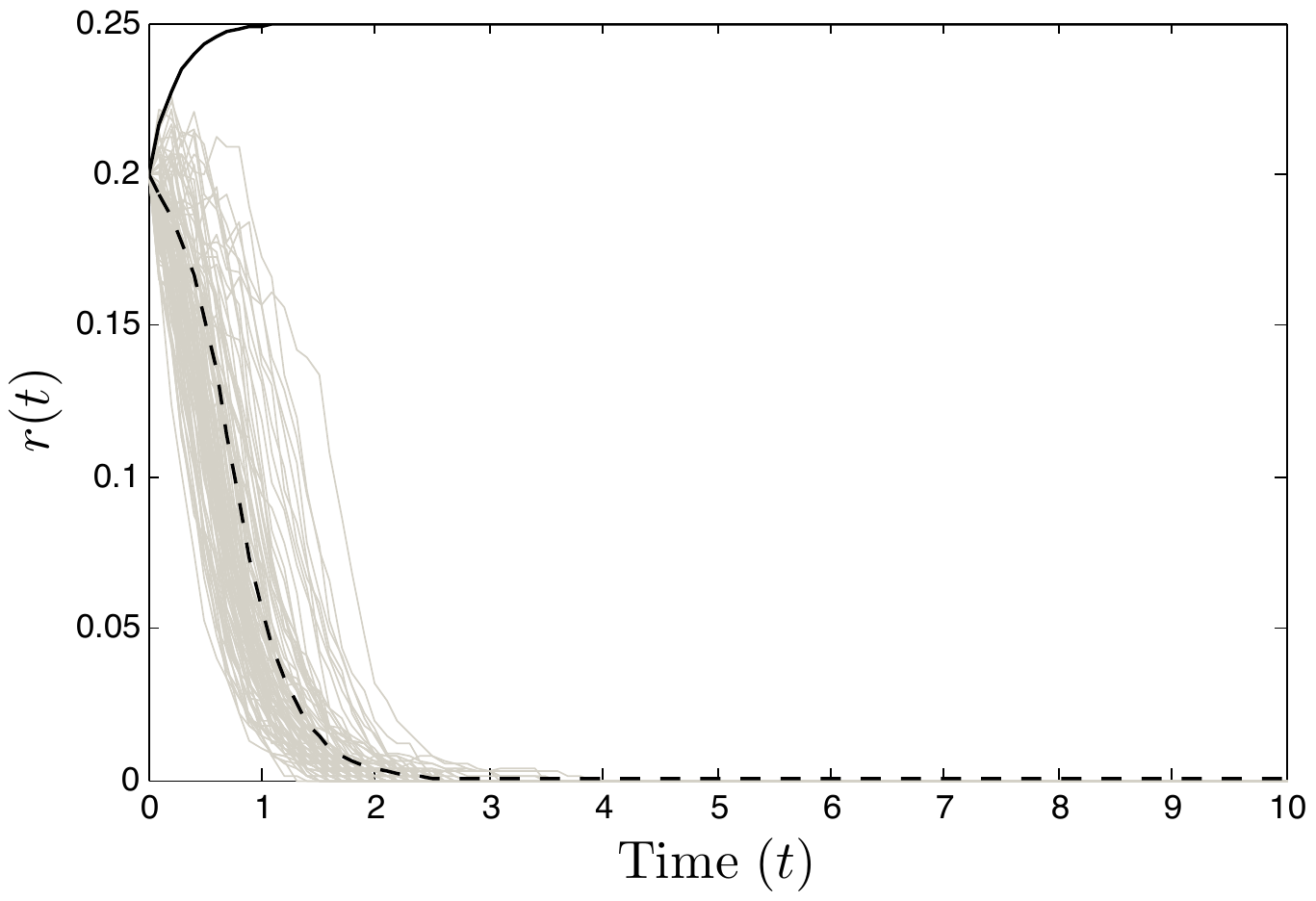}
	\caption{The SVF model can fail to approximate the ABM when the initial condition is near the boundary between the basins of attraction of $r=0$ and $r=1$, i.e. when $r_0\approx1-\alpha$. (Left) Facebook social network $G_F$ ($1-\hat{\alpha} = 0.179$) and (right) physical contact network $G_P$ ($1-\hat{\alpha} = 0.180$): \eq{SVF} has solution in Region IIIe, while ABM has solution in Region III0. Time traces of $rep=100$ ABM simulations (grey), solution to the SVF model (solid black), and solution to the BVF model (dashed black) with parameters $\theta=0.17$, $\beta=0.3$, $c_1=1$, and $c_2=3$, and initial condition $r_0=0.2$ (Region IIIe, $1-\hat{\alpha} <  c^*< \beta$). The BVF model performs much better than the SVF model.}
	\label{fig:fit_fail_r0_regions_alpha_IIIe1}
\end{figure}

\begin{figure}[h]
	\centering
		\includegraphics[width=0.48\linewidth]{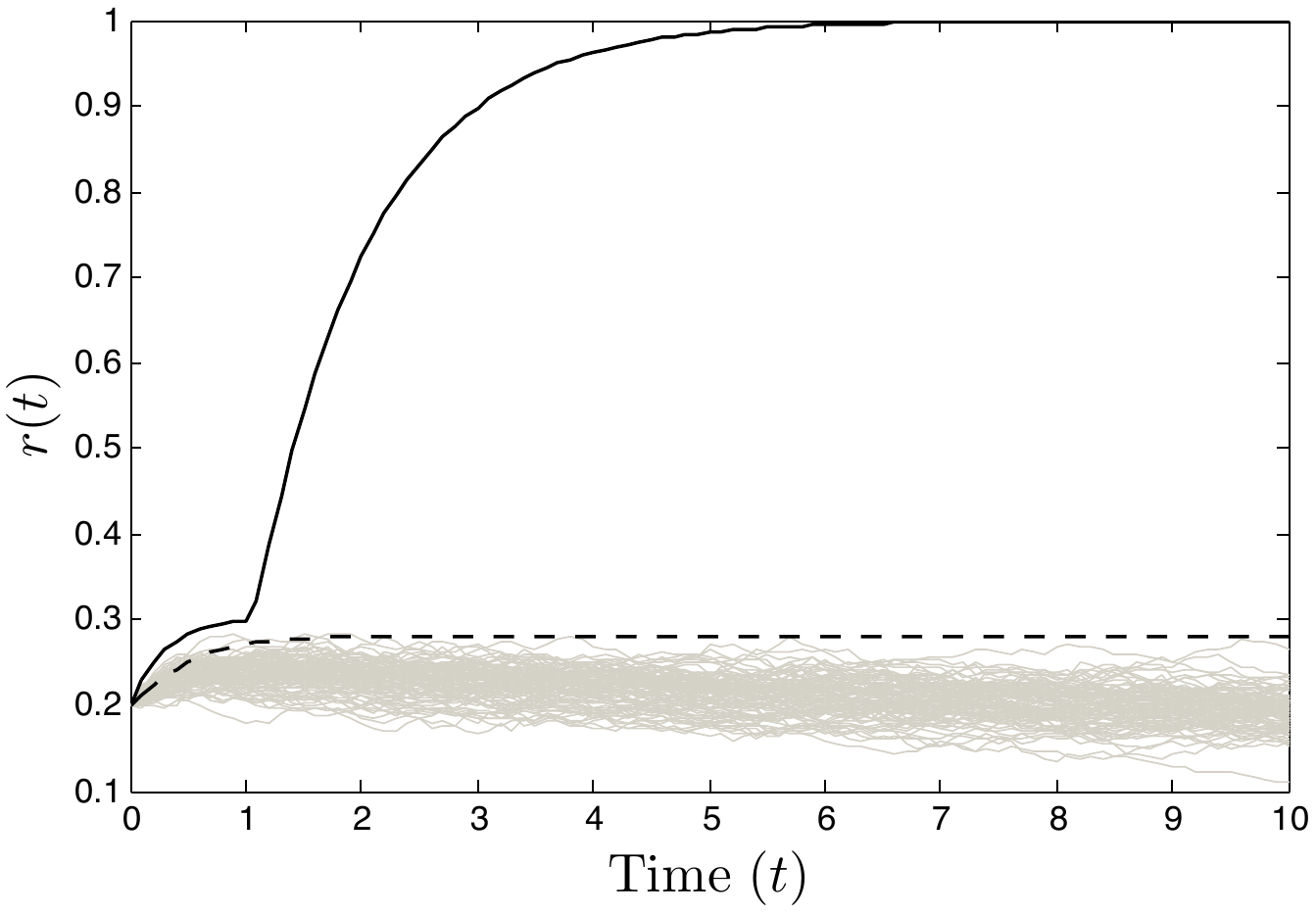}\hfill
		\includegraphics[width = 0.48\linewidth]{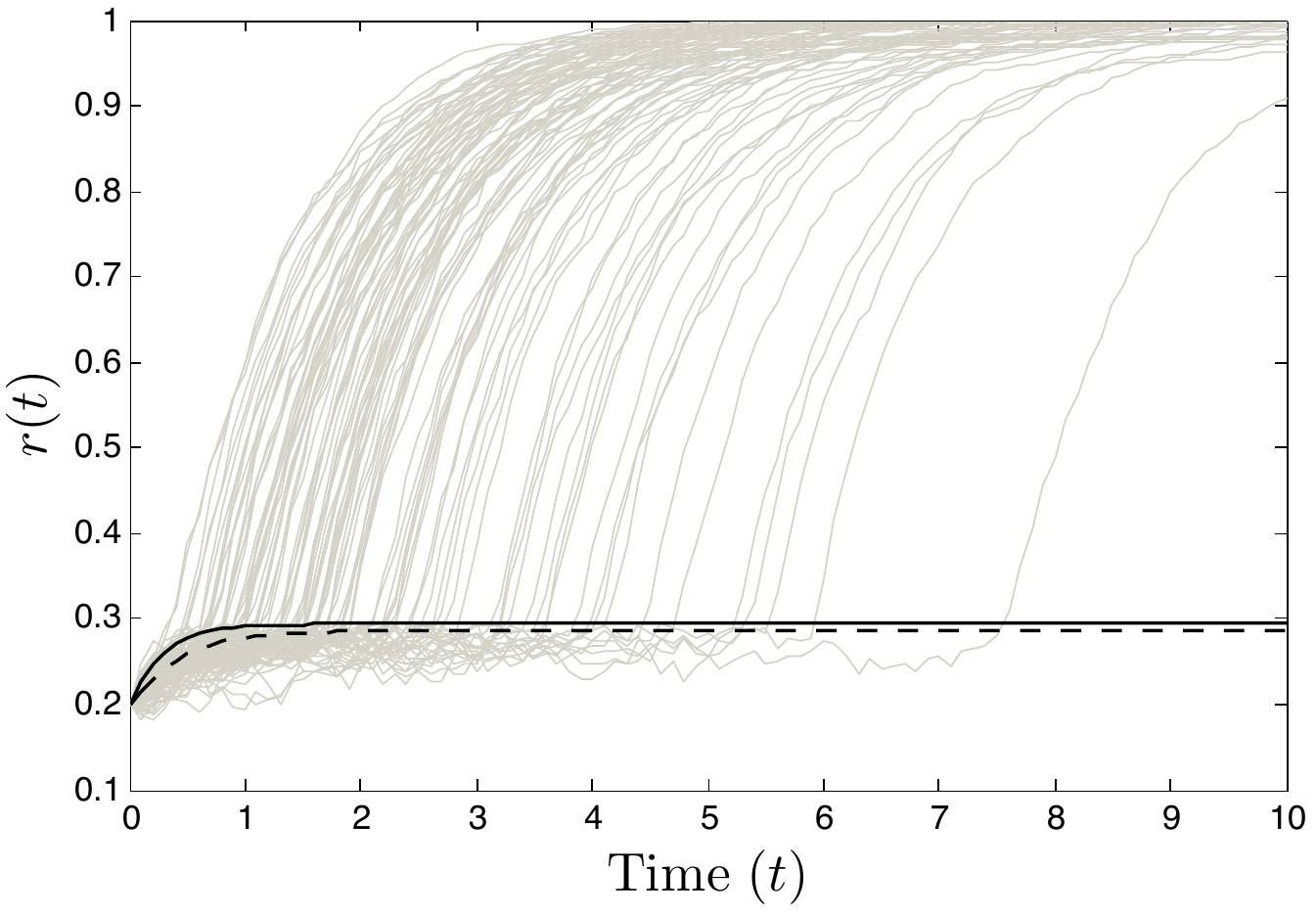}
	\caption{The SVF model can fail to approximate the ABM when parameters are close to the boundary between regions. (Left) Facebook social network $G_F$ ($1-\hat{\alpha} = 0.163$) with parameters in Region III1 ($c_2=2.3$, $1-\hat\alpha < \beta \leq c^*$): \eq{SVF} has solution in Region III1, while ABM has solution in Region IIIe. (Right) Physical contact network $G_P$ ($1-\hat{\alpha} = 0.165$) with parameters in Region IIIe ($c_2=2.4$, $1-\hat{\alpha} < c^* < \beta$): The SVF has solution in Region IIIe, while ABM has solution in Region III1. Time traces of $rep=100$ ABM simulations (grey), solution to the SVF model (solid black), and solution to the BVF model (dashed black) with parameters $\theta=0.15$, $\beta=0.3$, and $c_1=1$ and with intial condition $r_0=0.2$. The BVF model performs much better than the SVF model.}
	\label{fig:fit_fail_regions_alpha_IIIe1}
\end{figure}

Observe that, for the parameters employed in Figs.~\ref{fig:fit_alpha_ODE}-\ref{fig:fit_multi_theta_alpha} ($\theta=0.15$, $\beta=0.3$, $c_1=1$, and $c_2=9$), the SVF model lies in the stable police state dynamical regime (Region III0, $c^*=c_1/(c_1+c_2)  \leq 1-\hat{\alpha} < \beta $). We note that this is critical to the fitting procedure, since the parameter $\alpha$ appears in the solution to \eq{SVF} only when the parameters $\alpha$, $\beta$, $c_1$, and $c_2$ lie in Region III0, see Appendix \ref{Sec:SolnSVF}. We then also attempt to approximate the ABM by the SVF model in Regions IIIe and III1 by 
using the fitted $\hat{\alpha}$ displayed in Fig.~\ref{fig:fit_multi_theta_alpha}.
Figures \ref{fig:fit_alpha_IIIe1}-\ref{fig:fit_fail_regions_alpha_IIIe1} show some representative results. As illustrated in Fig.~\ref{fig:fit_alpha_IIIe1}, for many combinations of parameters the SVF model is able to approximate the behaviour of the ABM in parameter Region IIIe (left panels of Fig.~\ref{fig:fit_alpha_IIIe1}) and parameter Region III1 (right panels of Fig.~\ref{fig:fit_alpha_IIIe1}). However, we observe that under certain circumstances stochastic effects and details of the network structure become important and the SVF ODE model fails to approximate, even in a qualitative sense, the behaviour of the ABM. Specifically, the SVF model fails to capture the behaviour of the ABM when the initial condition is small ($r_0<<1$), when the initial condition is near the boundary between the basins of attraction of $r=0$ and $r=1$, and when parameters $c_1$, $c_2$, $\alpha$, and $\beta$ are close to the boundary between parameter Regions III0, IIIe, or III1 in Fig.\ \ref{fig:SVF_regimes}, see Figs.~\ref{fig:fit_fail_r0_alpha_IIIe1}, \ref{fig:fit_fail_r0_regions_alpha_IIIe1}, and \ref{fig:fit_fail_regions_alpha_IIIe1}, respectively. These are well-known limitations of population-level ODE models that attempt to approximate complex dynamics \cite{AmesEtAl11, Bansal07}. Note that these plots also provide comparisons with the binomial visibility function (BVF) model, which we introduce in Section \ref{Sec:AVF}.

In the previous section we observed that setting $\nu(r_a;\theta) = v_s(r_a;\alpha(\theta))$ in \eq{Approx} exactly recovers the SVF model. Therefore, a natural strategy for improving on the ability of the SVF model to approximate the aggregate behaviour of the ABM is to specify an alternative visibility function that better captures the structure of the underlying network. Section \ref{Sec:AVF} proposes two such visibility functions and compares their ability to approximate the ABM to the ability of the SVF model to approximate the ABM.

%
%


\FloatBarrier
\section{Alternative Visibility Functions for Population-Level ODE}
\label{Sec:AVF}


Due to the difficulty of analyzing and computational cost of simulating the ABM, as well as the limitations of the SVF in approximating the ABM, we propose two alternative choices for visibility functions (different from the step-visibility function $v_s(r;\alpha)$) that can lend insight into how the ABM process behaves on different networks. Specifically, Sections \ref{Sec:BVF}-\ref{Sec:EVF} introduce the binomial and empirical visibility functions, respectively (BVF and EVF). These two visibility functions can be seen to be equivalent in the limit of large network and sample size, as shown in Appendix \ref{Sec:Comp_BVF_EVF}. Section \ref{Sec:Comp_ABM_BVF_EVF} demonstrates numerically that the use of the BVF/EVF in \eq{Approx} represents an improvement over the use of the step visibility function $v_s(r;\alpha)$ (SVF) when approximating the ABM for real networks.


\subsection{Binomial Visibility Function}
\label{Sec:BVF}

\begin{figure}[h]
	\centering
	\begin{subfigure}[t]{0.48\linewidth}
		\includegraphics[width=\linewidth]{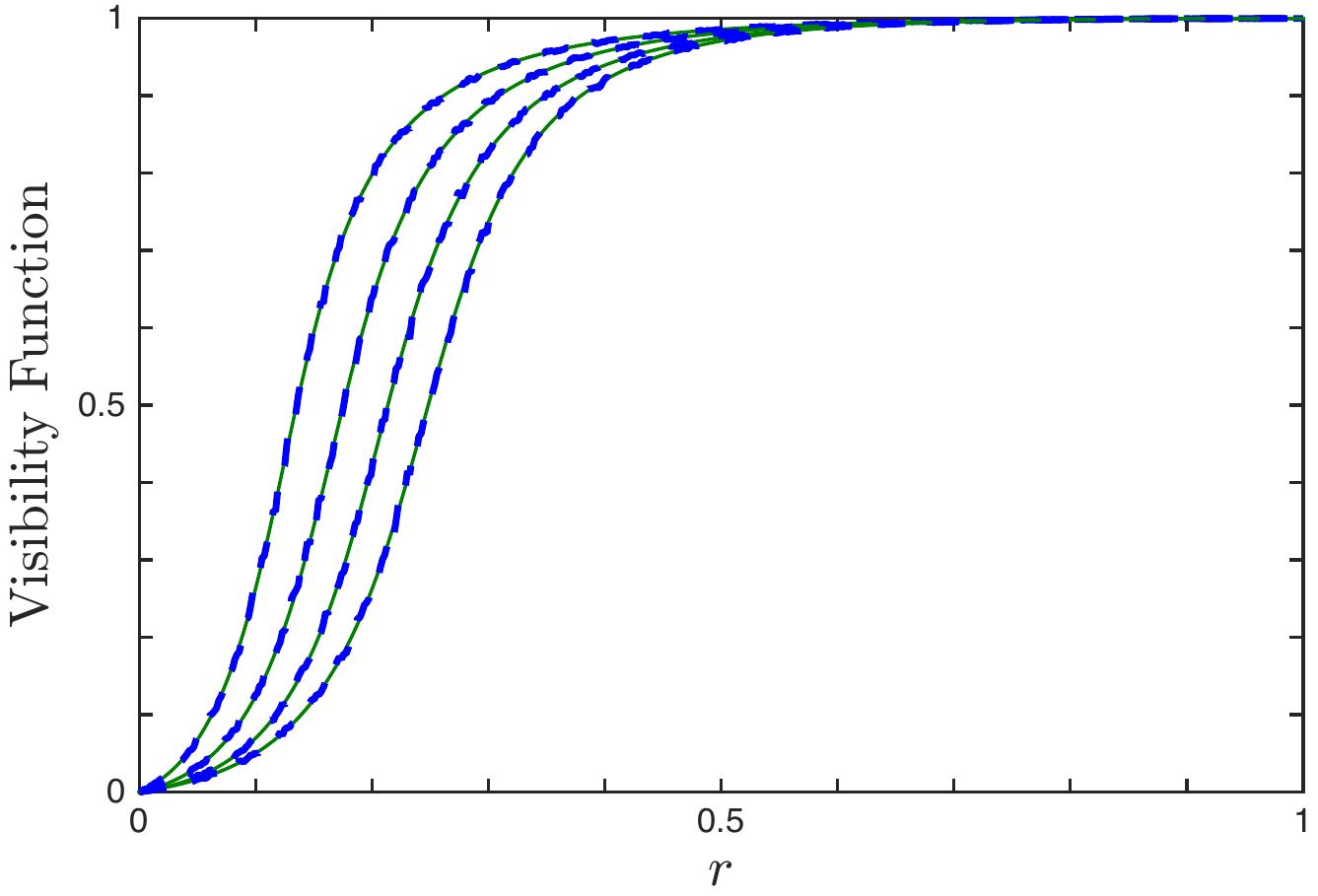}
		\caption{Facebook social network $G_F$}
	\end{subfigure}
	\hfill
	\begin{subfigure}[t]{0.48\linewidth}
		\includegraphics[width=\linewidth]{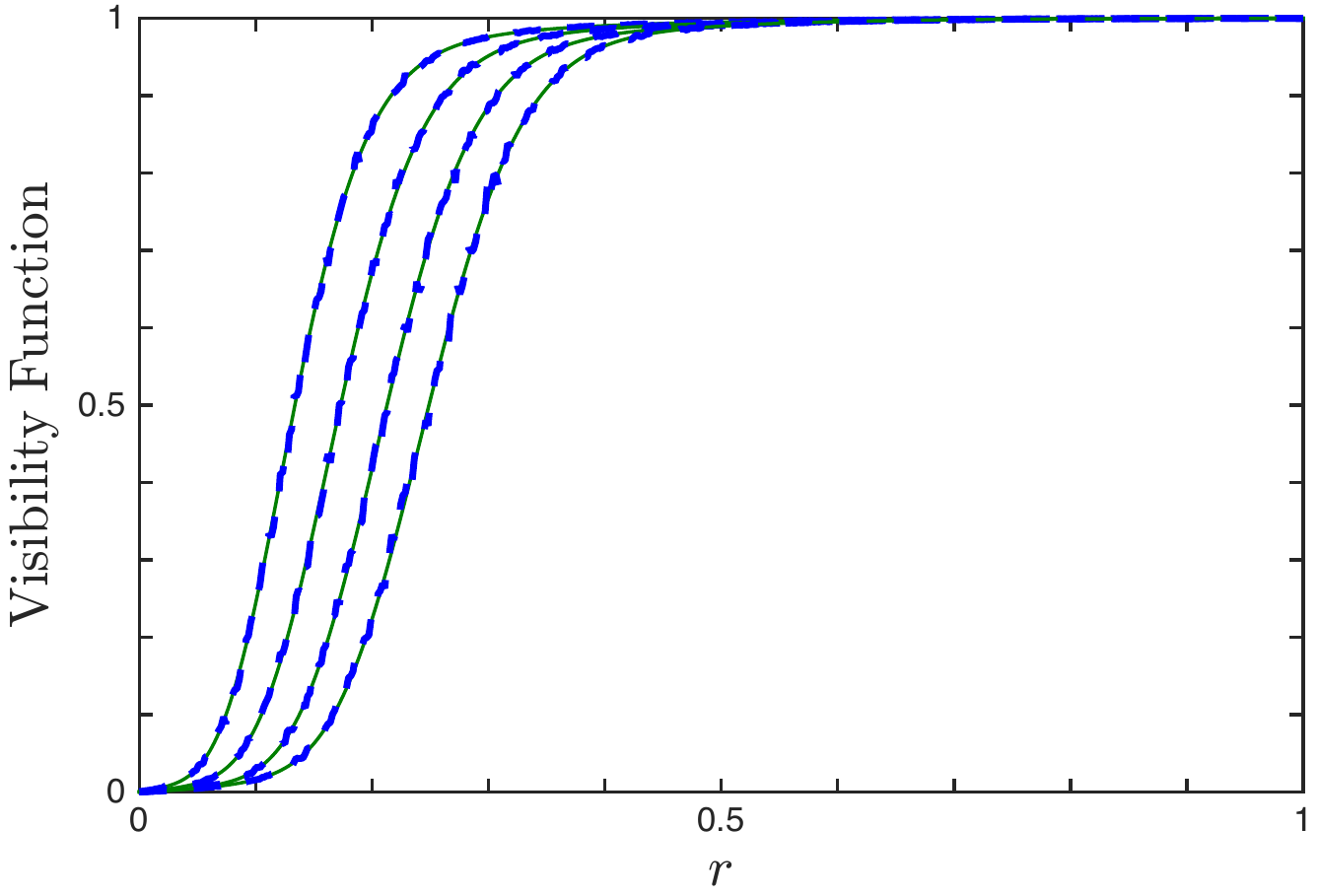}
		\caption{Physical Contact network $G_P$}
	\end{subfigure}
	\caption{Binomial visibility function (solid green) and empirical visibility function (dashed blue) for Facebook and physical contact networks generated for linear thresholds $\theta = \{0.13, 0.17, 0.21, 0.25\}$ (curves from left to right). The binomial and empirical visibility functions are computed on the grid $r\in\{0, \frac{1}{512}, \frac{2}{512}, \ldots, 1\}.$}
	\label{fig:BVF}
\end{figure}

Consider the following derivation. Suppose that $v\in V$ is a node in the network $G=G(V,E)$ with degree $k$, and that the fraction of nodes active in the revolution is $r$. Also, assume that the states of $v$'s neighbours are active with probability $r$ independently so that the probability of $v$ having $j$ neighbours active in the revolution is 
$$	
	\binom{k}{j}r^j(1-r)^{k-j}.
$$
The probability that the fraction of $v$'s neighbours exceeds the linear threhsold $\theta$ is, therefore,
\begin{equation}
	\label{eq:ProtoBVF}
	\sum_{j = \cl{\theta k}}^k \binom{k}{j} r^j(1-r)^{k-j} = \mbox{BinCDF}(k-\cl{\theta k};k, 1-r),
\end{equation}
where $\mbox{BinCDF}(x;n,p)$ is the cumulative distribution function for the binomial distribution with $n$ trials and probability of success $p$ evaluated at $x$. This quantity can be used to approximate the function $\nu(r; \theta)$ (the expected fraction of the population that can see the revolution). In \cite{LangDeSterck14} we considered $k$ in \eq{ProtoBVF} to be the average degree over all nodes and argued that \eq{ProtoBVF} is a sigmoidal function that can be approximated by step-visibility function $v_s(r;\alpha)$ for some appropriate visibility parameter $\alpha$
(see Appendix \ref{Sec:JustSVF} for a brief summary of this argument). One possible improvement on adopting $\nu(r;\theta) = v_s(r;\alpha(\theta))$ in \eq{Approx} (resulting in the SVF) is to actually adopt $\nu(r;\theta) = \mbox{BinCDF}(\fl{k}-\cl{\theta k};\fl{k}, 1-r)$ instead, where $k$ is the average degree of the network. However, we can improve on this even further in terms of using more relevant information about the real graph by choosing $\nu(r;\theta)$ to be equal to
\begin{equation}
	\label{eq:BVF}
	v_b(r;\theta,\rho_k) = \sum_k \rho_k \mbox{BinCDF}(k - \lceil \theta k \rceil; k, 1-r) = \sum_k \sum_{j=0}^{k-\cl{\theta k}} \rho_k \binom{k}{j} (1-r)^jr^{k-j}.
\end{equation}	
We call the function $v_b(r;\theta,\rho)$ in \eq{BVF} the \emph{binomial visibility function} (BVF), see Fig.~\ref{fig:BVF}. 
Note the steep sigmoidal shape of $v_b(r;\theta,\rho_k)$ for the two empirical networks of Fig.~\ref{fig:BVF}, further
justifying the choice of a step visibility function as an appropriate approximation in \cite{LangDeSterck14}.
We note that computing the binomial visibility function on some grid of $\mu+1$ discrete $r$-values $r\in\{r_i\}_{i=1}^{\mu+1}$ requires calculating a double sum for every $i=1,\ldots\mu+1$. In practice, this can be very expensive, especially for degree distributions $\rho_k$ with fat tails. For this reason, the following section introduces a complementary but less expensive approximation to the visibility function that is equivalent to the binomial visibility function of \eq{BVF} in the limit of large network and sample size.


\subsection{Empirical Visibility Function}
\label{Sec:EVF}


We now propose to reconstruct the binomial visibility function $v_b(r;\theta,\rho_k)$ empirically by sampling the underlying network directly. Specifically, for fixed $\theta$ we approximate $v_b$ by a discrete function $v_e\in \mathbb{R}^{\mu+1}$ where $\forall j=1,\ldots, \mu+1: v_{e,j} \approx v_b(\frac{j-1}{\mu})$. Since $v_b(0)=0$ and $v_b(1)=1$, we set $v_{e,1}=0$ and $v_{e, \mu+1}=1$. We calculate $v_{e,j}$ for $j=2,\ldots,\mu$ as follows.
\begin{enumerate}
	\item Seed the network with $r_j=\frac{j-1}{\mu}$ active nodes in expectation. (Generate $N$ random numbers drawn from the uniform distribution on $[0,1]$: $rand_i$ for $i=1,\ldots,N$. If $rand_i < \frac{j-1}{\mu}$ then node $i$ is active, otherwise node $i$ is inactive.)
	\item Determine the fraction of nodes that can see the revolution.
	\item Repeat 1-2 a total of $rep$ times and set $v_{e,j}$ to be the average over the realizations. (We choose $rep=100$ in our numerical experiments.)
\end{enumerate}
We denote the linear interpolation between the $v_{e,j}$ by the function $v_e(r;\theta)$
\begin{equation}
	\label{eq:EVF}
	v_e(r;\theta) = \left\{ \begin{array}{ll} v_{e,j} & \mbox{if } r = \frac{j-1}{\mu} \mbox{ for some } j=1,\ldots,\mu+1 \\
		 v_{e,j}\frac{j/\mu - r}{1/\mu} + v_{e,j+1}\frac{r - (j-1)/\mu}{1/\mu} & \mbox{if } r\in [\frac{j-1}{\mu}, \frac{j}{\mu}] \mbox{ for some } j=1,\ldots,\mu\end{array}\right. 
\end{equation}
which we call the \emph{empirical visibility function}, see Fig.~\ref{fig:BVF}.

We now make several brief remarks about the procedure for calculating the empirical visibility function, outlined above. Firstly, we remark that this procedure is easily extended to the case where $\theta_v$ are specified for each individual node. However, to keep our analysis as simple as possible we continue to restrict ourselves to the case where $\theta_v\equiv\theta$. 
In Appendix \ref{Sec:Comp_BVF_EVF} we show that the empirical and binomial visibility functions are equivalent in the limit of large network and sample size. Figure \ref{fig:BVF} shows that they can also be expected to give similar results for finite network and sample size.
Finally, we remark that compared to the binomial visibility function, the procedure for calculating the empirical visibility function is in practice significantly less costly to implement, especially for networks with fat-tailed degree distributions.


\subsection{Comparison of ABM and BVF/EVF Models}
\label{Sec:Comp_ABM_BVF_EVF}


In this section we compare the ability of the BVF model to approximate the ABM with that of the SVF model. We note that since the BVF and EVF are equivalent in the limit of large network and sample size
and are very close in practice (see Fig.~\ref{fig:BVF}), the observations we make about the behaviour of the BVF model also apply to the behaviour of the EVF model. We further note that the behaviour of the BVF/EVF model has parameters regimes that induce behaviour analogous to those observed in the SVF model.
Since the BVF/EVF have sigmoidal shape, this follows directly from the analysis presented in Appendix B of \cite{LangDeSterck14} for visibility and policing functions of general sigmoidal shape (the results from this analysis are summarized in Appendix \ref{Sec:Ext}).

\begin{figure}[h]
	\centering
	\begin{subfigure}[t]{0.48\linewidth}
		\includegraphics[width=\linewidth]{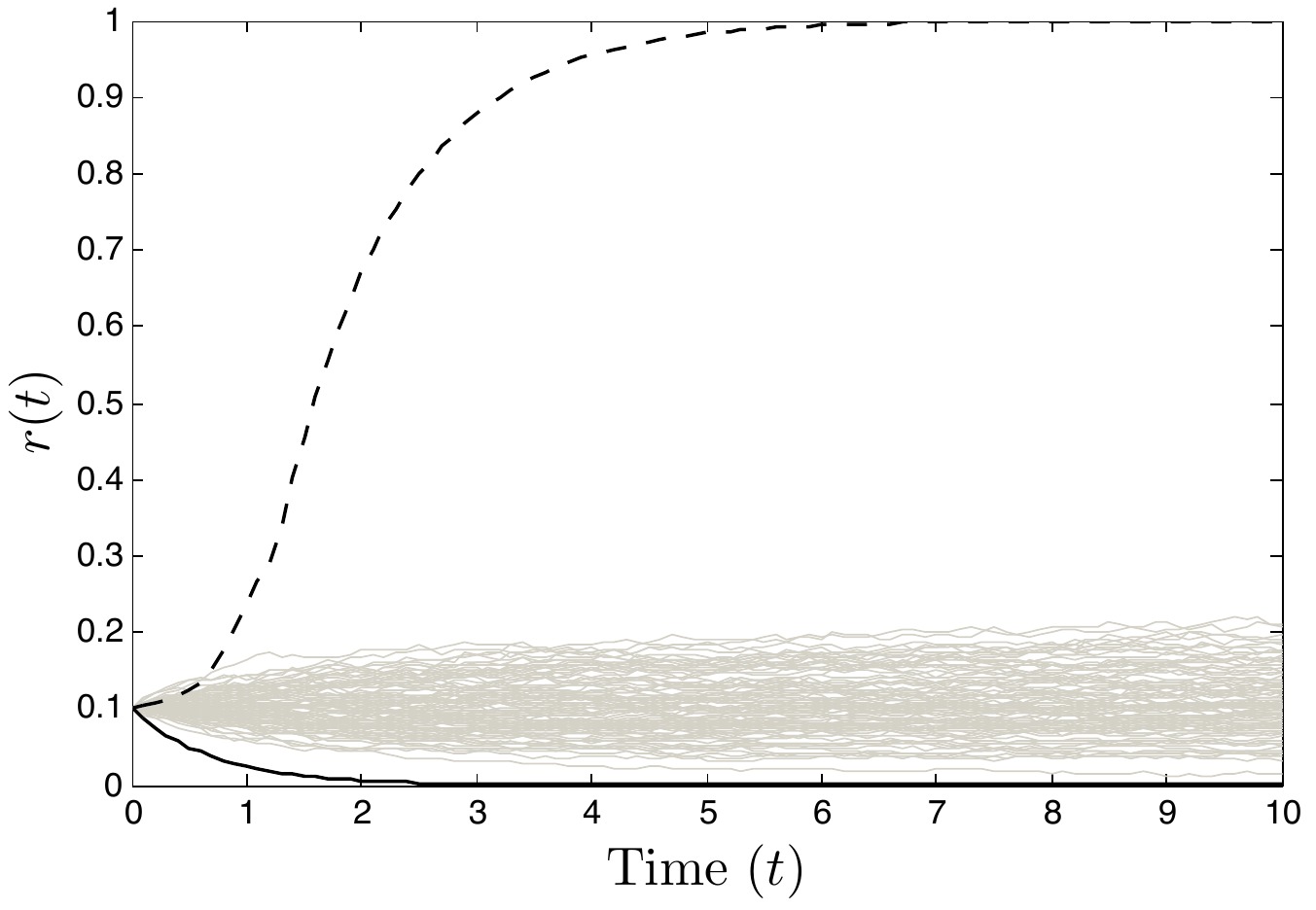}
		\caption{Both SVF and BVF models fail to approximate the dynamics of the ABM on the Facebook social network $G_F$ ($1-\hat{\alpha}_{ODE} = 0.163$) with parameters in Region III0 ($c_2=1.4$, $1-\hat\alpha_{ODE} < \beta \leq c^*\approx0.417$).}
		\label{fig:fail}
	\end{subfigure}
	\hfill
	\begin{subfigure}[t]{0.48\linewidth}
		\includegraphics[width=\linewidth]{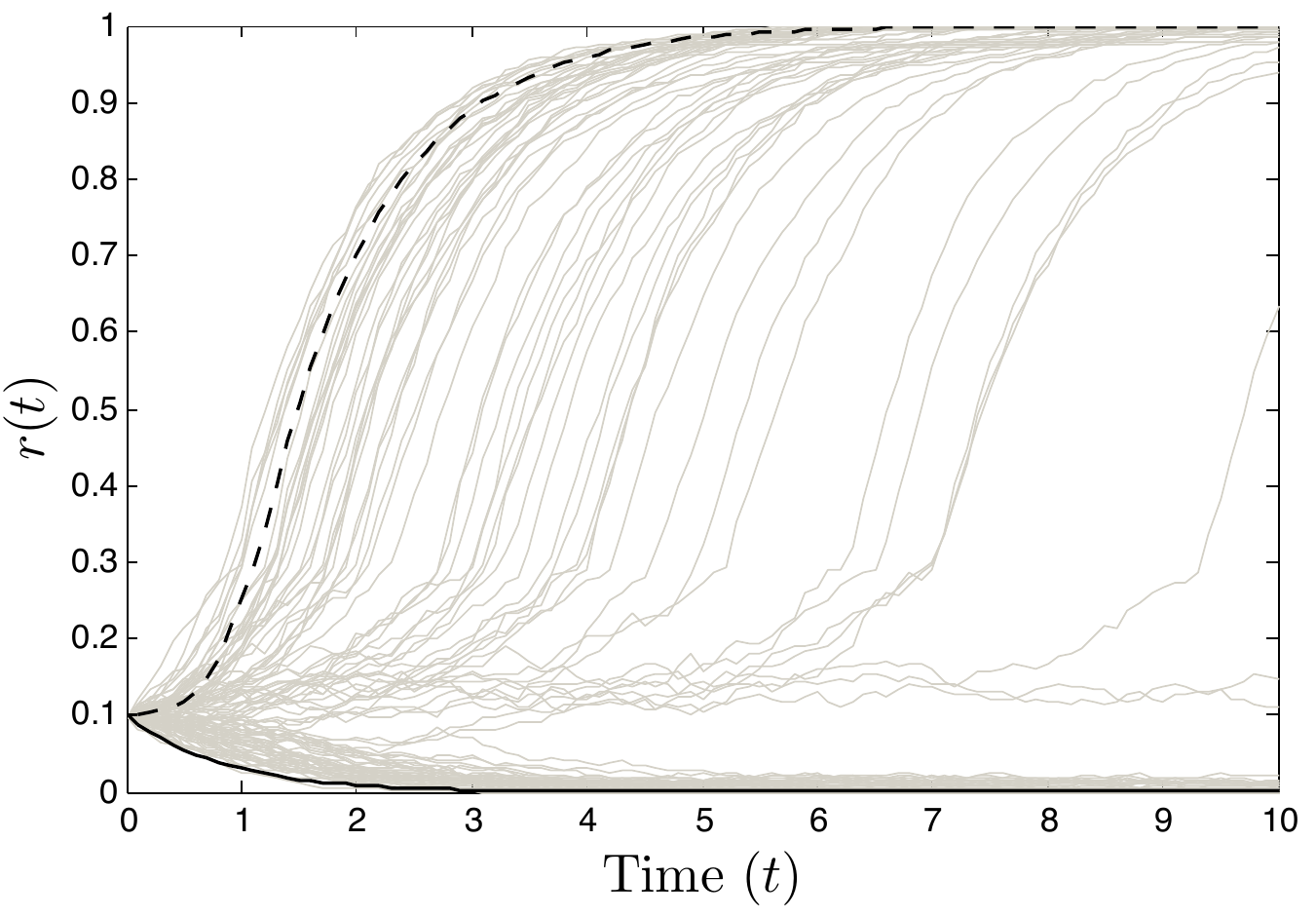}
		\caption{The SVF and BVF models provide complementary approximations of the dynamics of the ABM on the physical contact network $G_P$ ($1-\hat{\alpha}_{ODE} = 0.165$) with parameters in Region III0 ($c_2=1.2$, $1-\hat{\alpha}_{ODE} < c^*\approx0.455 < \beta$).}
		\label{fig:complement}
	\end{subfigure}
	\caption{The SVF and the BVF models can (a) both fail to approximate the dynamics of the ABM, or (b) act as complementary approximations to the dynamics of the ABM. Time traces of $rep=100$ ABM simulations (grey), solution to the SVF model (black), and solution to the BVF model (dashed black), with parameters $\theta=0.15$, $\beta=0.3$, and $c_1=1$ and with intial condition $r_0=0.1$. }
\end{figure}

We observe four types of outcomes when comparing the ability of the BVF model to approximate the ABM to the ability of the SVF model to approximate the ABM:
\begin{enumerate}
	\item The BVF model may succeed better in approximating the dynamics of the ABM than the SVF model, even if the improvement is only qualitative in nature, see Figs.~\ref{fig:fit_fail_r0_alpha_IIIe1}a and b (right panel), \ref{fig:fit_fail_r0_regions_alpha_IIIe1}, and \ref{fig:fit_fail_regions_alpha_IIIe1} (left panel), 
	\item The BVF model may complement the dynamics predicted by the SVF model, see Fig.~\ref{fig:complement},
	\item The BVF model may produce qualitatively similar predictions to the SVF model, see Figs.~\ref{fig:fit_alpha_IIIe1}, \ref{fig:fit_fail_r0_alpha_IIIe1}b (left panel) and \ref{fig:fit_fail_regions_alpha_IIIe1} (right panel), or
	\item Both the BVF and the SVF models fail to approximate the dynamics of the ABM either qualitatively or quantitatively, see Figs.~\ref{fig:fit_fail_r0_alpha_IIIe1}b (left panel) and \ref{fig:fail}. 
\end{enumerate}
For each of these outcomes, the approximation of the ABM by the BVF model is no worse than the approximation by the SVF model, and, in many cases the approximation of the ABM by the BVF model is much better than the approximation by the SVF model. Therefore, we conclude that the BVF model, and hence the EVF model, represents an improvement over the SVF model with respect to their ability to approximate the behaviour of the ABM. This is no surprise, since the BVF/EVF model takes important information of the real network into account.


\section{Higher Order ODE Models}
\label{Sec:HOT}

In the previous section we introduced an extension to the SVF model. In particular, we showed that the BVF/EVF model is better able to approximate the ABM than the SVF model. However, both the BVF/EVF and SVF models are one-dimensional ordinary differential equation. This section now compares the performance of the BVF/EVF models as approximations to the ABM to the performance of a higher order (i.e., higher dimensional) compartmental model which we modify from \cite{NekoveeEtAl07}. We call this model the \emph{degree approximation} (DA) because it compartmentalizes individuals based on their state as well as their degree. We demonstrate numerically that the BVF model, and hence the EVF model, is no worse at approximating the aggregate behaviour of the ABM than the DA. This highlights the usefulness of the BVF/EVF model, especially since the DA is in practice much more computationally expensive to solve and much harder to analyze than the BVF/EVF model.

\subsection{Degree Approximation Model}

Before we detail the approach which we modify from \cite{NekoveeEtAl07}, it is useful to introduce some notation. We define $V_k$ to be the set of nodes with degree $k$ and $N_k$ to be the set of nodes that have at least one neighbour of degree $k$,
\begin{align*}
	V_k &= \{v\in V: d_v = k\}, \mbox{ and}\\
	N_k &= \{w\in V:\exists v\in V_k \mbox{ such that } (v,w)\in E\},
\end{align*}
respectively, where we denote the degree distribution of $N_k$ by $\rho_{k,j}$, i.e.
$$	
	\rho_{k,j} = \frac{\mbox{number of nodes of degree $j$ that have at least one neighbour of degree $k$}}{\mbox{number of nodes that have at least one neighbour of degree $k$}} = \frac{|V_j \cap N_k|}{|N_k|}.
$$
As mentioned above, the degree approximation aggregates individuals according to their state and degree. Thus, conditioning on the state of the system at time $t_0$, we define
$$
	 r^{(k)}(t) =r^{(k)} (t|t_0) = \frac{1}{|V_k|}\sum_{v\in V_k}\ex[ s_v(t)|t_0]
$$
to be the fraction of nodes with degree $k$ that are expected to be in state 1 at time $t$. Since nodes and edges are neither created nor destroyed, the fraction of nodes with degree $k$ that are expected to be in state 0 at time $t$ is given by $1-r^{(k)}(t)$. Analogously, we define
$$
	 r_d(t) = r_d(t|t_0) = \frac{1}{N} \sum_k |V_k|r^{(k)}(t)
$$
to be the fraction of nodes expected to be in state 1 at time $t$.

Applying the notation introduced above with the rules characterizing the dynamics of the process we find an equation analogous to \eq{preApprox}

\begin{equation}
	\label{eq:preApproxDA}
	\Delta r^{(k)}(t) = r^{(k)}(t) - r^{(k)}(t_0) = g^{(k)}(t|t_0) - d^{(k)}(t|t_0),
\end{equation}
where the expected growth and decay of the fraction of active nodes is modelled by the nonnegative growth and decay functions $g^{(k)}(t)=g^{(k)}(t|t_0)$ and $d^{(k)}(t)=d^{(k)}(t|t_0)$, respectively. In order to close the degree approximation we need to approximate the quantities $g^{(k)}(t)$ and $d^{(k)}(t)$ in terms of $r^{(k)}(t)$, $\{\theta_v\}_{v\in V}$, and $\beta$.

For notational convenience we first suppress the time argument, and consider the case where $\theta_v\equiv\theta$. In order to approximate $g^{(k)}(t)$ or $d^{(k)}(t)$ in terms of $r^{(k)}$, $\theta$, and $\beta$, we are now required to make an assumption about how the states of nodes sharing a common neighbour are correlated. Specifically, as for the BVF and EVF models, for any fixed $v\in V_k$ we assume that the states of any two neighbours of $v$ are independent. This assumption implies that the probability of $v$ having exactly $j$ active neighbours is $\binom{k}{j}\bar{n}_k^j(1-\bar{n}_k)^{k-j}$, where $\bar{n}_k = \sum_l \rho_{k,l}r^{(l)}$ is the fraction of nodes in $ N_k$ with state 1. So, the probability of $v$ having at least $\lceil \theta k\rceil$ active neighbours is
$$
	\sum_{j=\lceil \theta k \rceil}^k \binom{k}{j} \bar{n}_k^j (1 - \bar{n}_k)^{k-j} = 1 - \Binom(\lceil \theta k \rceil -1;k,\bar{n}_k) = \Binom(k - \lceil \theta k \rceil; k,1-\bar{n}_k),
$$
It follows that
\begin{equation}
	\label{eq:A0DA}
	g^{(k)}(t) = c_1 \spc (1- r^{(k)}(t_0)) \spc \Binom(k - \lceil \theta k \rceil; k,1-\bar{n}_k) \Dt + o(\Dt).
\end{equation}
As in \eq{A1}, we choose
\begin{equation}
	\label{eq:A1DA}
	d^{(k)}(t) = c_2 \spc r^{(k)}(t_0) \spc p(r_d(t_0);\beta) \Dt + o(\Dt).
\end{equation}
Combining equations \eqref{eq:preApproxDA}-\eqref{eq:A1DA} and dividing by $\Delta t$ gives
$$
	\frac{\Delta r^{(k)}}{\Delta t} = c_1 \spc (1 - r^{(k)}(t_0) ) \spc \Binom(k - \lceil \theta k \rceil; k,1-\bar{n}_k) 
		- c_2 \spc r^{(k)}(t_0) \spc p(r^{(k)}(t_0);\beta) + o(1).
$$
Now, taking the limit as $\Delta t\rightarrow 0$ yields
\begin{equation}
	\label{eq:ApproxDA}
	\frac{dr_k}{dt} = c_1 (1 - r^{(k)}) \Binom(k - \lceil \theta k \rceil; k,1-\bar{n}_k) - c_2 \spc r^{(k)} \spc p(r^{(k)}; \beta).
\end{equation}

\begin{figure}[h]
	\centering
	\begin{subfigure}[t]{\linewidth}
		\includegraphics[width=0.3\linewidth]{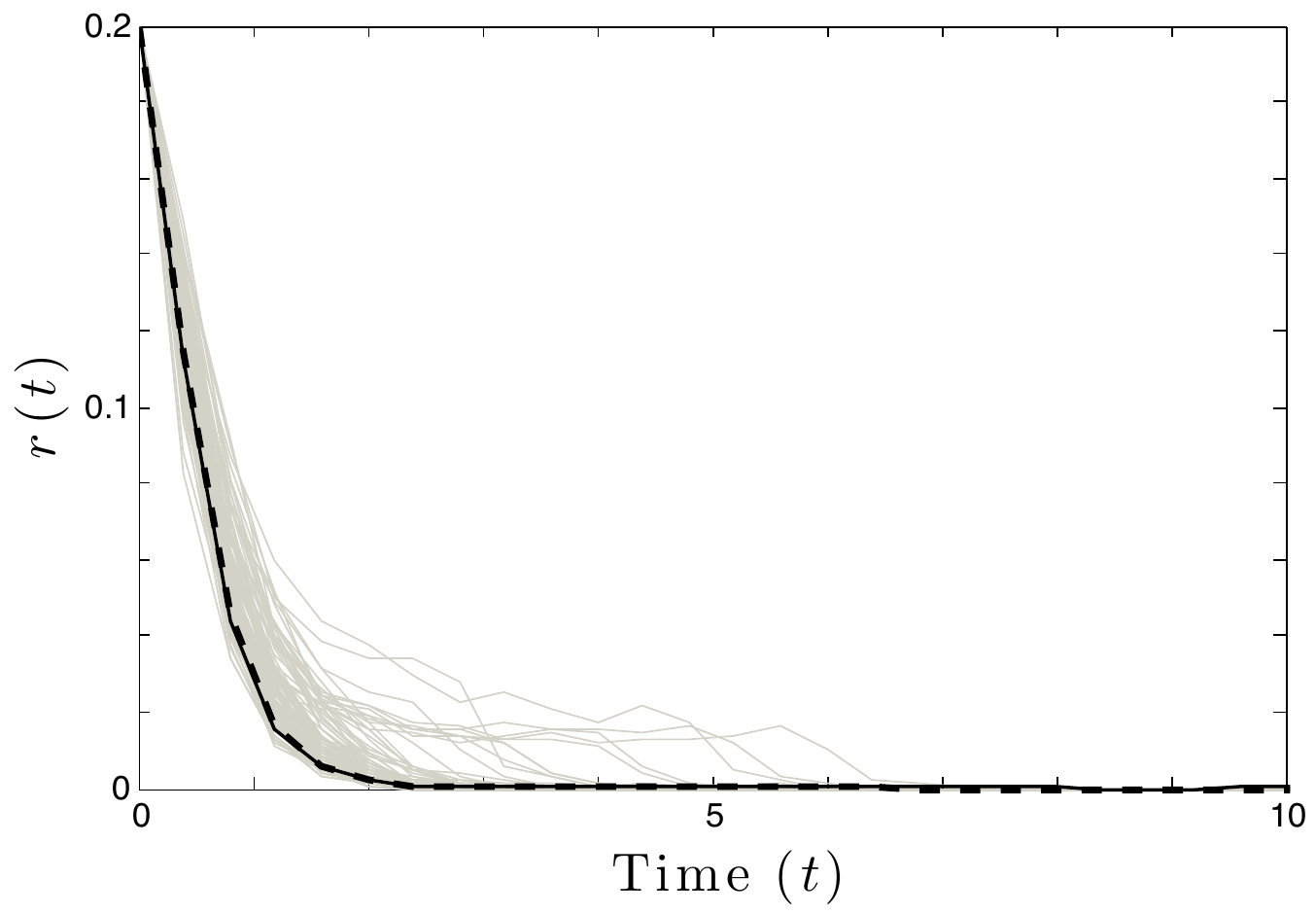}
		\hfill
		\includegraphics[width=0.3\linewidth]{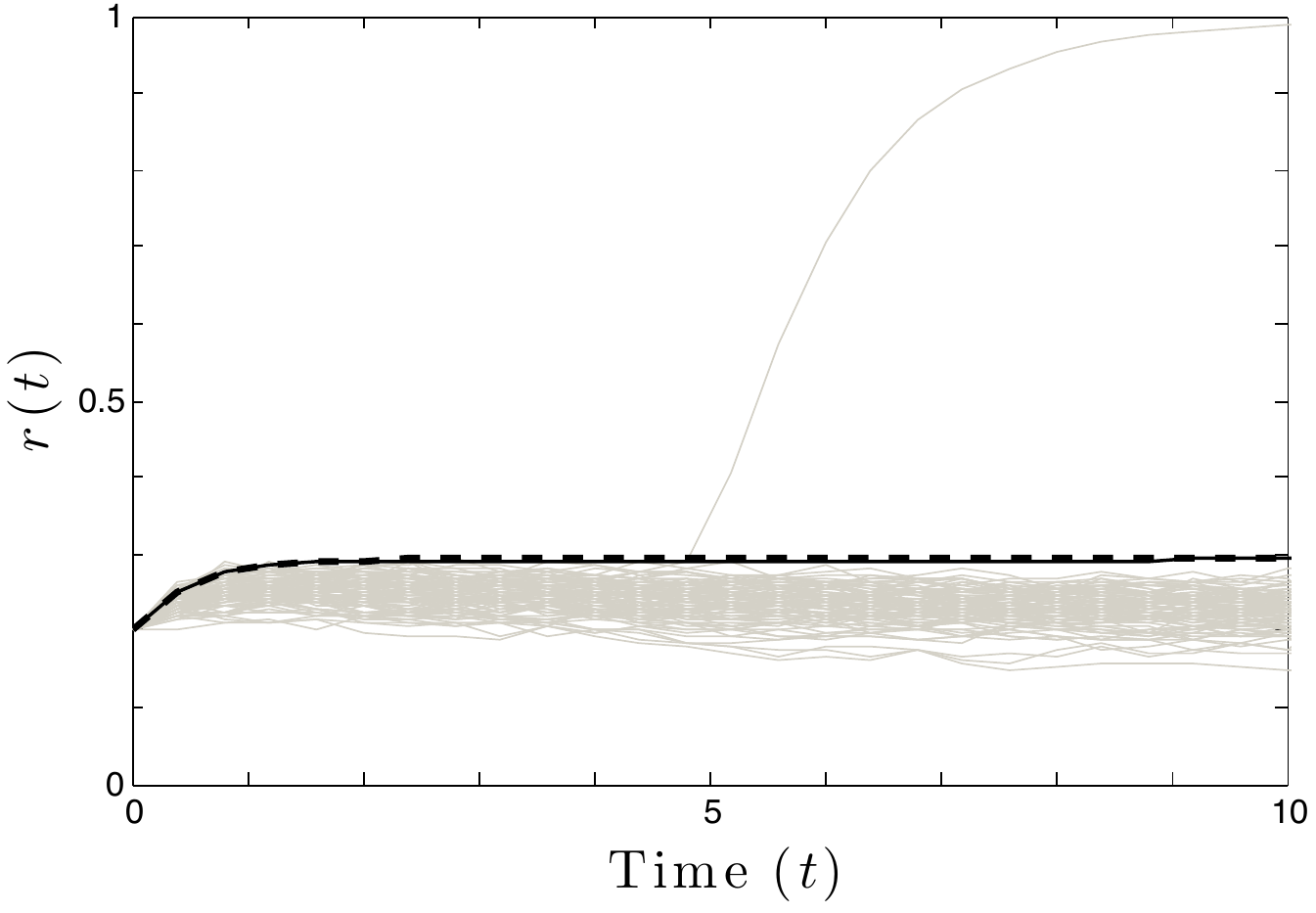}
		\hfill
		\includegraphics[width=0.3\linewidth]{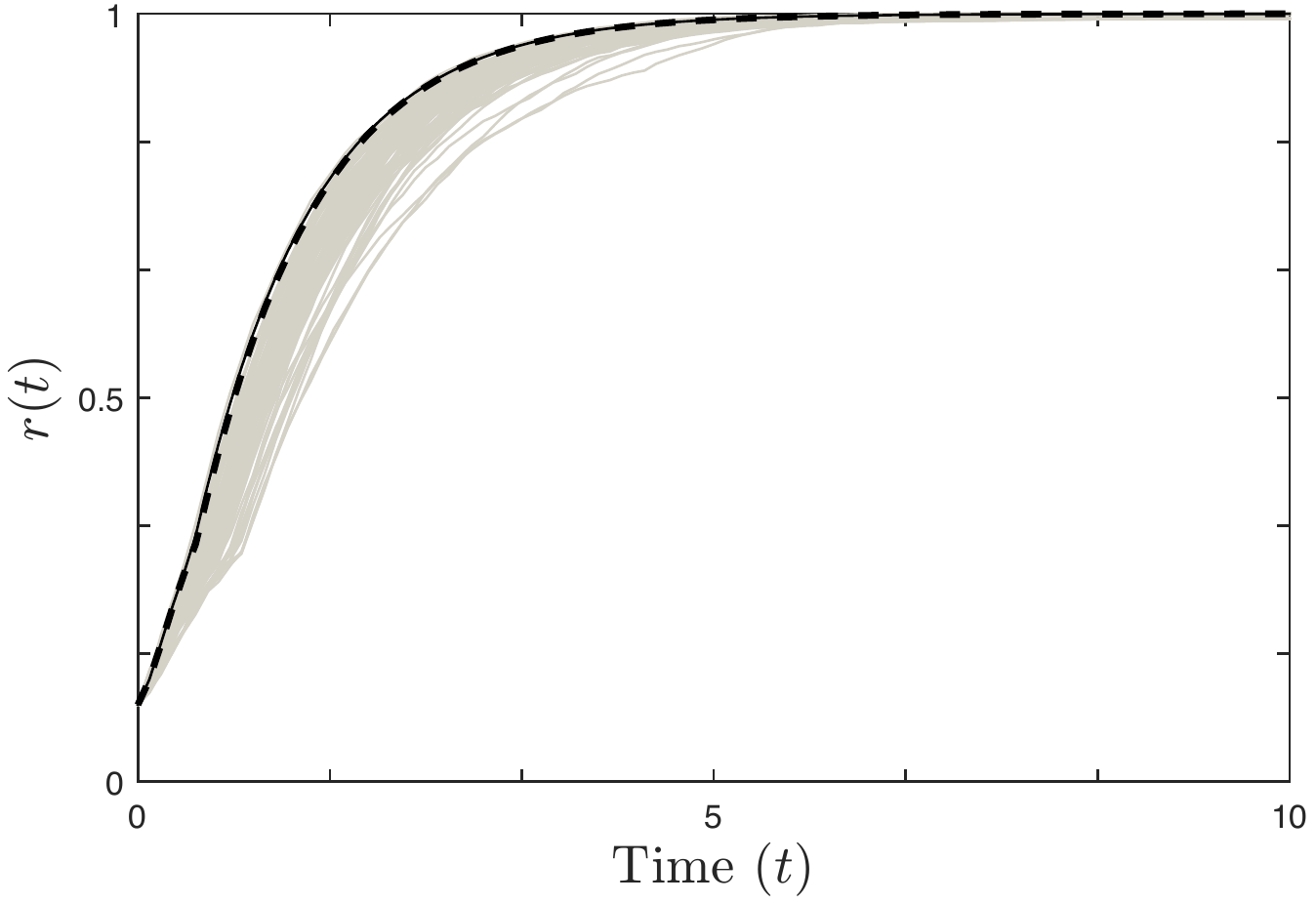}
		\caption{Facebook social network $G_F$. Parameters $c_1=1$, $\beta=0.3$, and (Left) $\theta=0.19$, $c_2=3.0$, $r_0=0.2$ (Middle) $\theta=0.15$, $c_2=2.2$, $r_0=0.2$ (Right) $\theta=0.11$, $c_2=1.0$, $r_0=0.1$.}
	\end{subfigure}
	\\
	\begin{subfigure}[t]{\linewidth}
		\includegraphics[width=0.3\linewidth]{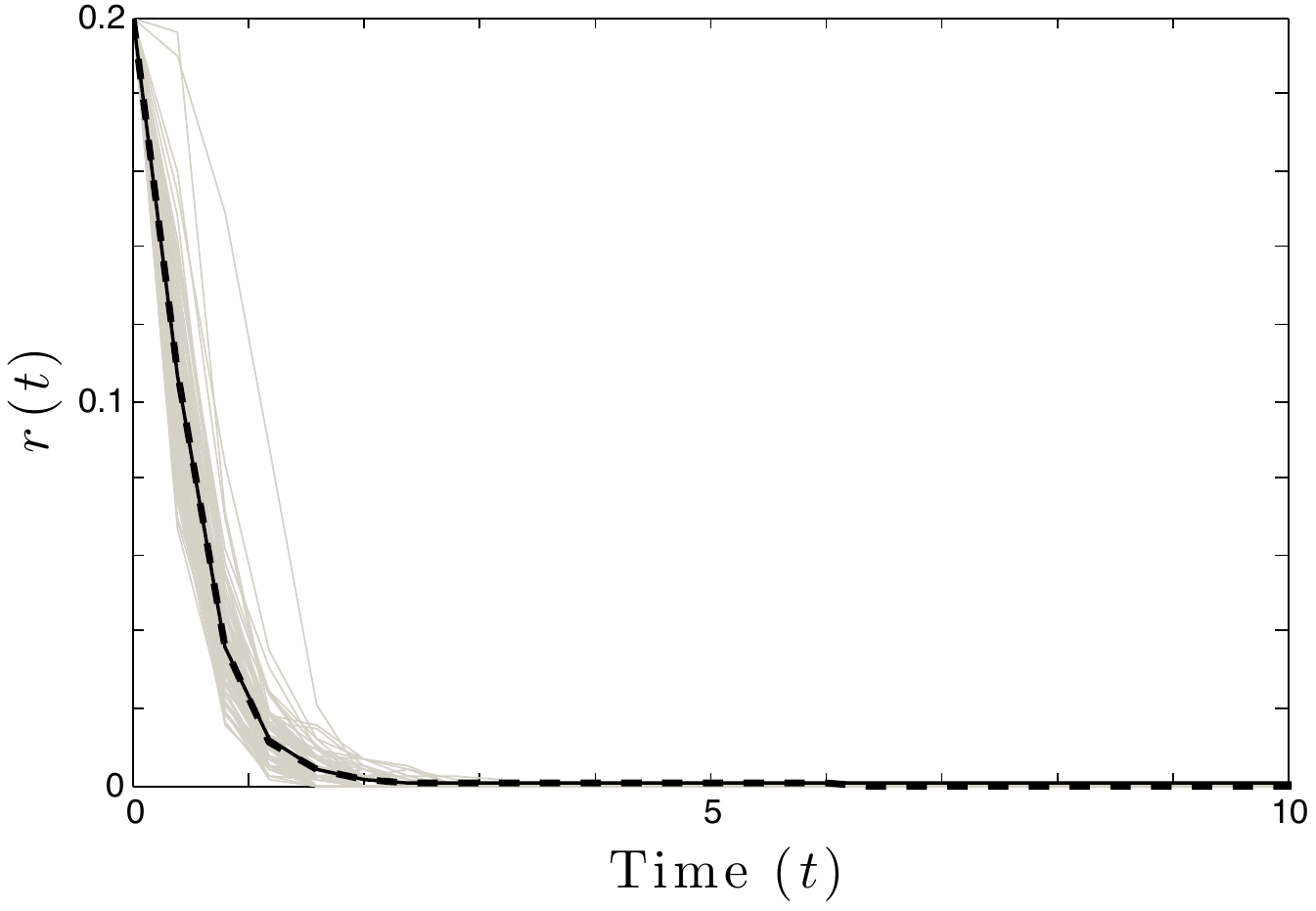}
		\hfill
		\includegraphics[width=0.3\linewidth]{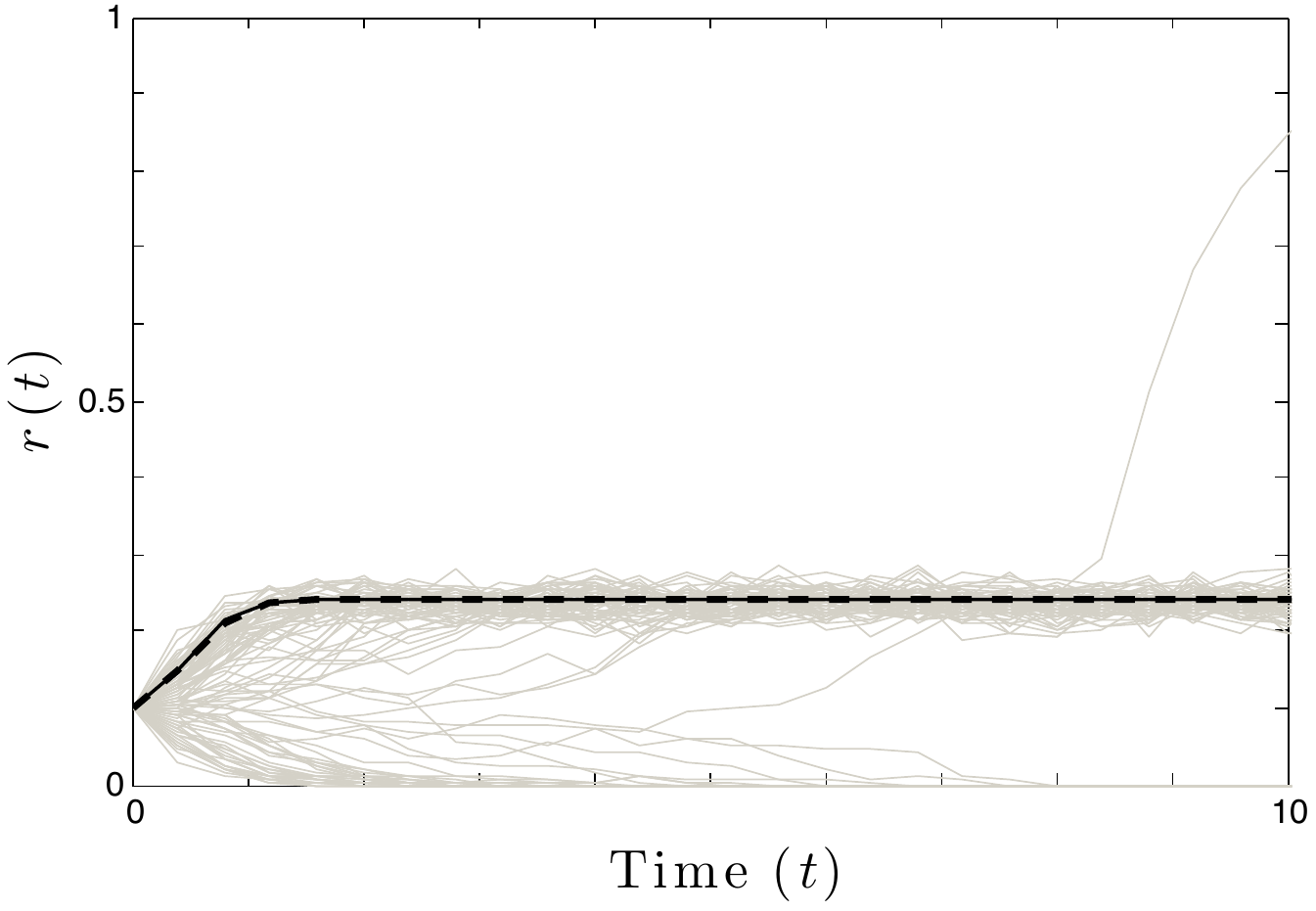}
		\hfill
		\includegraphics[width=0.3\linewidth]{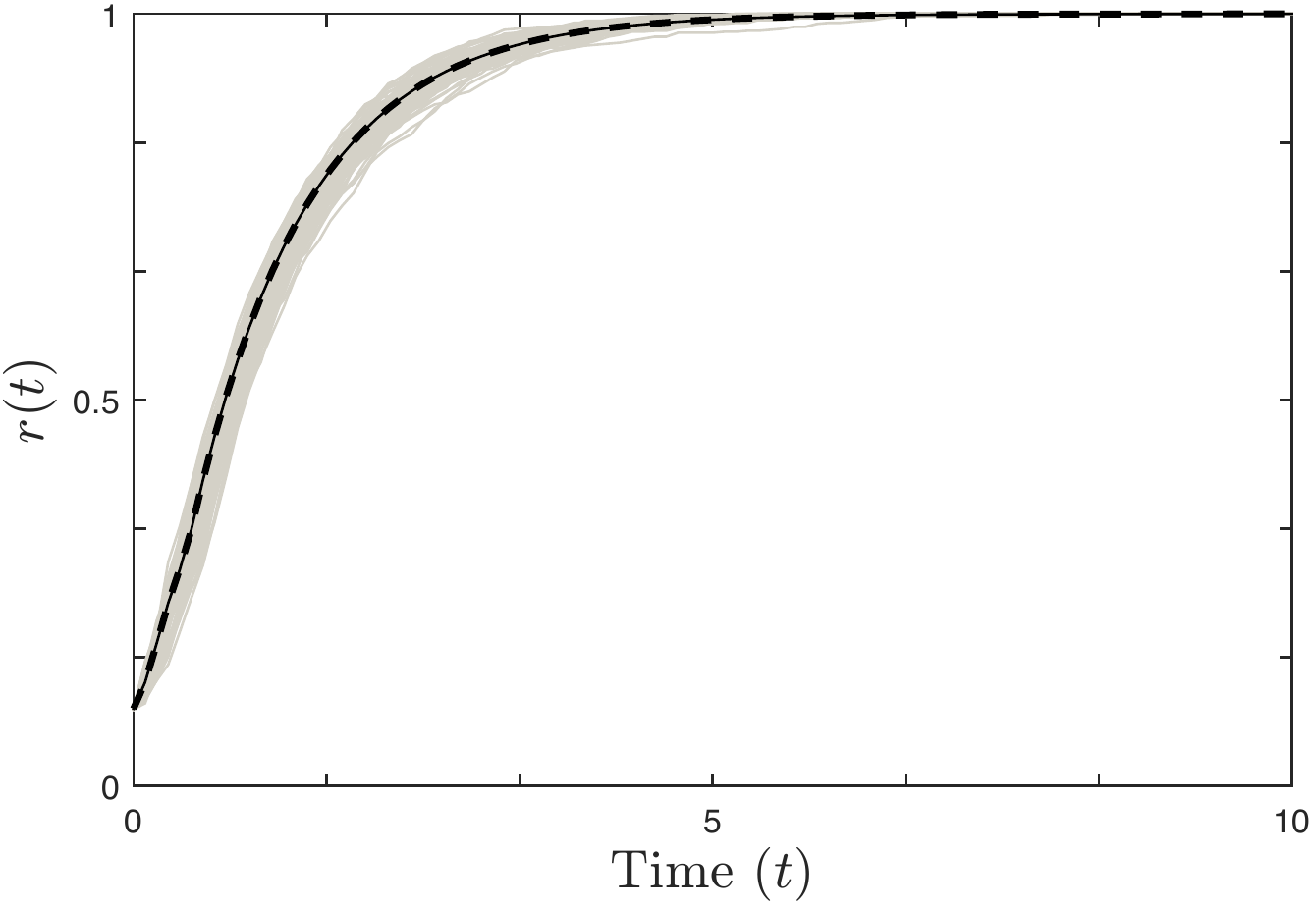}		
		\caption{Physical contact network $G_P$. Parameters $c_1=1$, $\beta=0.3$, and (Left) $\theta=0.19$, $c_2=3.0$, $r_0=0.2$ (Middle) $\theta=0.11$, $c_2=3.0$, $r_0=0.1$ (Right) $\theta=0.11$, $c_2=1.0$, $r_0=0.1$.}
	\end{subfigure}
	\caption{The BVF model and the DA produce qualitatively similar predictions. From left to right ABM simulations die out (Region III0), reach an intermediate equilibrium (Region IIIe), or reach the fully realized revolution (Region III1). Time traces of $rep=100$ ABM simulations (grey), solution to BVF model (dashed black) and DA (solid black).}
	\label{fig:EVF_DA}
\end{figure}

\subsection{Comparison of ABM and Degree Approximation}
\begin{figure}[h]
	\centering
	\begin{subfigure}[t]{0.48\linewidth}
		\includegraphics[width=\linewidth]{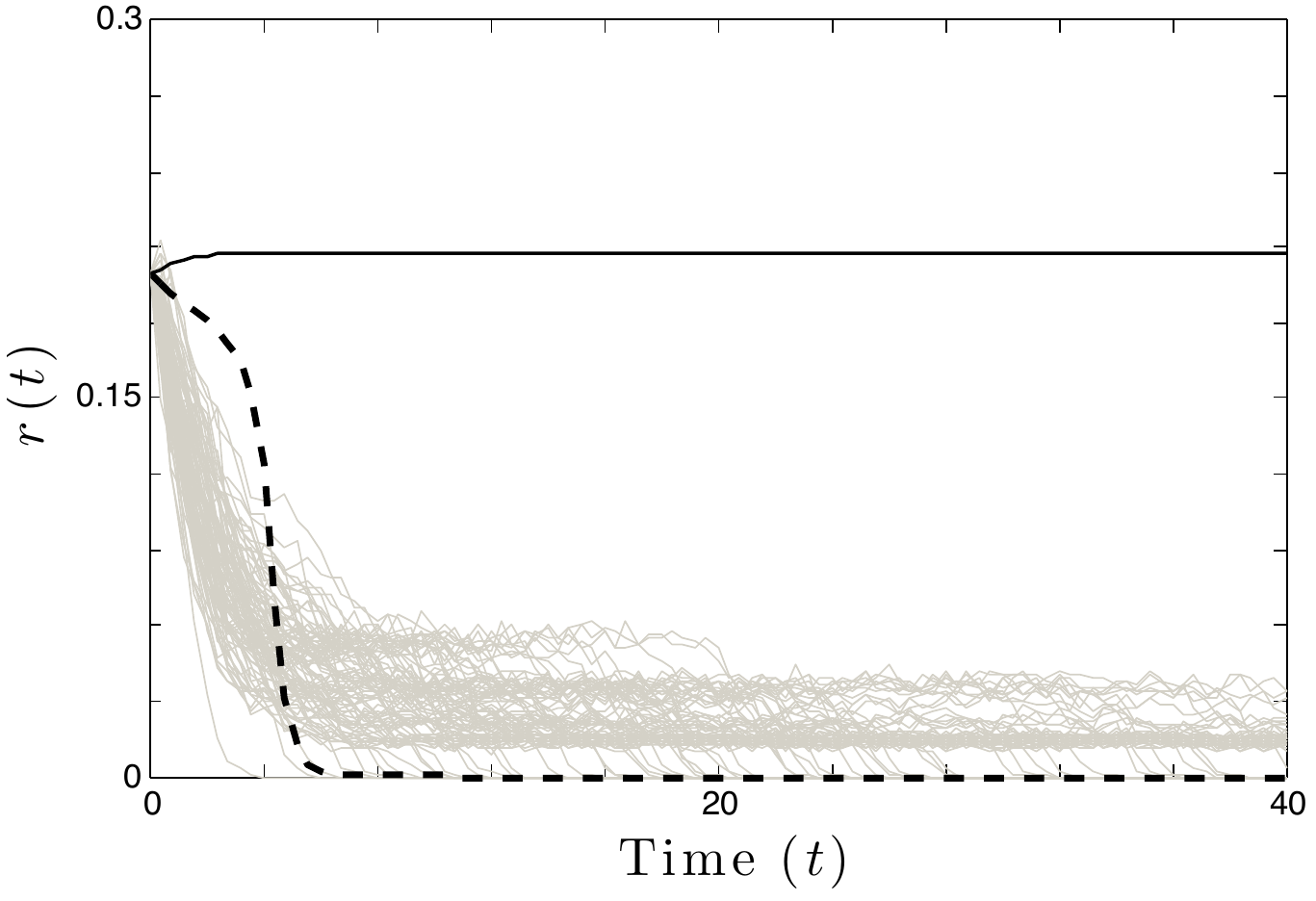}
		\caption{Facebook social network $G_F$ with parameters $c_1=1$, $c_2=3$, $\theta=0.15$, and $\beta=0.3$, with initial condition $r_0=0.2$. The BVF model is a better approximation of ABM realizations than the DA.}
	\end{subfigure}
	\hfill
	\begin{subfigure}[t]{0.48\linewidth}
		\includegraphics[width=\linewidth]{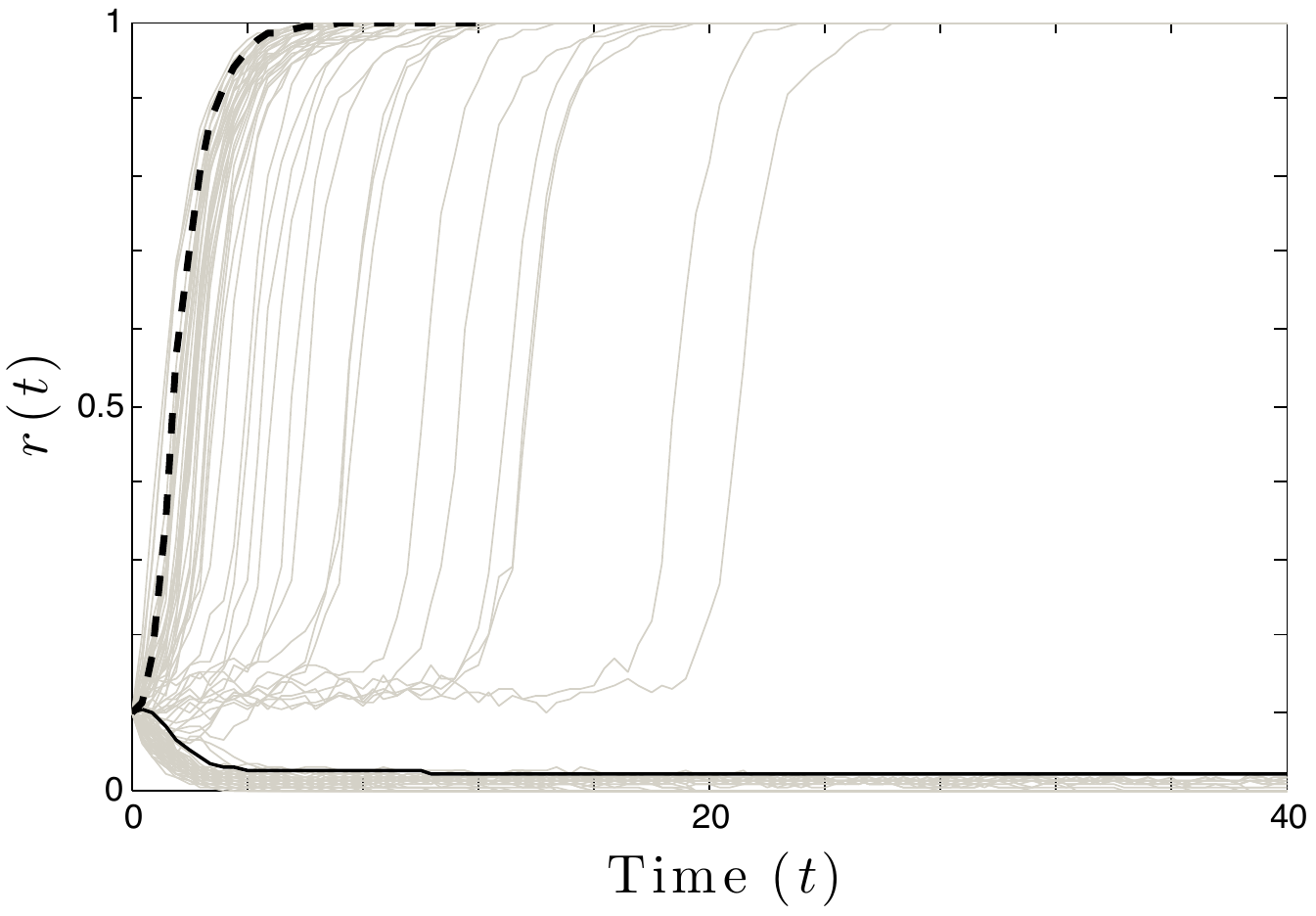}
		\caption{Physical contact network $G_P$ with parameters $c_1=1$, $c_2=1.2$, $\theta=0.15$, and $\beta=0.3$, with initial condition $r_0=0.1$. The BVF model and the DA provide complementary information on how the ABM realizations evolve.}
	\end{subfigure}
	\caption{Example of parameters for which (a) The BVF model outperforms the DA on the Facebook social network, and (b) the BVF model and the DA complement each other on the physical contact network. Time traces of $rep=100$ ABM simulations (grey), solution to the BVF model (dashed black) and the DA (solid black).}
	\label{fig:EVF_DA_2}
\end{figure}

In Section \ref{Sec:Comp_ABM_BVF_EVF} we showed that the approximation of the ABM by the BVF/EVF model is no worse, and usually much better, than the approximation of the ABM by the SVF model. In this section we show that the approximation of the ABM by the BVF/EVF model is no worse than the approximation of the ABM by the DA given in \eq{ApproxDA}. Specifically, we observe that the BVF and DA model produce similar qualitative predictions for most choices of parameters $\theta$, $\beta$, $c_1$, and $c_2$, see for example Fig.~\ref{fig:EVF_DA}. Also, although this is an atypical result, we present one set of parameters where the BVF model outperforms the DA on the Facebook social network $G_F$, and where the BVF model complements the DA on the physical contact network $G_P$, see Fig.~\ref{fig:EVF_DA_2}. We observe that, for the specific parameters and networks used in the simulations of Figs.\ \ref{fig:EVF_DA}-\ref{fig:EVF_DA_2}, the DA does not significantly outperform the BVF model. 
It was confirmed in extensive additional simulations that this occurs generically for large parts of the parameter space of the models.
This is worth emphasizing once more, since the BVF model is much easier to analyze and much less costly to solve than the DA. This supports our proposition that the BVF (or EVF) model is a powerful tool for approximating and analyzing the ABM and a general technique for formulating tractable models for spreading processes on social networks that take real network characteristics into account.

\section{The Basic Reproduction Number ($R_0$) and the Initial Slope of the BVF/EVF}
\label{Sec:R0}

In epidemiological models the \emph{basic reproduction number} $R_0$ is defined to be the number of secondary infections caused by a single infected individual introduced into a population that is entirely susceptible \cite{Hethcote00}. Note that in this context exposure to one infected individual is sufficient for a new infection. If $R_0 < 1$ then an infected individual is, on average, unable to replace himself and the outbreak terminates on its own in short order. On the other hand, if $R_0>1$ then an infected individual is, on average, more than able to replace himself and the outbreak may spread into a full-blown epidemic. We define the basic reproduction number for our linear threshold model of political revolution analogously: the basic reproduction number is the average number of individuals that become active in the revolution due directly to the introduction of a single active individual into a population that is otherwise completely inactive. In this section, we show that the basic reproduction number in our linear threshold ABM is related to the initial slope of the BVF/EVF by the equation
\begin{equation}
	\label{eq:R0visp}
	R_0 = \underbrace{\frac{c_1}{c_1+c_2} }_{=c^*} v_b'(0;\theta,\rho_k),
\end{equation}
and discuss efficient ways to compute $R_0$.

We begin by deriving an expression for $R_0$. Suppose that we activate individual $v\in V$ in the population, and suppose that individual $v$ has degree $k$. Further, suppose that this individual has $\mathcal{V}$ neighbours who can see the revolution (i.e., the linear threshold criterion is satisfied) once individual $v$ is activated. Let $w\in V$ be a neighbour of $v$'s that can see the revolution, and let $\tau_1$ and $\tau_2$ be the first arrival time of Poisson processes with rates $c_1$ and $c_2$, respectively. The probability that individual $w$ becomes active before individual $v$ becomes inactive is
\begin{align*}
	\pr{\tau_1 < \tau_2} &= \int_0^\infty \pr{\tau_1 < k | \tau_2 = t}\pr{\tau_2 = t}dt\\
	&= \int_0^\infty [1-\exp(-c_1 t)] c_2\exp(-c_2 t) dt\\
	&= \frac{c_1}{c_1+c_2} = c^*.
\end{align*}
Since we only consider activations that result directly from $v$'s activation (and not activations caused by a combination of the activation of $v$ and the subsequent activation of $v$'s neighbours), for our purposes the behaviour of each of $v$'s neighbours is independent. Thus, the expected number of $v$'s neighbours which become active is $\mathcal{V}c^*$.

Recall that the social network $G = G(V,E)$ has $N$ nodes, $M$ edges, degree distribution $\rho_k$ and secondary degree distribution $\rho_{k,j}$.
%
%
For the single activated individual $v$ with degree $k$ the expected number of neighbours who can see the revolution is
$$
	\mathbb{E}[\mathcal{V}] = k \sum_j \rho_{k,j}\ind{ j\theta \leq 1}.
$$
Averaging over all possible degrees $k$ now yields the basic reproduction number
$$
	R_0 =c^*\ex[\mathcal{V}] = \frac{c_1}{c_1+c_2} \sum_{k,j} k \rho_k \rho_{k,j}\ind{ j\theta \leq 1}
	= c^*\sum_{k=1}^\infty \sum_{j=1}^{\lfloor \theta^{-1} \rfloor} k \rho_k \rho_{k,j}.
$$
It is now possible to prove the relationship given in \eq{R0visp}. Writing
$$
	R_0 =\frac{c_1}{c_1+c_2} \sum_{j=1}^{\lfloor \theta^{-1} \rfloor} \sum_{k=1}^\infty k \rho_k \rho_{k,j},
$$
it suffices to show that
$$
	j \rho_j = \sum_{k=1}^\infty k \rho_k \rho_{k,j},
$$
since
\begin{align}
	v_b(r;\theta,\rho_k) &= \sum_{k=1}^\infty \rho_k \mbox{BinCDF}(k - \lceil \theta k \rceil; k, 1-r) \nonumber \\
	& = \sum_{k=1}^\infty \sum_{j=0}^{k-\lceil \theta k \rceil} \rho_k \binom{k}{j} (1-r)^j r^{k-j} \nonumber \\
	& = \sum_{k=1}^\infty \rho_k \left[ \binom{k}{0} r^k + \ldots +\binom{k}{k-\lceil \theta k \rceil} (1-r)^{k-\lceil \theta k \rceil}r^{\lceil \theta k \rceil}\right] \nonumber \\
	&\implies v_b'(0;\theta,\rho_k) = \sum_{k=1}^\infty k \rho_k \spc \ind{\theta k \leq 1} = \sum_{k=1}^{\lfloor \theta^{-1}\rfloor} k \rho_k.
\label{eq:vbprime}
\end{align}
We note that together with \eq{R0visp}, \eq{vbprime} gives an exact and cheap way to compute $R_0$.

From the undirected social network $G=G(V,E)$ form the directed social network $G' = G(V,E')$, where
%
$
	E' = \{(v,w): \{v,w\}\in E\}.
$
%
In this case, the number of edges emanating from all nodes of degree $k$ is $k \rho_k N$. So, the number of edges emanating from nodes of degree $k$ and incident on nodes of degree $j$ is $k\rho_k N \rho_{k,j}$. Thus, the number of edges incident on nodes of degree $j$ is
$$
	 N \sum_{k=1}^\infty k \rho_k \rho_{k,j}.
$$
Equivalently, the number of edges incident on nodes of degree $j$ is 
$$
	j\rho_j N.
$$
Putting these two expressions together, we find
$$
	jp_j = \sum_{k=1}^\infty k\rho_k\rho_{k,j},
$$
which completes the proof.

Intuitively, this makes sense since the number of nodes that can see the revolution when one node is activated can also be approximated by
$$
	N \, v_b\left(\frac{1}{N}; \theta, \rho_k\right) = \frac{v_b\left(\frac{1}{N}; \theta, \rho_k\right) - 
	\overbrace{v_b\left(0; \theta, \rho_k\right)}^{=0}}{\frac{1}{N}}
	\approx v_b'\left(0; \theta, \rho_k\right).
$$


\section{Application: Linear Threshold Propagation on Online versus Offline Social Networks}
\label{Sec:Interp}
In this section we investigate experimentally the differences in spreading behaviour that occur under the linear threshold model for the spread of political revolutions when applied to some empirical online and offline social networks, searching for some initial quantitative evidence that political revolutions may be facilitated by the network structure of online social networks of social media.

\subsection{Network Structure: Online versus Offline Social Networks}

%
%

It is often assumed that the connectivity of modern online social networks was an important factor in the spread of political revolutions in the past decade, e.g., in the Arab Spring revolutions of 2011 \cite{eltantawy2011arab,khondker2011role,tufekci2012social,KhamisVaughn11}, while traditional offline social networks (using in-person physical contact, or phone or mail interaction for safe communication) featured a different connectivity structure that was often severely restricted by censorship of the regime.
Unfortunately, representative samples of the offline, traditional communication networks that were in existence prior to the adoption of new media technologies are unavailable at sufficient scale for countries affected by the Arab Spring, or indeed for any country. As opposed to the online networks of social media which are, by their nature, digitally stored and available, the offline social networks of pre-Internet societies have not been recorded at scale, simply because it was impractical in terms of cost and effort. This is a serious roadblock when investigating the effects of the structure of new media networks on the dynamics of political revolution, and if one wants to compare with propagation on pre-Internet social networks, it is necessary to identify proxy networks that are likely to be reasonable approximations to pre-Internet social networks, in terms of network structure.

In this section, we use simulations of our ABM model on two empirical networks as a starting point  to investigate differences in propagation properties that may arise between online and offline social networks within the linear threshold propagation model. We choose the small physical contact network $G_P$ between individuals from \cite{SalatheEtAl10} as a representative for offline social networks, and we choose the Facebook network $G_F$ as a representative for online social networks.




%
\begin{figure}[h]
	\centering
	\begin{subfigure}[t]{0.48\linewidth}
		\includegraphics[width=\linewidth]{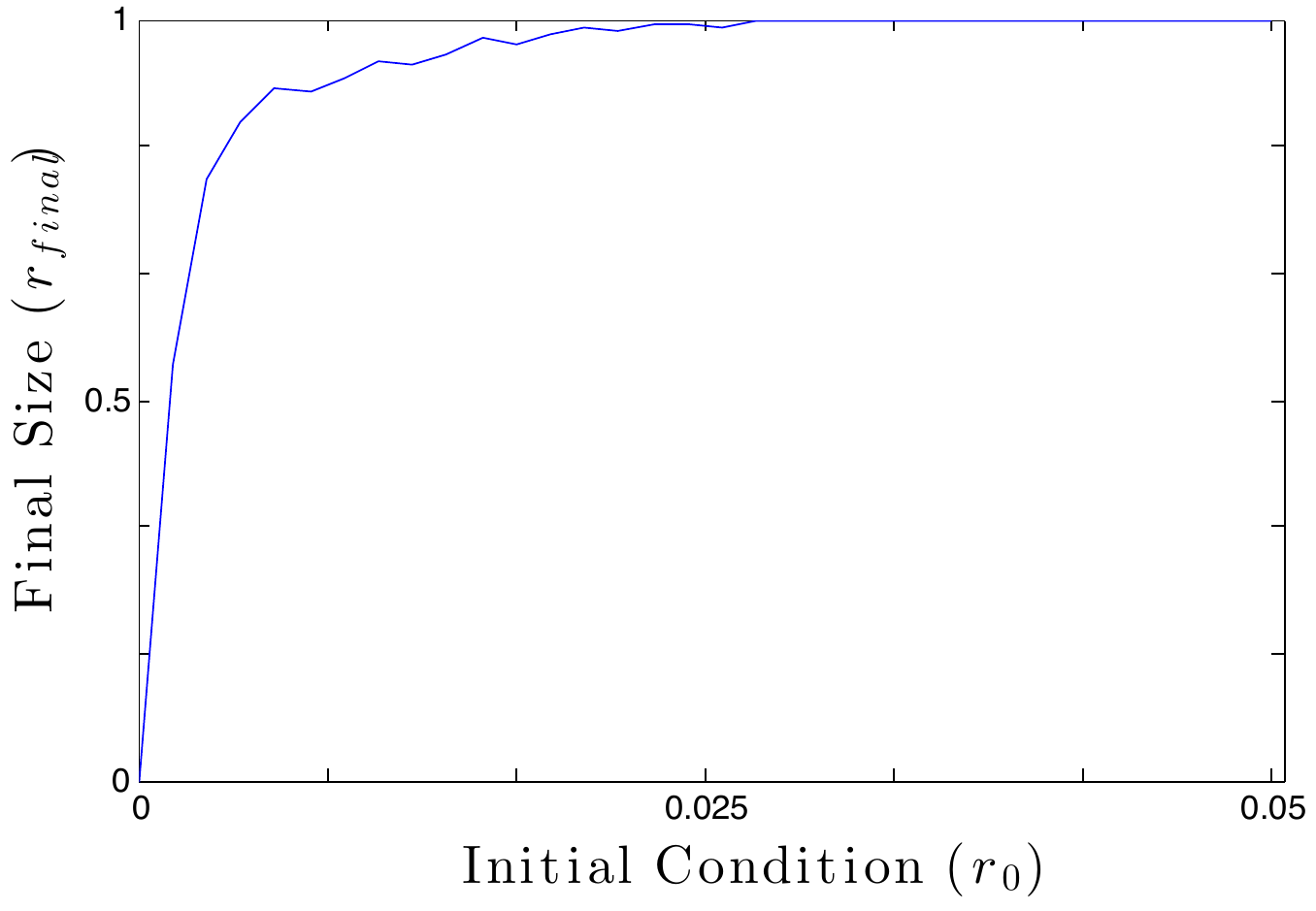}
		\caption{Facebook social network $G_F$}
	\end{subfigure}
	\hfill
	\begin{subfigure}[t]{0.48\linewidth}
		\includegraphics[width=\linewidth]{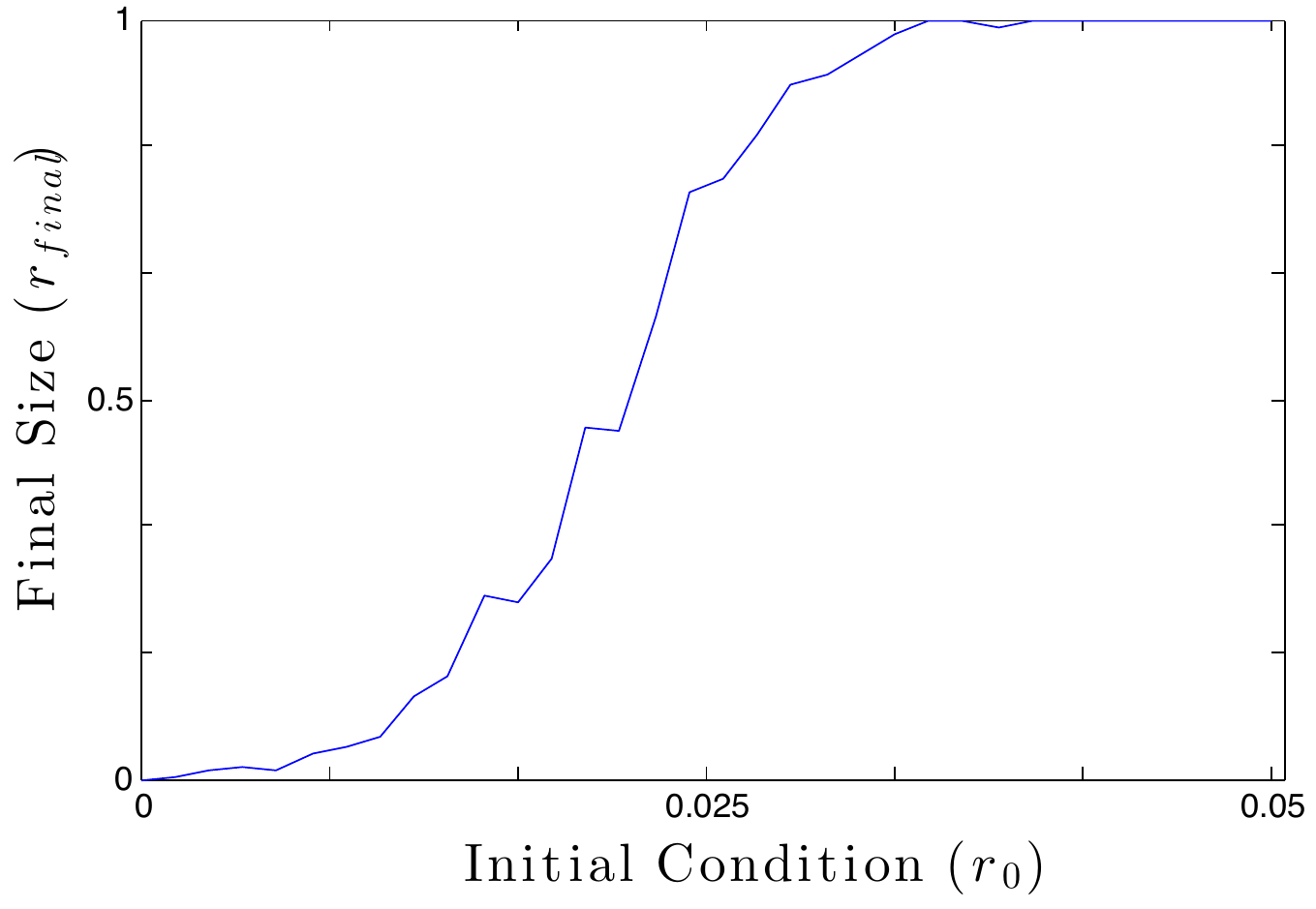}
		\caption{Physical contact network $G_P$}
	\end{subfigure}
	\caption{Average final size $r_{final}$ of $rep=100$ realizations of ABM simulations on Facebook and physical contact networks with $r_0\in[0,0.05]$, $\theta=0.1$, $\beta=0.3$, $c_1=1$, and $c_2=0.2$ (Region III1, $1-\hat{\alpha} < \beta\leq c^*$). Realizations are allowed to run until time $t_{final}=20$. Basic reproduction numbers $R_0$ for Facebook social and physical contact networks are $R_0 = 1.12>1$ and $R_0 = 0.35<1$, respectively.}
	\label{fig:r0_rfinal}
\end{figure}

\subsection{ABM on Facebook $G_F$ and Physical Contact $G_P$ Networks}
We now briefly examine the differences in how the political revolution spreads on the Facebook social and physical contact networks via direct simulation of the ABM. In particular, choosing parameters $\theta=0.1$, $\beta=0.3$, $c_1=1$, and $c_2=0.2$ in the unstable police state region (Region III1, $1-\hat{\alpha} < \beta \leq c^*$), we simulate $rep=100$ realizations of the ABM for each initial condition $r_0\in\{0, 0.0015, 0.003, \ldots, 0.0495\}$ from time $t=0$ until time $t=20$. The average final size (at $t=20$) for each initial condition $\langle r_{final}|r_0\rangle = \left.\langle r_a(20) \rangle\right|_{r_0}$ is recorded and displayed in Fig.~\ref{fig:r0_rfinal}, which demonstrates that for these parameters the political revolution described by our ABM propagates to a much greater extent on the Facebook social network, i.e. the online proxy network, than on the physical contact network, i.e. the offline proxy network.

As an important illustration of the usefulness of the basic reproduction number $R_0$ we defined in Sec.\ \ref{Sec:R0}, we find that this is consistent with a difference in $R_0$ for the two networks: the basic reproduction number is $R_0=1.12$ (above the value of 1 which is expected to be required for propagation) for the Facebook network, and $R_0=0.35<1$ for the physical contact network.

These results indicate that the offline social network is less conducive to spreading the revolution in the ABM than the online social network: it has a smaller basic reproduction number, and in simulations a larger initial population of revolutionaries is required to spread the revolution. This provides some initial quantitative evidence that the spread of revolutions under a linear threshold process may occur more easily on modern online social networks than on traditional offline networks. While this is an interesting first observation, the next section discusses limitations and further investigations that are required to address this intriguing but complex question more comprehensively.


\FloatBarrier

\section{Discussion and Conclusion}

In this paper we developed a linear threshold agent based model (ABM) to model the spread of a political revolution in a dictatorial regime. We showed that this model is consistent with previous simple step visibility function (SVF) ODE model developed in \cite{LangDeSterck14}. Using the relationship between these two models as a template we developed a hierarchy of models of varying complexity that approximate the behaviour of the ABM, see Table~\ref{tab:hierarchy}. Of all the models we have identified, we find that the BVF and EVF models (models of moderate complexity) offer the optimal combination of low computational complexity (cost), ease of analysis, and ability to approximate the behaviour of the ABM. Specifically, we find that for most parameters and initial conditions the BVF/EVF model is better able to approximate average ABM behaviour than the SVF. Also, for most networks the BVF/EVF model is much less costly to solve and much easier to analyze than the degree approximation from \cite{NekoveeEtAl07}. Importantly, the analysis of the simple ODE models in terms of stability of solutions for various parameter regimes directly gives insight in the qualitative dynamics of the linear threshold ABM for the spread of political revolutions.

We extended the concept of the basic reproduction number $R_0$ from epidemiology \cite{Hethcote00} to the linear threshold ABM, we showed how it is related to the slope of the empirical or binomial visibility functions at $r=0$, and we provided efficient ways to compute or estimate it. Analogously to epidemiological models, when $R_0>1$ we expect the political revolution to spread, and when $R_0<1$ we expect it to die out. Thus, computing this quantity for a network can give an indication of how the ABM will behave on that network without the need to perform simulations, as we have demonstrated for empirical networks. 

The Facebook and physical contact networks we consider as case studies, i.e. the online and offline proxy networks, provide initial support to the hypothesis that the adoption of online social media may facilitate the spread of political revolutions by effectively changing the connectivity structure of the population in a way that makes linear threshold spreading more effective. Specifically, we find that for certain parameters the online proxy network is more susceptible to the linear threshold spreading process than the offline proxy network. Moreover, we find that the different behaviour of these two networks is consistent with the basic reproduction number calculated for these two networks.

In addition to studying the ABM and its approximations on more and larger online and offline networks, much work remains to be done in the actual modelling of the political revolution process. For example, throughout this manuscript we have assumed that the linear threshold of individuals is constant for the population, that the underlying communication networks are static, that the graph is undirected, and that the nodes that are initially activated are chosen uniformly at random. Each of these assumptions represents a major simplification of reality that needs to be addressed to study further aspects of the spread of political revolutions on online social networks.  For example, consider that in our model individuals estimate the current participation in the revolution by sampling their neighbours. Since individuals with greater sample size, i.e. larger degree, can form more accurate estimates of the current participation in the revolution, they should be willing to join the revolution at a lower linear threshold $\theta$ than individuals who are more uncertain in their estimate of revolution size. We would therefore expect the linear threshold to vary from individual to individual as a function of their network degree. Or, consider that one of the principal strengths of new media is the ability of individuals to search for both content and like-minded individuals. Thus, we would expect that the underlying communication network should be changing on the same time scale as the revolution. Finally, one might consider that the nodes that are predisposed to be active at the initial stages of a revolution, e.g. nodes that represent activists, may also have larger degree (if, for example, they are charismatic and allowed to accrue followers) or smaller degree (if, for example, they are the specific target of regime censorship). There is significant value in the kind of parsimonious ABM model on static networks that we have considered in this paper, because much can be learned from this type of model and it is easier to analyze and interpret than more complex models. Nevertheless, extensions along the lines sketched above are important avenues for further study.

The simulations in this paper were performed in Matlab. All Matlab code and data files will be made publicly available online upon publication of this paper.


\appendix

\section{Physical Contact Network Data}
\label{Sec:ext_net}
In this appendix, we briefly discuss the network extraction protocol for the physical contact network $G_P$ from \cite{SalatheEtAl10}.

The network presented in \cite{SalatheEtAl10} was constructed by distributing wireless sensors to students, teachers, and staff at a U.S. high school during a one day period from approximately 08:00 to 16:30. When two wireless sensors are in proximity of one another, i.e. when they are less than approximately 3m apart, they register an interaction with a temporal resolution of 20s. Data are given for each seperate interaction in CSV format with three columns: ID of first wireless sensor, ID of second wireless sensor, duration of interaction (measured in 20s increments). A weighted undirected network $G_{P,0}$ is formed by connecting each pair of individuals with an edge whose weight is given by the total amount of time they spent in proximity to one another. In order to admit a comparison with the unweighted undirected Facebook subnetwork, we de-weight network $G_{P,0}$ by discarding all edges whose weight is less than the minimum duration $w_P$, and by weighting all remaining edges equally. We denote the largest connected component of the resulting network by $G_{P}$, and for convenience we refer to this network as the \emph{physical contact network}.


The choice of the minimum duration $w_P$ is important since it determines, for example, the average degree of the physical contact network $G_P$. We choose $w_P$ such that the average degree of $G_P$ is close to the average degree of $G_F$, for the following reason.
In the simulations carried out in this paper to compare propagation under the linear threshold ABM on $G_P$ and $G_F$, we seed the networks with a fixed percentage of active nodes and use the same threshold $\theta$ on both networks. We want to calibrate the average degree of $G_P$ to the average degree of $G_F$, such that, on average, nodes in $G_P$ have the same chance as nodes in $G_F$ to satisfy the linear threshold criterion and see the revolution. After this calibration, differences in propagation between $G_P$ and $G_F$ (for the same $\theta$) are only due to the differences in network structure that go beyond the average degree.
This motivates us to choose $w_P = 34$ sensor measurements, equivalent to 11 minutes and 20 seconds, in order to match the average degrees of $G_P$ and the Facebook subnetwork as closely as possible.

\section{Justification of Step Visibility Function}
\label{Sec:JustSVF}

\begin{figure}[h]
	\centering
	\hfill
	\includegraphics[width=2.5in]{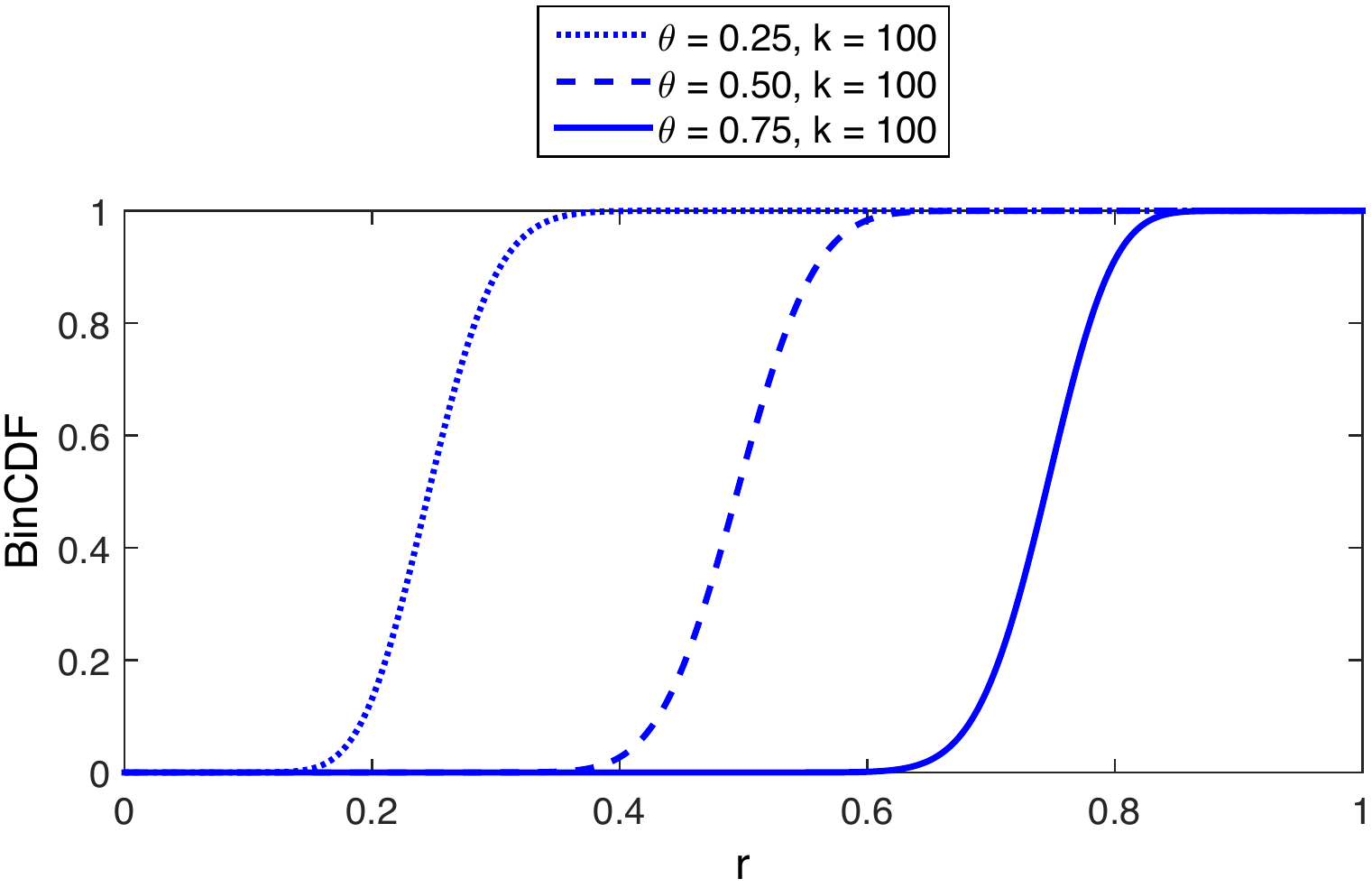}
	\hfill
	\includegraphics[width=2.5in]{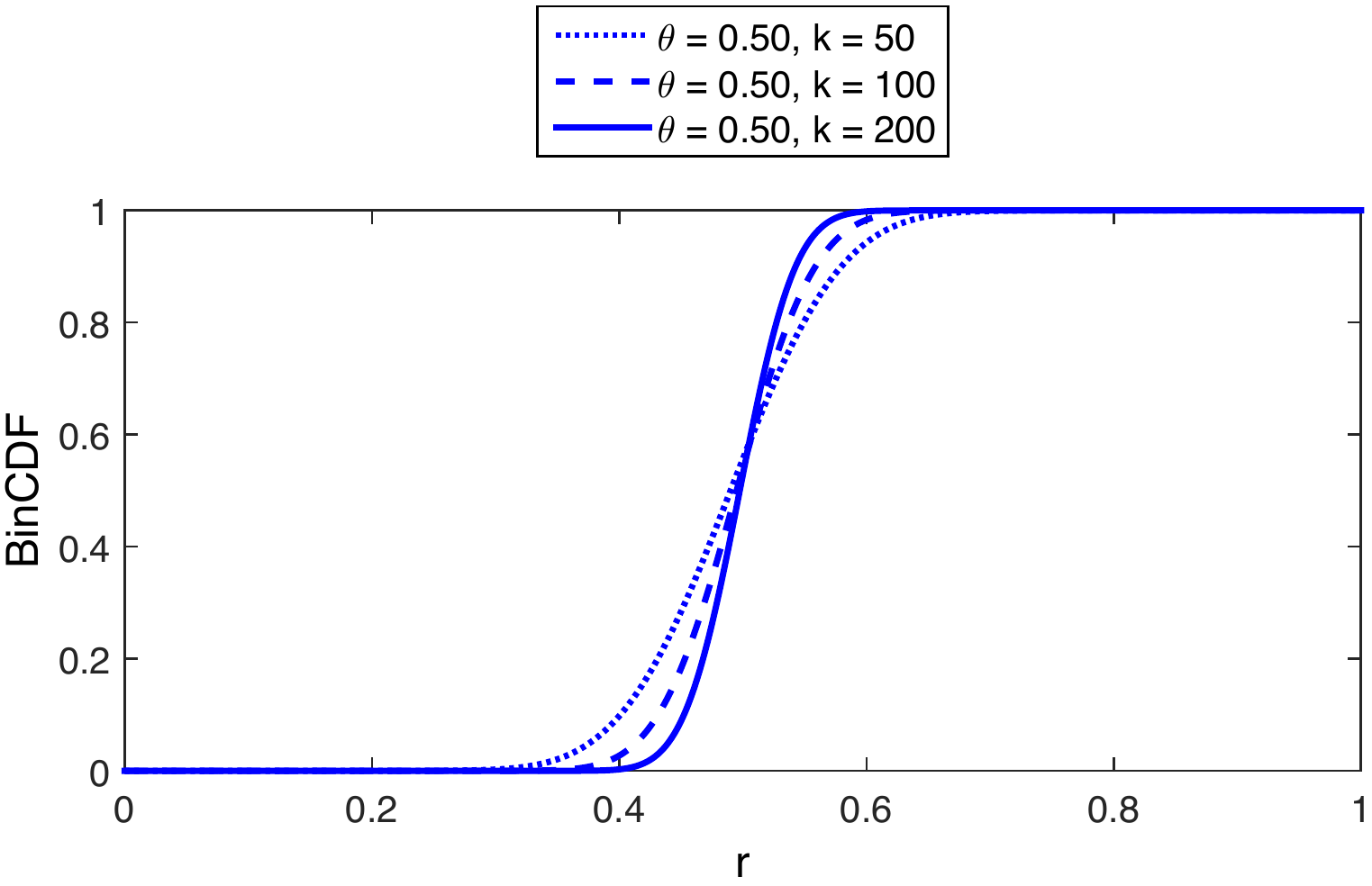}
	\hfill\spc
	\\
	\includegraphics[width=2.5in]{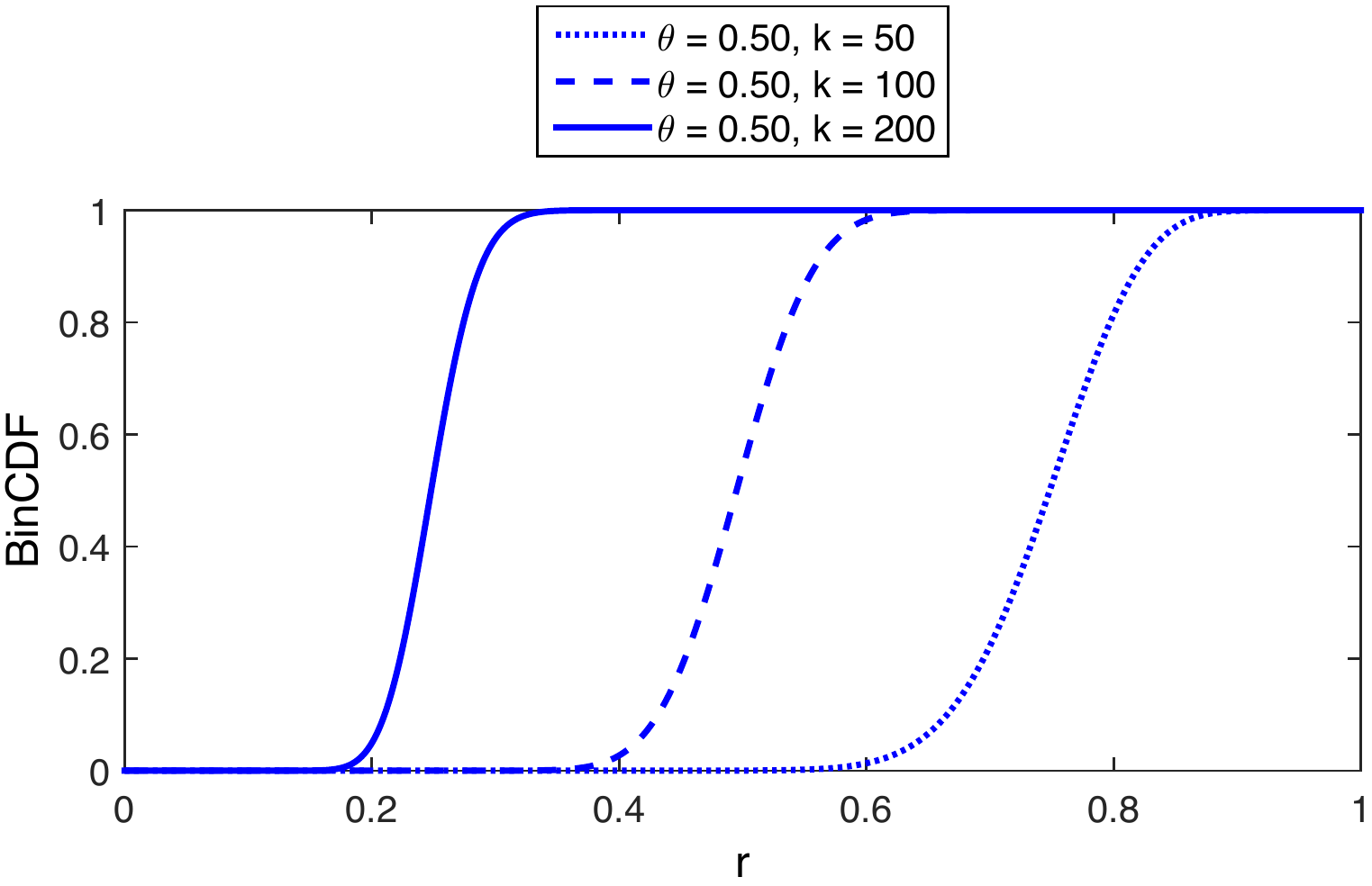}
	\caption{Dependence of $\mbox{BinCDF}(\lfloor k\rfloor - \lceil\theta k\rceil; \lfloor k\rfloor,1-r)$ on (top left) $\theta$, (top right) $k$, and (bottom) negatively correlated $\theta$ and $k$.}
	\label{fig:Binom}
\end{figure}

As argued in Appendix A of \cite{LangDeSterck14}, the expression for the visibility function given in \eq{ProtoBVF}, with $k$ taken to be the average degree of the network, can be expected to have a steep sigmoidal shape, which can be approximated by a step function.
Indeed, Fig.\ \ref{fig:Binom} shows that $\mbox{BinCDF}(\lfloor k\rfloor - \lceil\theta k\rceil; \lfloor k\rfloor,1-r)$ has a clear sigmoidal shape, with a steep transition from 0 to 1. In \cite{LangDeSterck14} it was also argued that increased average degree $k$ can be expected to co-occur with decreased threshold $\theta$, since the more neighbors one has the more certain one can be about the true extent of the revolution because large samples are more reliable. Under this assumption, 
$\mbox{BinCDF}(\lfloor k\rfloor - \lceil\theta k\rceil; \lfloor k\rfloor,1-r)$ can be considered a single-parameter sigmoidal function, which can be approximated by the single-parameter step visibility function (SVF) $v_s(r;\alpha)$, with parameter $\alpha$ indicating where the step transition occurs.

We conclude this section with a few remarks regarding the assumption made in Sec.\ \ref{Sec:BVF} and in \cite{LangDeSterck14} that the states of an individual's neighbors are uncorrelated. 
While this assumption is applied in many models of spreading processes, for example rumour spreading processes \cite{NekoveeEtAl07} and epidemiological processes \cite{KissEtAl06, PastorSatorrasVespignani02,LindquistEtAl10},
this assumption was not verified in \cite{LangDeSterck14}. In this paper we show that this assumption results in a valid approximation by comparing simulation results of the SVF model, and also the BVF/EVF models (which were derived in
Sec.\ \ref{Sec:AVF} under the same assumption), with ABM simulations for empirical networks (which do not make this assumption), finding consistent results for large parts of parameter space.

\section{Analysis of BVF/EVF Model}
\label{Sec:Ext}
\begin{figure}[h]
	\centering
	\includegraphics[width=0.4\linewidth]{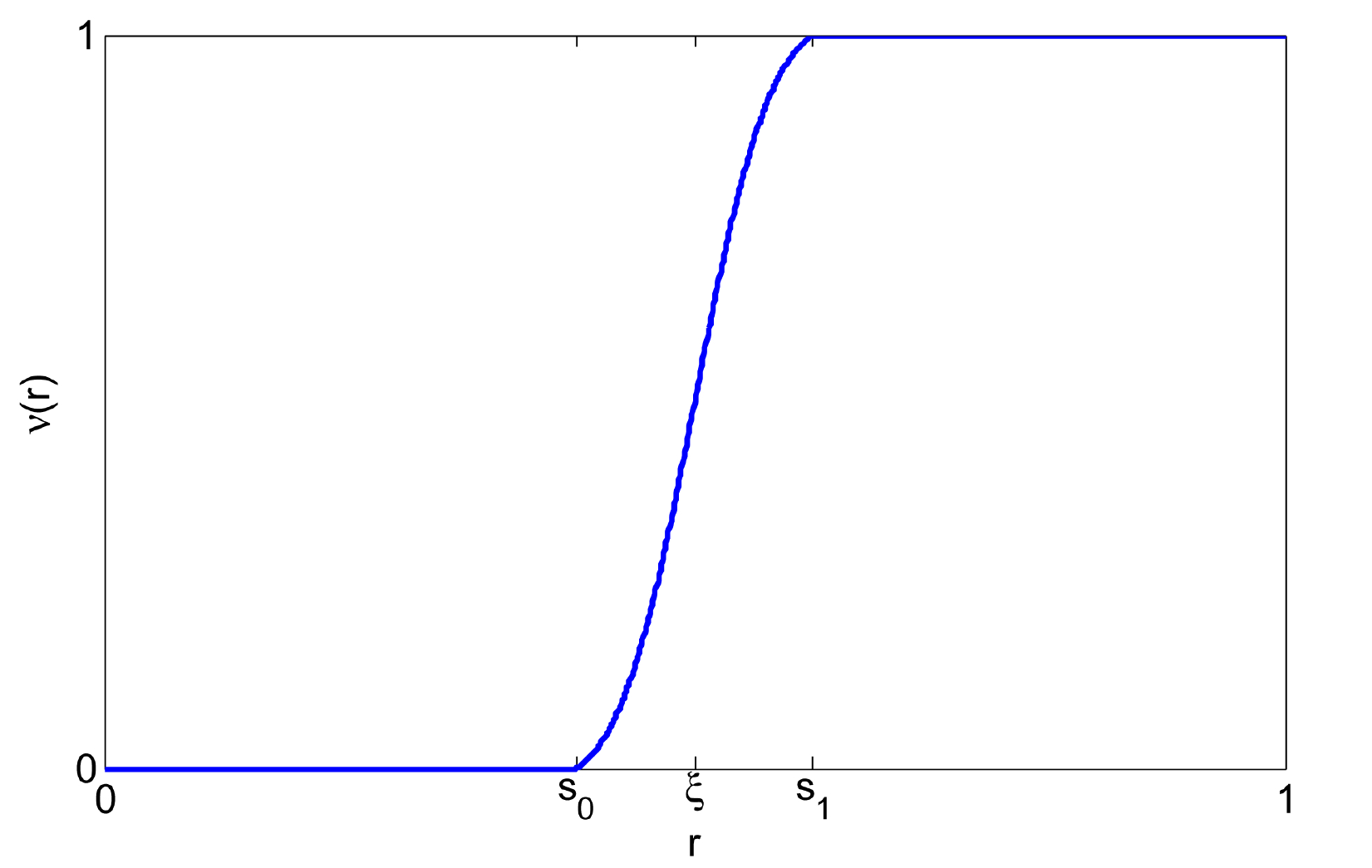}
	\caption{Visibility function $\nu(r)$ with a fast transition from 0 to 1 that follows a sigmoidal profile.}
	\label{fig:Sigmoid}
\end{figure}

In this section, we summarize the parameter regime and solution stability analysis that was developed in Appendix B of
\cite{LangDeSterck14} for the one-dimensional ODE model 
\begin{equation}
	\label{eq:Ext}
	\frac{dr}{dt} = \dot{r} = \underbrace{c_1\spc (1-r) \spc \nu(r)}_{\gamma(r)} - \underbrace{c_2 \spc r\spc \rho(r)}_{\delta(r)},
\end{equation}
with generic sigmoidal visibility function $\nu(r)$ as in Fig.~\ref{fig:Sigmoid}. Since the newly derived BVF/EVF in this paper have sigmoidal shape, the stability analysis of \cite{LangDeSterck14} applies to the BVF/EVF models.

\begin{figure}[h]
	\centering
	\includegraphics[width=0.4\linewidth]{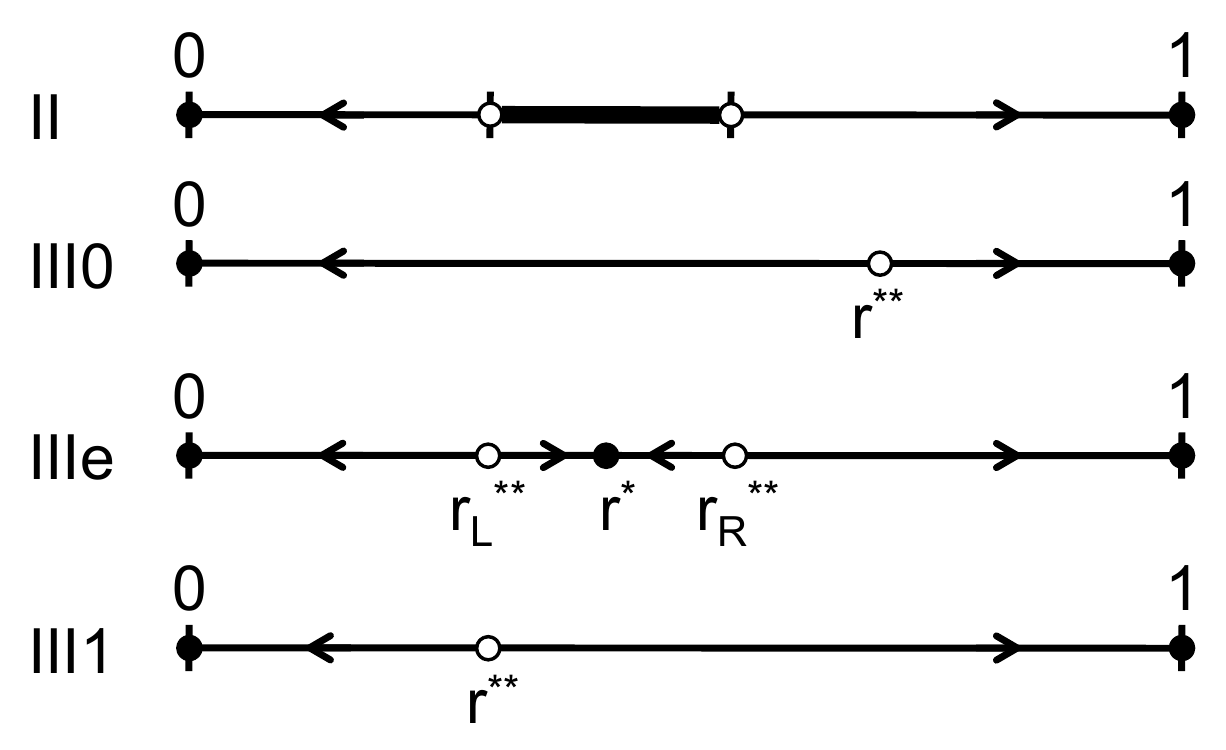}
	\caption{Equilibria, stability and basins of attraction on the $r$-axis ($r \in [0,1]$) for model \eqref{eq:Ext}. Closed (open) circles represent locally asymptotically stable (unstable) equilibria. Left (right) arrows indicate regions where $\dot{r}<0$ ($\dot{r}>0$). The phase portraits are similar to the case of the SVF model in Fig.\ \ref{fig:SVF_stability}.}
	\label{fig:SummaryExt}
\end{figure}

As is shown in \cite{LangDeSterck14}, the dynamics of model \eqref{eq:Ext} with generic sigmoidal visibility and policing functions closely follows the corresponding dynamics of model \eqref{eq:SVF} in Region II (an open interval of equlibria $\subset (0,1)$), Regions III0 and III1 (one equilibrium $\in (0,1)$), and Region IIIe (three equilibria $\in (0,1)$), thus establishing the equivalence in terms of dynamic behaviour of models \eqref{eq:SVF} (with step functions) and \eqref{eq:Ext} (with sigmoidal functions, including the BVF/EVF models). We summarize the phase portraits of the different regions for model \eqref{eq:Ext} in Fig.\ \ref{fig:SummaryExt}, which are analogous to the phase portraits of the SVF model shown in Fig.\ \ref{fig:SVF_stability}. See Appendix B of \cite{LangDeSterck14} for a detailed derivation and further explanation.
For the BVF/EVF models, the values of equilibrium quantities like $r^*$ and $r^{**}$ depend on $\theta$, $\beta$ and the network structure.

It is also important to point out how the SVF stability region plot of Fig.\ \ref{fig:SVF_regimes} generalizes to the BVF/EVF models. For a given empirical network, the BVF/EVF model does not feature a range of visibility parameters $\alpha$ like the SVF model, but each choice of $\theta$ results in a different BVF/EVF function. For a fixed $\theta$, the different phase portrait behaviours of Fig.\ \ref{fig:SummaryExt} occur depending on the values of $\beta$, $c_1$ and $c_2$. As a result, for a given empirical network, the parameter space for the BFV/EVF models as a function of $\theta$, $\beta$, $c_1$ and $c_2$ can be divided in regions as for the SVF model in Fig.\ \ref{fig:SVF_regimes}, where $\theta$ replaces $\alpha$ as the parameter on the horizontal axis. The precise shape of the BVF/EVF parameter regions would depend on the actual empirical network considered. For example, the boundary between the II and III regions is in general not a straight line, but needs to be determined numerically.

\section{Gillespie's Algorithm}
\label{Sec:Gillespie}

Following \cite{AmesEtAl11}, we numerically simulate the ABM by implementing Gillespie's Algorithm.
In this appendix we give a brief overview of this algorithm. 
We first introduce some notation. As above, let $r_a$ denote the fraction of individuals in the population that are expected to be active in the revolution at time $t$. For each node $v \in V$ we let $\gamma_v(t)=1$ if $v$ can see the revolution at time $t$, i.e. if $v$ satisfies \eq{0to1}, and let $\gamma_{v}(t)=0$ otherwise. Furthermore, we let $\xi_{1,v}$ and $\xi_{2,v}$ denote the first arrival times of independent Poisson processes with rates $c_1$ and $c_2$, respectively. 

Gillespie's Algorithm is based on the fact that the sum of two independent Poisson variables is also a Poisson variable with rate equal to the sum of the rates of the original processes. It follows that
$$
	\sum_{v\in V} \left[ \xi_{1,v}\spc (1-s_{v}(t)) \spc \gamma_{v}(t)+ \xi_{2,v} \spc s_{v}(t) \spc p(r_a(t); \beta)\right]
$$
is a Poisson process with rate
$$	
	\Lambda = \sum_{v\in V} \left[ c_1 \spc(1-\spc s_{v}(t)) \spc \gamma_{v}(t) + c_2 \spc s_{v}(t) \spc p(r_a(t); \beta) \right].
$$
The first arrival time of this process, therefore, is an exponential random variable $\tau$ with rate $\Lambda$. At time $t+\tau$ the state of exactly one of the nodes will change. Moreover, since $\xi_{1,v}$ and $\xi_{2,v}$ are independent, the probability that the state of node $v$ will change is
$$
	\mathbb{P}_v = \frac{ c_1 \spc (1-\spc s_{v}(t)) \spc \gamma_{v}(t)+ c_2 \spc s_{v}(t) \spc p(r_a(t);\beta) }{\Lambda}
$$
The Gillespie Algorithm then proceeds iteratively in three steps.
\begin{enumerate}
	\item Find the time $\tau$ of the next event by drawing $\tau$ from an exponential distribution with rate $\Lambda$.
	\item Determine which node changes state by drawing one node from $V=\{v_i\}_{i=1}^N$, where node $v\in V$ is drawn with probability $\mathbb{P}_v$.
	\item Update $t \leftarrow t + \tau$ and re-calculate $s_{v}(t)$, $r(t)$, $\gamma_{v}(t)$.
\end{enumerate}

\section{Complete Analytic Solution to SVF Model}
\label{Sec:SolnSVF}

Since the complete analytic solution to the SVF model was not provided in \cite{LangDeSterck14} but is useful in the current paper, we provide it here.

The solution to \eq{SVF} is as follows. In Region I ($1-\alpha = \beta$) and Region II ($1-\alpha > \beta$)
$$
	r(t) = \left\{ \begin{array}{ll} r_0 e^{-c_2(t-t_0)} & \mbox{if } r_0 < \beta\\
		 r_0 & \mbox{if } r_0 \in [\beta, 1-\alpha]\\
		1 - (1-r_0)e^{-c_1(t-t_0)} & \mbox{if } r_0 > 1-\alpha \end{array} \right. .
$$
In Regions III0 ($c^* \leq 1-\alpha < \beta$), IIIe ($1-\alpha < c^* < \beta$), and III1 ($1-\alpha < \beta \leq c^*$)
$$
	r(t) = \left\{ \begin{array}{ll} r_0 e^{-c_2(t-t_0)} & \mbox{if } r_0 < 1-\alpha\\ \\
		 \left[ c^*+ (r_0-c^*)e^{-(c_1+c_2)(t-t_0)} \right] \mathbb{I}_{\{t<t_{\alpha}\}}  \\
		\hspace{5mm} + (1- \alpha) e^{-c_2(t-t_\alpha)} \mathbb{I}_{\{t\geq t_{\alpha}\}}
			& \mbox{if } r_0 \in [\beta, 1-\alpha] \mbox{ and } c^* \leq 1-\alpha < \beta\\ \\
		c^*+ (r_0-c^*)e^{-(c_1+c_2)(t-t_0)} & \mbox{if }  r_0 \in [\beta, 1-\alpha] \mbox{ and } 1-\alpha< c^* < \beta\\\\
		\left[ c^*+ (r_0-c^*)e^{-(c_1+c_2)(t-t_0)} \right] \mathbb{I}_{\{t<t_{\beta}\}}  \\
		\hspace{5mm} + \left[ 1 - (1- \beta) e^{-c_1(t-t_\beta)} \right] \mathbb{I}_{\{t\geq t_{\beta}\}}
			& \mbox{if } r_0 \in [\beta, 1-\alpha] \mbox{ and } 1-\alpha < \beta \leq c^*\\\\
		1 - (1-r_0)e^{-c_1(t-t_0)} & \mbox{if } r_0 > \beta \end{array} \right. ,
$$
where
\begin{align*}
	t_\alpha & = t_0 - \frac{1}{c_1+c_2}\log\left(\frac{1-\alpha-c^*}{r_0-c^*}\right), \mbox{ and}\\
	t_\beta & = t_0 - \frac{1}{c_1 + c_2}\log\left(\frac{\beta-c^*}{r_0-c^*}\right).
\end{align*}

\section{Equivalence of Binomial and Empirical Visibility Functions}
\label{Sec:Comp_BVF_EVF}
We briefly argue that the empirical and binomial visibility functions are equivalent in the limit of large network and sample size, i.e. as $N, rep \rightarrow\infty$, see also Fig.~\ref{fig:BVF}. Let $v\in V$ be a node in the network $G=G(V,E)$ with degree $k$. Now, fix $\theta$ and suppose that we are calculating $v_{e,j} \approx v_b(\frac{j-1}{\mu}) = v_b(r_j)$ via the algorithm presented in Sec.~\ref{Sec:EVF}. The $l$\textsuperscript{th} iteration of Step 1 of this algorithm selects $\mathcal{N}_l$ nodes uniformly at random to be active and results in $\mathcal{V}_l$ nodes that can see the revolution. Note that by the Law of Large Numbers
\begin{align*}
	\frac{\mathcal{N}_l}{N} &\stackrel{p}{\rightarrow} r_j = \frac{j-1}{\mu} \mbox{ as $N\rightarrow\infty$, and }\\
	v_{e,j} = \frac{1}{rep}\sum_{l=1}^{rep} \frac{\mathcal{V}_l}{N}  &\stackrel{p}{\rightarrow} \mathbb{E}\left[\frac{\mathcal{V}_l}{N}\right] = \mathbb{E}\left[\frac{\mathcal{V}_1}{N}\right] \mbox{ as $rep\rightarrow\infty$,}
\end{align*}
where $\stackrel{p}{\rightarrow}$ denotes convergence in probability. Since the $\mathcal{N}_l$ activated nodes are chosen uniformly at random, the statuses of the neighbours of $v$ are independent. Therefore, the probability that $v$ can see the revolution is
$$
	\sum_{l = \cl{\theta k}}^k \binom{k}{l} \left( \frac{\mathcal{N}_l}{N}\right )^l \left(1-\frac{\mathcal{N}_l}{N} \right)^{k-j} 
		= \mbox{BinCDF}\left(k-\cl{\theta k};k, 1-\frac{\mathcal{N}_l}{N}\right).
$$
It follows that the expected fraction of nodes that have degree $k$ and can see the revolution is
$$
	\mathbb{E}\left[ \rho_k \mbox{BinCDF}\left(k-\cl{\theta k};k, 1-\frac{\mathcal{N}_l}{N}\right) \right],
$$
and hence, the expected fraction of nodes that can see the revolution is
$$
	\mathbb{E}\left[ \frac{\mathcal{V}_l}{N}\right] 
	= \mathbb{E} \left[\sum_k\rho_k \mbox{BinCDF}\left(k-\cl{\theta k};k, 1-\frac{\mathcal{N}_l}{N}\right)\right].
$$
We conclude the proof by observing that by the Continuous Mapping Theorem $\frac{\mathcal{N}_l}{N}\stackrel{p}{\rightarrow}r_j$ and $N\rightarrow\infty$ implies
$$
	\mathbb{E}\left[ \frac{\mathcal{V}_l}{N}\right]  \stackrel{p}{\rightarrow}
	\sum_k\rho_k \mbox{BinCDF}\left(k-\cl{\theta k};k, 1-r_j\right) = v_b(r;\theta,\rho_k)\mbox{ as }N\rightarrow\infty,
$$
and hence,
$$
	v_{e,j} \rightarrow v_b(r_j;\theta,\rho_k)\mbox{ as }N\rightarrow\infty, rep\rightarrow\infty.
$$

\end{document}